%% file: paper.tex
 \documentclass[final,5p,times,twocolumn]{elsarticle}

\usepackage{booktabs}
\usepackage{multirow}

\usepackage{amssymb}
\usepackage{hyperref}

\usepackage[usenames,dvipsnames]{xcolor}

\usepackage{framed}
\usepackage{url}
\usepackage{picture}
\usepackage{cespvect}
\usepackage{subfigure}
\usepackage{picins}
\usepackage{balance}
\usepackage{amsmath}

\def\scale2fig{0.68}

\def\N{N^\prime}

\graphicspath{{./figures/}{./figures_final/}{./figures/gnuplot/}}

\newcommand{\be}{\begin{equation}}
\newcommand{\ee}{\end{equation}}
\newcommand{\R}{\mathcal{R}}
\newcommand{\I}{\mathcal{I}}

\journal{Ad Hoc Networks}

\begin{document}

\begin{frontmatter}



\title{To Compress or Not To Compress: Processing vs Transmission \\ Tradeoffs for Energy Constrained Sensor Networking}

%%\author{Borja Martinez$^\star$, Davide Zordan$^\dag$, Ignasi Vilajosana$^\star$, Michele Rossi$^\dag$\\
%%\thanks{$^\star$Affiliation and address here. $^\dag$Department of Information Engineering, University of Padova. Via Gradenigo 6/B, 35131 Padova, Italy.}} 


\author[DEI]{Davide Zordan}
\author[WS]{Borja Martinez}
\author[WS]{Ignasi Vilajosana}
\author[DEI]{Michele Rossi}

\address[DEI]{Department of Information Engineering, University of Padova.\\ Via Gradenigo 6/B, 35131 Padova, Italy.}
\address[WS]{Worldsensing, Baixada de Gomis 1, 08023 Barcelona, Spain}

%%\thanks{$^\star$Affiliation and address here. $^\dag$}} 
%%\address{}

%%---------------------------------------------------------------------------------------------------------------------
% ABSTRACT
%%---------------------------------------------------------------------------------------------------------------------
\input{abstract}

%%---------------------------------------------------------------------------------------------------------------------
%%---------------------------------------------------------------------------------------------------------------------


%%---------------------------------------------------------------------------------------------------------------------
%%---------------------------------------------------------------------------------------------------------------------

\end{frontmatter}

%%\linenumbers

%-------------------------------------------------------------------------------
% INTRODUCTION
%-------------------------------------------------------------------------------
\input{introduction}
\input{methods}
%\input{methods_rev}

%-------------------------------------------------------------------------------
% RESULTS
%-------------------------------------------------------------------------------
\input{results}
%\input{results_rev}

%-------------------------------------------------------------------------------
% CONCLUSION
%-------------------------------------------------------------------------------
\input{conclusion}

\section*{Acknowledgment}
The work in this paper has been supported in part by the MOSAICS project, �MOnitoring Sensor and Actuator networks through Integrated Compressive Sensing and data gathering�, funded by the University of Padova under grant no. CPDA094077 and by the European Commission under the 7th Framework Programme (SWAP project, GA 251557 and CLAM project, GA 258359). We gratefully acknowledge Paolo Casari for helpful discussions on underwater acoustic communications. The work of Ignasi Vilajosana and Borja Martinez has been supported, in part, by Spanish grants PTQ-08-03-08109 and INN-TU-1558.


%% References

\section*{References}
\bibliographystyle{model1-num-names}
\bibliography{paper}

\input{biographies}
%\balance

\end{document}

%% file: abstract.tex
\begin{abstract}

In the past few years, lossy compression has been widely applied in the field of wireless sensor networks (WSN), where energy
efficiency is a crucial concern due to the constrained nature of the transmission devices. Often, the common thinking among researchers and implementers is that compression is always a good choice, because the major source of energy consumption in a sensor node comes from the transmission of the data. Lossy compression is deemed a viable solution as the imperfect reconstruction of the signal is often acceptable in WSN, subject to some application dependent maximum error tolerance. Nevertheless, this is seldom supported by quantitative evidence. In this paper, we thoroughly review a number of lossy compression methods from the literature, and analyze their performance in terms of compression efficiency, computational complexity and energy consumption. We consider two different scenarios, namely, wireless and underwater communications, and show that signal compression may or may not help in the reduction of the overall energy consumption, depending on factors such as the compression algorithm, the signal statistics and the hardware characteristics, i.e., micro-controller and transmission technology. The lesson that we have learned, is that signal compression may in fact provide some energy savings. However, its usage should be carefully evaluated, as in quite a few cases processing and transmission costs are of the same order of magnitude, whereas, in some other cases, the former may even dominate the latter. In this paper, we show quantitative comparisons to assess these tradeoffs in the above mentioned scenarios (i.e., wireless versus underwater). In addition, we consider recently proposed and lightweight algorithms such as Lightweight Temporal Compression (LTC) as well as more sophisticated FFT- or DCT-based schemes and show that the former are the best option in wireless settings, whereas the latter solutions are preferable for underwater networks. Finally, we provide formulas, obtained through numerical fittings, to gauge the computational complexity, the overall energy consumption and the signal representation accuracy of the best performing algorithms as a function of the most relevant system parameters.

\end{abstract}

%% file: introduction.tex
\section{Introduction}
\label{sec:introduction}

% BMH: Start the introduction with a definition Of Internet Of Thinks
In recent years, wireless sensors and mobile technologies have experienced a tremendous upsurge. Advances in hardware design and micro-fabrication have made it possible to potentially embed sensing and communication devices in every object, from banknotes to bicycles, leading to the vision of the Internet of Things (IoT)~\cite{IoT-Magazine-2010}. It is expected that physical objects in the near future will create an unprecedented network of interconnected physical things able to communicate information about themselves and/or their surroundings and also capable of interacting with the physical environment where they operate.

Wireless Sensor Network (WSN) technology has now reached a good level of maturity and is one of the main enablers for the IoT vision: notable WSN application examples include environmental monitoring~\cite{Szewczyk-2004}, geology~\cite{Werner-2006} structural monitoring~\cite{Xu-2004}, smart grid and household energy metering~\cite{Kappler-2004,SmartMetering-2011}. The basic differences between WSN and IoT are the number of devices, that is expected to be very large for IoT, and their capability of seamlessly communicating via the Internet Protocol, that will make IoT technology pervasive.

The above mentioned applications require the collection and the subsequent analysis of large amounts of data, which are to be sent through suitable routing protocols to some data collection point(s). One of the main problems of IoTs is thus related to the foreseen large number of devices: if this number will keep increasing as predicted in~\cite{Dodson-2003}, and all signs point toward this direction, the amount of data to be managed by the network will become prohibitive. Further issues are due to the constrained nature of IoT devices in terms of limited energy resources (devices are often battery operated) and to the fact that data transmission is their main source of energy consumption. This, together with the fact that IoT nodes are required to remain unattended (and operational) for long periods of time, poses severe constrains on their transmitting capabilities. 

Recently, several strategies have been developed to prolong the lifetime of battery operated IoT nodes. These comprise processing techniques such as data aggregation~\cite{Fasolo-2007}, distributed~\cite{Pattem-2004} or temporal~\cite{Sharaf-2003}  compression as well as battery replenishment through energy harvesting~\cite{Harvesting-Magazine-2010}. The rationale behind data compression is that we can trade some additional energy for compression for some reduction in the energy spent for transmission. As we shall see in the remainder of this paper, if the energy spent for processing is sufficiently smaller than that needed for transmission, energy savings are in fact possible. 

In this paper we focus on the energy saving opportunities offered by data processing and, in particular, on the effectiveness of the {\it lossy temporal compression} of data. With lossy techniques, the original data is compressed by however discarding some of the original information in it, so that at the receiver side the decompressor can reconstruct the original data up to a certain accuracy. Lossy compression makes it possible to trade some reconstruction accuracy for some additional gains in terms of compression ratio with respect to lossless schemes. Note that 1) these gains correspond to further savings in terms of transmission needs, 2) depending on the application, some small inaccuracy in the reconstructed signal can in fact be acceptable, 3) given this, lossy compression schemes introduce some additional flexibility as one can tune the compression ratio as a function of energy consumption criteria.

We note that much of the existing literature has been devoted to the systematic study of lossless compression methods. \cite{Marcelloni-2009} proposes a simple Lossless Entropy Compression (LEC) algorithm, comparing LEC with standard techniques such as gzip, bzip2, rar and classical Huffman and arithmetic encoders. A simple lossy compression scheme, called Lightweight Temporal Compression (LTC)~\cite{2004-Schoellhammer}, was also considered. However, the main focus of this comparison has been on the achievable compression ratio, whereas considerations on energy savings are only given for LEC.
\cite{Byl-2009} examines Huffman, Run Length Encoding (RLE) and Delta Encoding (DE), comparing the energy spent for compression for these schemes. \cite{Liang-2011} treats lossy (LTC) as well as lossless (LEC and Lempel-Ziv-Welch) compression methods, but only focusing on their compression performance. Further work is carried out in~\cite{Sadler-2006}, where the energy savings from lossless compression algorithms are evaluated for different radio setups, in single- as well as multi-hop networks. Along the same lines,~\cite{Barr-2006} compares several lossless compression schemes for a StrongArm CPU architecture, showing the unexpected result that data compression may actually cause an increase in the overall energy expenditure. A comprehensive survey of practical lossless compression schemes for WSN can be found in~\cite{Compression-Survey-2012}. The lesson that we learn from these papers is that lossless compression can provide some energy savings. These are however smaller than one might expect because, for the hardware in use nowadays (CPU and radio), the energy spent for the execution of the compression algorithms (CPU) is of the same order of magnitude of that spent for transmission (radio).

Some further work has been carried out for what concerns lossy compression schemes. LTC~\cite{Liang-2011}, PLAMLiS~\cite{Liu-2007} and the algorithm of~\cite{Pham-2008} are all based on Piecewise Linear Approximation (PLA). Adaptive Auto-Regressive Moving Average (A-ARMA)~\cite{Lu-2010} is instead based on ARMA models (these schemes will be extensively reviewed in the following Section~\ref{sec:compression_methods}). Nevertheless, we remark that no systematic energy comparison has been carried out so far for lossy schemes. In this case, it is not clear whether lossy compression can be advantageous in terms of energy savings  and what the involved tradeoffs are in terms of compression ratio {\it vs} representation accuracy and yet how these affect the overall energy expenditure. In addition, it is unclear whether the above mentioned linear and autoregressive schemes can provide at all advantages as compared with more sophisticated techniques such as Fourier-based transforms, which have been effectively used to compress audio and video signals and for which fast and computationally efficient algorithms exist. In this paper, we fill these gaps by systematically comparing existing lossy compression methods among each other and against polynomial and Fourier-based (FFT and DCT) compression schemes. Our comparison is carried out for two wireless communication setups, i.e., for radio and acoustic modems (used for underwater sensor networking, see, e.g.,~\cite{Casari-2011}) and fitting formulas for the relevant performance metrics are obtained for the best performing algorithms in both cases.

Specifically, the main contributions of this paper are:

\begin{itemize}
\item we thoroughly review lossy compression methods from the literature as well as polynomial, FFT- and DCT-based schemes, quantifying their performance in terms of compression efficiency, computational complexity (i.e., processing energy) and energy consumption for two radio setups, namely, wireless (IEEE~802.15.4) and underwater (acoustic modems) radios. For FFT- and DCT-based methods we propose our own algorithms, which exploit the properties of these transformations.
\item We assess whether signal compression may actually help in the reduction of the overall energy consumption, depending on the compression algorithm, the chosen reconstruction accuracy, the signal statistics and the transmission technology (i.e., wireless versus underwater). In fact, we conclude that signal compression may be helpful; however, in quite a few cases processing and transmission costs are of the same order of magnitude. Also, in some other cases, the former may even dominate the latter. Notably, our results indicate that PLA methods (and in particular among them LTC) are the best option for wireless radios, whereas DCT-based compression is the algorithm of choice for acoustic modems. Thus, the choice of the compression algorithm itself is highly dependent on the energy consumption associated with radio transmission.
\item We provide formulas, obtained through numerical fittings and validated against real datasets, to gauge the computational complexity, the overall energy consumption and the signal representation accuracy of the best performing algorithms in each scenario, as a function of the most relevant system parameters. These can be used to generalize the results obtained here for the selected radio setups to other architectures.
\end{itemize}

The rest of the paper is organized as follows. Section~\ref{sec:compression_methods} discusses modeling techniques from the literature, along with some lossy compression schemes that we introduce in this paper. In Section~\ref{sec:results} we carry out a thorough performance evaluation of all considered methods, whereas our conclusions are drawn in Section~\ref{sec:conclusions}. 

%% file: methods.tex
\section{Lossy Compression for Constrained Sensing Devices}
\label{sec:compression_methods}

In the following, we review existing lossy signal compression methods for constrained sensor nodes, we present an improved ARMA-based compressor and we apply well known FFT and DCT techniques to achieve efficient lossy compression algorithms. We start with the description of what we refer to here as ``adaptive modeling techniques'' in Section~\ref{sec:adaptive_modeling}. Hence, in Section~\ref{sec:fft_based_techniques} we discuss techniques based on Fourier transforms.

\subsection{Compression Methods Based on Adaptive Modeling}
\label{sec:adaptive_modeling}

For the Adaptive Modeling schemes, some signal model is iteratively updated over pre-determined time windows, exploiting the correlation structure of the signal through linear, polynomial or autoregressive methods; thus, signal compression is achieved by sending the model parameters in place of the original data.

%------------------------------------------------------------------------------
% Segment Fitting Methods
%------------------------------------------------------------------------------

\subsubsection{Piecewise Linear Approximations (PLA)}
\label{sec:PLA_intro}

The term Piecewise Linear Approximation (PLA) refers to a family of linear approximation techniques. These build on the fact that, for most time series consisting of environmental measures such as temperature and humidity, linear approximations work well enough over short time frames. The idea is to use a sequence of line segments to represent an input time series $x(n)$ with a bounded approximation error. Further, since a line segment can be determined by only two end points, PLA leads to quite efficient representations of time series in terms of memory and transmission requirements. 

\begin{figure}[ht]
    \centering
    \includegraphics[width=0.3\textwidth]{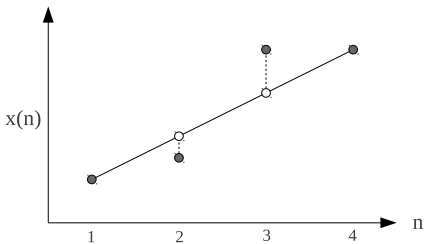}
    \caption{Approximation of a time series $x(n)$ by a segment.}
    \label{fig:segment_fit}
\end{figure}

For the reconstruction at the receiver side, at the generic time $n$ observations are approximated through the vertical projection of the actual samples over the corresponding line segment (i.e., the white-filled dots in Fig.~\ref{fig:segment_fit}). The approximated signal in what follows is referred to as $\hat{x}(n)$. The error introduced is the distance from the actual samples (i.e., the black dots in the figure) to the segment along this vertical projection, i.e., $|\hat{x}(n)-x(n)|$. 
Most PLA algorithms use standard least squares fitting to calculate the approximating line segments. Often, a further simplification is introduced to reduce the computation complexity, which consists of forcing the end points of each line segment to be points of the original time series $x(n)$. This makes least squares fitting unnecessary as the line segments are fully identified by the extreme points of $x(n)$ in the considered time window. Following this simple idea, several methods have been proposed in the literature. Below we review the most significant among them.\\

%------------------------------------------------------------------------------
% LTC Method
%------------------------------------------------------------------------------

\begin{figure}[ht]
\centering
        \subfigure[]{%
            \includegraphics[width=0.3\textwidth]{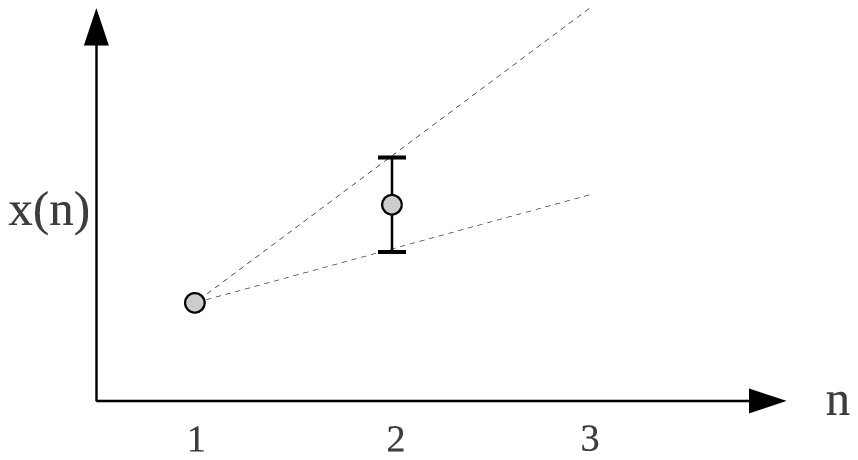}%  
            \label{fig:LTC_a}
        }
        \subfigure[]{%
            \includegraphics[width=0.3\textwidth]{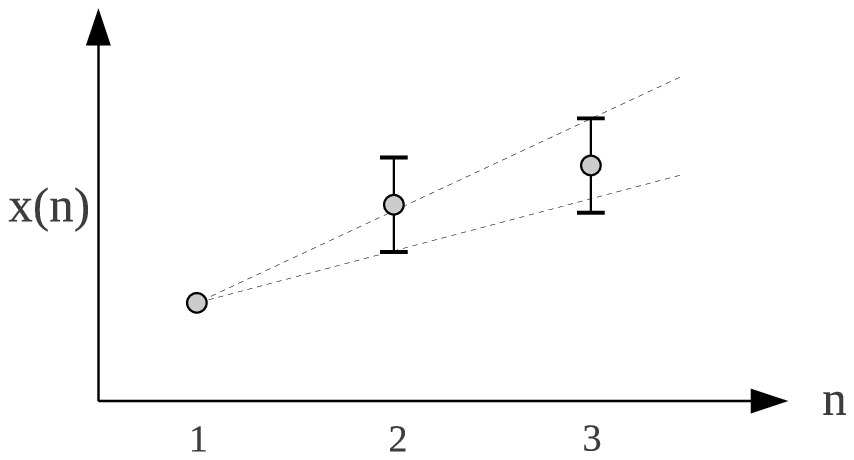}%
            \label{fig:LTC_b}
        }
        \subfigure[]{%
            \includegraphics[width=0.3\textwidth]{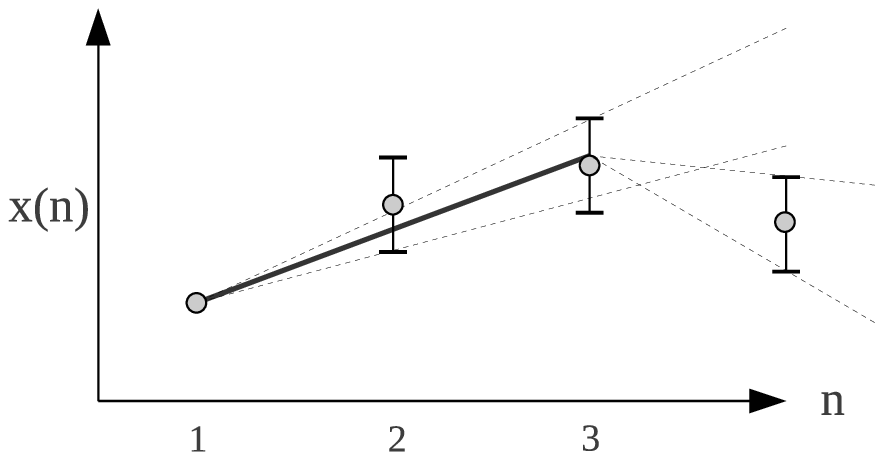}%  
            \label{fig:LTC_c}
        }
    \caption{Lightweight Temporal Compression example.}
    \label{fig:LTC}
\end{figure}

\noindent \textbf{Lightweight Temporal Compression (LTC)~\cite{2004-Schoellhammer}:} the LTC algorithm is a low complexity PLA technique. Specifically, let $x(n)$ be the points of a time series with $n=1,2,\dots,N$. The LTC algorithm starts with $n=1$ and fixes the first point of the approximating line segment to $x(1)$. The second point $x(2)$ is transformed into a vertical line segment that determines the set of all ``acceptable'' lines $\Omega_{1,2}$ with starting point $x(1)$. This vertical segment is centered at $x(2)$ and covers all values meeting a maximum tolerance $\varepsilon \geq 0$, i.e., lying within the interval $[x(2)-\varepsilon, x(2)+\varepsilon]$, see Fig.~\ref{fig:LTC_a}. The set of acceptable lines for $n=3$, $\Omega_{1,2,3}$, is obtained by the intersection of $\Omega_{1,2}$ and the set of lines with starting point $x(1)$ that are acceptable for $x(3)$, see Fig.~\ref{fig:LTC_b}. If $x(3)$ falls within $\Omega_{1,2,3}$ the algorithm continues with the next point $x(4)$ and the new set of acceptable lines $\Omega_{1,2,3,4}$ is obtained as the intersection of $\Omega_{1,2,3}$ and the set of lines with starting point $x(1)$ that are acceptable for $x(4)$. The procedures is iterated adding one point at a time until, at a given step $s$, $x(s)$ is not contained in $\Omega_{1,2,\dots,s}$. Thus, the algorithm sets $x(1)$ and $x(s-1)$ as the starting and ending points of the approximating line segment for $n=1,2,\dots,s-1$ and starts over with $x(s-1)$ considering it as the first point of the next approximating line segment. In our example, $s=4$, see Fig.~\ref{fig:LTC_c}.

When the inclusion of a new sample does not comply with the allowed maximum tolerance, the algorithm starts over looking for a new line segment. Thus, it self-adapts to the characteristics of $x(n)$ without having to fix beforehand the lapse of time between subsequent updates.\\

%------------------------------------------------------------------------------
% PLAMLiS Method
%------------------------------------------------------------------------------

\noindent \textbf{PLAMLiS~\cite{Liu-2007}:} as LTC, PLAMLiS represents the input data series $x(n)$ through a sequence of line segments. Here, the linear fitting problem is converted into a set-covering problem, trying to find the minimum number of segments that cover the entire set of values over a given time window. This problem is then solved through a greedy algorithm that works as follows: let $x(n)$ be the input time series over a window $n=1,2,\dots,N$, with $\mathcal X = \{ x(1),x(2),\dots,x(N) \}$. For each $x(i) \in \mathcal X$ segments are built associating $x(i)$ with $x(j)$ ($j > i$), which is the farthest away from $x(i)$ such that the line segment $(x(i),x(j))$ meets the given error bound $\varepsilon$. That is, the difference between the compressed signal $\hat{x}(k)$ and $x(k)$ is no larger than $\varepsilon$ for $i<k<j$ . Let $\mathcal F_i$ denote the subset consisting of all the points covered by this line segment, formally:
\begin{eqnarray}
\mathcal F_i & = & \big \{x(k) \in X, i \leq k \leq j, s.t., j-i \textrm{ is maximized}, \nonumber \\
 & & \textrm{given } |\hat{x}(k) - x(k) | \leq \varepsilon, \forall \; i < k < j \big \} \, . \nonumber
\end{eqnarray}
After having iterated this for all points in $\mathcal X$ we obtain set $\mathcal F = \{\mathcal F_1,\mathcal F_2,\dots, \mathcal F_N\}$. Now, the PLAMLiS problem amounts to picking the least number of subsets from $\mathcal F$ that cover all the elements in $\mathcal X$, which is the {\it minimum set cover problem} and is known to be NP-complete. The authors of~\cite{Liu-2007} suggest an approximate solution to it through a greedy algorithm. Set $\mathcal F$ is scanned by picking the subset $\mathcal F_i$ that covers the largest number of still uncovered points in $\mathcal X$, this set is then removed from $\mathcal F$ and added to the empty set $\mathcal S$, i.e., $\mathcal F \leftarrow \mathcal F \setminus \mathcal F_i$, $\mathcal S \leftarrow \mathcal S \cup \mathcal F_i$ and the algorithm is reiterated with the new set $\mathcal F$ until all points in $\mathcal X$ are covered. The subsets in $\mathcal S$ define the approximating segments for $x(n)$. %This algorithm runs in $O(N^2 \log(N))$ time and is memory space is $O(N)$.
\\

%------------------------------------------------------------------------------
% Enhanced PLAMLiS METHOD (with Top-Down approach)
%------------------------------------------------------------------------------

\noindent \textbf{Enhanced PLAMLiS~\cite{Pham-2008}:} several refinements to PLAMLiS have been proposed in the literature to reduce its computational cost. In~\cite{Pham-2008} a top-down recursive segmentation algorithm is proposed. As above, consider the input time series $x(n)$ and a time window $n=1,2,\dots,N$. The algorithm starts by taking a first segment $(x(1),x(N))$, if the maximum allowed tolerance $\varepsilon$ is met for all points along this segment the algorithm ends. Otherwise, the segment is split in two segments at the point $x(i)$, $1<i<N$, where the error is maximum, obtaining the two segments $(x(1),x(i))$ and $(x(i),x(N))$. The same procedure is recursively applied on the resulting segments until the maximum error tolerance is met for all points.
%------------------------------------------------------------------------------

\subsubsection{Polynomial Regression (PR)}
\label{sec:PR}

The above methods can be modified by relaxing the constraint that the endpoints of the segments $x(1)$ and $x(N)$ must be actual points from $x(n)$. In this case, polynomials of given order $p \geq 1$ are used as the approximating functions, whose coefficients are found through standard regression methods based on least squares fitting~\cite{Polynomial-book-2003}. Specifically, we start with a window of $p$ samples, for which we obtain the best fitting polynomial coefficients. Thus, we keep increasing the window length of one sample at a time, computing the new coefficients, while the target error tolerance is met.

However, tracing a line between two fixed points as done by LTC and PLAMLiS has a very low computational complexity, while least squares fitting can have a significant cost. Polynomial regression obtains better results in terms of approximation at the cost of higher computational complexities, which increase with the polynomial order, with respect to the linear models of Section~\ref{sec:PLA_intro}. 

%------------------------------------------------------------------------------

\subsubsection{Auto-Regressive (AR) Methods}
\label{sec:auto_regressive}

Auto Regressive (AR) models in their multiple flavors (AR, ARMA, ARIMA, etc.) have been widely used for time series modeling and forecasting in fields like macro-economics or market analysis. The basic idea is to build up a model based on the history of the sampled data, i.e., on its correlation structure.
Many environmental and physical quantities can be modeled through AR, and hence these models are specially indicated for WSN monitoring applications. When used for signal compression AR obtains a model from the input data and sends it to the receiver in place of the actual time series. The reconstructed model is thus used at the data collection point  (the sink) for data prediction until it receives model updates from the sensor nodes. Specifically, each node locally verifies the accuracy of the predicted data values with respect to the collected samples. If the accuracy is within a prescribed error tolerance, the node assumes that the sink can rebuild the data correctly and it does not transmit any data. Otherwise, it computes a new model and communicates the corresponding parameters to the sink. %For each model update, only the model parameters are communicated to the sink. 
\\

\noindent \textbf{Adaptive Auto-Regressive Moving Average (A-ARMA)~\cite{Lu-2010}:} the basic idea of A-ARMA~\cite{Lu-2010} is that of having each sensor node compute an ARMA model based on fixed-size windows of $\N < N$ consecutive samples. Compression is achieved through the transmission of the model parameters to the sink in place of the original data, as discussed above.
In order to reduce the complexity in the model estimation process, adaptive ARMA employs low-order models, whereby the validity of the model being used is checked through a moving window technique. 

The algorithm works as follows. First, once a WSN node has collected $\N$ samples starting from sample $n$, builds an ARMA Model $M^{(n)} = ARMA(p,q,\N,n)$, where the order of $p$ and $q$ of the ARMA process, and the window length $\N$ must be fixed a priori. For this model, the current estimation window goes from step $n$ to step $n+\N-1$ (covering samples $\{x(n),\dots,x(n+\N-1)\}$). Upon collecting the subsequent $K$ samples, $M^{(n)}$ is used to obtain the predicted values $\{\hat{x}(n+\N),\dots,\hat{x}(n+\N+K-1)\}$. Thus, the RMS error between predicted and actual values is computed. If this error is within the allowed error tolerance, the sensor node keeps using its current ARMA model for the next $K$ values, i.e., its prediction window is moved $K$ steps to the right, covering steps $n+\N+K-1$ to $n+\N+2K-1$. In this case, the decoder at the WSN sink uses $M^{(n)}$ to reconstruct the signal from step $n+\N$ to $n+\N+K-1$ obtaining $\{\hat{x}(n+\N),\dots,\hat{x}(n+\N+K-1)\}$. However, if the target tolerance is not met, the node moves its window and recomputes the new ARMA model parameters using the most recent $\N$ samples.\\

\noindent \textbf{Modified Adaptive Auto-Regressive (MA-AR):} the A-ARMA algorithm was designed with the main objective of reducing the complexity in the model estimation process. For each model update, only one estimation is performed at the beginning of the stage, and always over a fixed window of $\N$ samples. A drawback of this approach is that, especially for highly noisy environments, the estimation over a fixed window can lead to poor results when used for forecasting.

In order to avoid this, we hereby propose a modified version of A-ARMA which uses a $p$-order AR model, dividing the time in terms of {\it prediction cycles}, whose length $\N$ is variable and adapts to the characteristics of the signal. Let $n$ be the the time index at the beginning of a prediction cycle. The first $p$ collected samples $\{x(n),\dots,x(n+p-1)\}$ must be encoded and transmitted; these will be used at the receiver to initialize the predictor. Upon collecting sample $x(n+p)$, the $p$ parameters of a AR model $M^{(n,1)}=AR(n,p,1)$ are computed, where $n$ is the starting point of the estimation window and $\N=p+1$ is its window size, i.e., the support points for the estimation are $\{x(n),\dots,x(n+p)\}$). $M^{(n,1)}$ is thus used to predict $\hat{x}(n+p)$, considering $\{x(n),\dots,x(n+p-1)\}$ as initial values. If the target tolerance is met, that is, if $|\hat{x}(n+p)-x(n+p)|<\varepsilon$, the model is temporally stored as valid. When the next sample $x(n+p+1)$ becomes available, a new model $M^{(n,2)}=AR(n,p,2)$ is obtained over the new estimation window $\{x(n),\dots,x(n+p+1)\}$ of size $\N=p+2$. Then, $M^{(n,2)}$ is used to predict $\hat{x}(n+p)$ (one-step ahead) and $\hat{x}(n+p+1)$ (two-step ahead), with the model $M^{(n,2)}$ initialized with the values $\{x(n),\dots,x(n+p-1)\}$, and predicted values are compared with the real samples to check whether $|\hat{x}(n+p+i)-x(n+p+i)| < \varepsilon$ for $i=0,1$. This process is iterated until, for some value $k \geq 1$, $M^{(n,k)}$ is no longer capable of meeting the target reconstruction accuracy for at least one sample: that is, when $|\hat{x}(n+p+i)-x(n+p+i)| > \epsilon$ for at least one $i \in \{0,\dots,k-1\}$. In this case, the last valid model $M^{(n,k-1)}$, with $k-1 \geq 1$, is encoded and transmitted to the decoder at the receiver side, where $M^{(n,k-1)}$ is initialized with $\{x(n),\dots,x(n+p-1)\}$ and used to obtain the estimates $\{\hat{x}(n+p),\dots,\hat{x}(n+p+k-2)\}$. The length of the final estimation window is $\N=p+k-1$. At this point, a new prediction cycle starts with sample $x(n+p+k-1)$ and the new model $M^{(n+p+k-1,1)}$.
 
The key point in our algorithm is the incremental estimation of the AR parameters: only the contribution of the last sample is considered at each iteration in order to refine the AR model. In this way, the computational cost for recomputing a new model for each sample is minimized. The AR parameters can be obtained through least squares minimization. The one-step ahead predictor for an AR process is defined as:
$$\hat{x}(n)= \xi_1 x(n-1) + \dots + \xi_p x(n-p) \, .$$
To simplify the notation, in the following without loosing generality, we assume $n=0$. The least squares method minimizes the total error $\mathcal {E}$ defined as the sum of the individual errors of the one-step ahead predictor for each of the $\N$ samples in the estimation window:
$$\mathcal{E} = \sum_{i=p}^{\N} {(x(i)-\hat{x}(i))}^2 = \sum_{i=p}^{\N} [x(i)- (\xi_1 x(i-1) + \dots + \xi_p x(i-p)) ]^2 \, .$$
Minimizing for each $\xi_k$ yields a set of equations:
$$
{{\partial \mathcal{E}} \over {\partial \xi_k}} = -2 \sum_{i=p}^{\N} [x(i)-(\xi_1 x(i-1) + \dots + \xi_p x(i-p))]x(i-k)=0 \, ,
$$
with $k=1,2,\dots,p$, which can be expressed in terms of the following linear system of equations:
\be
\left(
\begin{array}{cccc}
f(1,1) & f(1,2)  & \cdots & f(1,p) \\
f(2,1) & f(2,2) & \cdots & f(2,p) \\
\vdots & & & \vdots \\
f(p,1) & f(p,2) & \cdots & f(p,p)
\end{array}
\right)
\left(
\begin{array}{c}
\xi_1 \\
\xi_2 \\
\vdots \\
\xi_p 
\end{array}
\right)
=
\left(
\begin{array}{c}
f(1,0) \\
f(2,0) \\ 
\vdots \\
f(p,0)
\end{array}
\right)
\label{eq:AR_LE}
\ee
where $f(r,s) \triangleq \sum_{i=p}^{\N} x(i-r) x(i-s)$.
This is a linear system of $p$ equations and $p$ unknown values (the $\xi_i$ coefficients) that can be solved through a standard Gaussian elimination method. Each entry of the matrix involves $\N-p$ multiplications and $\N-p$ additions. Thus, the estimation of the AR model has complexity $\mathcal O (p^2 (\N-p))$ associated with the matrix construction and $\mathcal O(p^3)$ for solving the linear system (\ref{eq:AR_LE}). Note that, for signals with high temporal correlation $\N-p$ is usually much larger than the AR order $p$ ($p \leq 5$ to bound the computational complexity). In this case, the dominating term is $\mathcal O (p^2 (\N-p))$ and increases with increasing $\N$ and thus, with increasing correlation length, as we will show shortly in Section~\ref{sec:results}.

%Our proposed algorithm performs an AR estimation each time a new sample is available and then, the estimation over the entire signal would have a total complexity of $\mathcal O (\N N p^3)$. However, if at each iteration the state of the matrix is saved, the $\N$ previous multiplications of each sum in the matrix can be reused, only introducing $p^2$ new operations for the new sample $x(n+1)$, leading to $\mathcal O (N p^2)$. %But this is precisely the dominant term, because for our purposes the order $p$ of the AR model will be very small, i.e., $p \leq 4$, whilst the window length $\N$ should be of hundreds of samples.

%%%%%%%%%%%%%%%%%%%%%%%%%%%%
% Fourier based methods
%%%%%%%%%%%%%%%%%%%%%%%%%%%%

\subsection{Compression Methods Based on Fourier Transforms} 
\label{sec:fft_based_techniques}

For Fourier-based techniques, compression is achieved through sending subsets of the FFT or DCT transformation coefficients. Below, we propose some possible methods that differ in how the transformation coefficients are picked.

\subsubsection{Fast Fourier Transform (FFT)}
\label{sub:fft}
The first method that we consider relies on the simplest way to use the Fourier transform for compression. Specifically, the input time series $x(n)$ is mapped to its frequency representation $X(f) \in \mathbb{C}$ through a Fast Fourier Transform (FFT). We define $X_{\R}(f) \triangleq \mathfrak{Re}\{X(f)\}$, and  $X_{\I}(f) \triangleq \mathfrak{Im}\{X(f)\}$ as the real and the imaginary part of $X(f)$, respectively. Since $x(n)$ is a real-valued time series, $X(f)$ is Hermitian, i.e., $X(-f) = \overline{X(f)}$. This symmetry allows the FFT to be stored using the same number of samples $N$ of the original signal. For $N$ even we take $f \in \{f_1,\dots,f_{N/2}\}$ for both $X_{\R}(\cdot)$ and $X_{\I}(\cdot)$, while if $N$ is odd we take $f \in \{f_1,\dots,f_{\lfloor N/2 \rfloor +1}\}$ for the real part and $f \in \{f_1,\dots,f_{\lfloor N/2 \rfloor}\}$ for the imaginary part.

The compressed representation $\hat X(f) \triangleq \hat X_{\R}(f) + j \hat X_{\I}(f)$ will also be in the frequency domain and it is built (for the case of $N$ even) as follows:
\begin{enumerate}
    \item initialize $\hat X_{\R}(f) = 0$ and $\hat X_{\I}(f) = 0$, $\forall \; f \in \{f_1,\dots, f_{N/2} \}$ ;
    \item select the coefficient with maximum absolute value from $X_{\R}$ and $X_{\I}$, i.e., $f^* \triangleq \arg \! \max_{f} \max \{|X_{\R}(f)|,|X_{\I}(f)|\}$ and $M \triangleq \arg \! \max_{i \in \{\R,\I\}}\{|X_{i}(f^*)|\}$.
    \item set $\hat X_{M}(f^*) = X_{M}(f^*)$ and then set $X_{M}(f^*)=0$.
    \item if $\hat x(n)$, the inverse FFT of $\hat X(f)$, meets the error tolerance constraint continue, otherwise repeat from step 2.;
    \item encode the values and the positions of the harmonics stored in $\hat X_{\R}$ and $\hat X_{\I}$. 
\end{enumerate}

Hence, the decompressor at the receiver obtains $\hat X_{\R}(f)$ and $\hat X_{\I}(f)$ and exploits the Hermitian symmetry to reconstruct $\hat X(f)$.

\subsubsection{Low Pass Filter (FFT-LPF)} % (fold)
\label{sub:low_pass_filter}
We implemented a second FFT-based lossy algorithm, which we have termed FFT-LPF. Since the input time series $x(n)$ are in may common cases slow varying signals (i.e., having large temporal correlation) with some high frequency noise superimposed, most of the significant coefficients of $X(f)$ reside in the low frequencies. For FFT-PLF, we start setting $\hat X_{\R}(f) = 0$ for all frequencies. Thus, $X_{\R}(f)$ is evaluated from $f_1$, incrementally moving toward higher frequencies, $f_2,f_3,\dots$. At each iteration $i$, $X_{\R}(f_i)$ is copied onto $\hat X_{\R}(f_i)$ (both real and imaginary part), the inverse FFT is computed taking $\hat X_{\R}(f)$ as input and the error tolerance constraint is checked on the so obtained $\hat x(n)$. If the given tolerance is met the algorithm stops, otherwise it is reiterated for the next frequency $f_{i+1}$.   
% subsubsection low_pass_filter (end)

\subsubsection{Windowing} % (fold)
\label{sub:windowing}
The two algorithms discussed above suffer from an edge discontinuity problem. In particular, when we take the FFT over a window of $N$ samples, if $x(1)$ and $x(N)$ differ substantially the information about this discontinuity is spread across the whole spectrum in the frequency domain. Hence, in order to meet the tolerance constraint for all the samples in the window, a high number of harmonics is selected by the previous algorithms, resulting in a poor compression and in a high number of operations. 

To solve this issue, we implemented a version of the FFT algorithm that considers overlapping windows of $N + 2W$ samples instead of disjoint windows of length $N$, where $W$ is the number of samples that overlap between subsequent windows. The first FFT is taken over the entire window and the selection of the coefficients goes on depending on the selected algorithm (either FFT or FFT-LPF), but the tolerance constraint is only checked on the $N$ samples in the central part of the window.
With this workaround we can get rid of the edge discontinuity problem and encode the information about the $N$ samples of interest with very few coefficients as it will be seen shortly in Section~\ref{sec:results}. As a drawback, the direct and inverse transforms have to be taken on longer windows, which results in a higher number of operations.

% subsection windowing (end)

\subsubsection{Discrete Cosine Transform (DCT)} % (fold)
\label{sub:dct}

We also considered the Discrete Cosine Transform (type II), mainly for three reasons: 1) its coefficients are real, so we did not have to cope with real and imaginary parts, thus saving memory and number of operations; 2) it has a strong ``energy compaction'' property~\cite{Rao-1990}, i.e., most of the signal information tends to be concentrated in a few low-frequency components; 3) the DCT of a signal with $N$ samples is equivalent to a DFT on a real signal of even symmetry with double length, so DCT does not suffer from the edge discontinuity problem.

% subsubsection dct (end)

% section fft_based_techniques (end)

%% file: results.tex
\newtheorem{mydef}{Definition}
\newcommand{\EV}{\mathrm{E}}
\newcommand{\argmin}{\mathrm{argmin}}

\section{Performance Comparison}
\label{sec:results}

The objects of our discussion in this section are:
\begin{itemize}
\item to provide a thorough performance comparison of the compression methods of Section~\ref{sec:compression_methods}. The selected performance metrics are: 1) compression ratio, 2) computational and transmission energy and 3) reconstruction error at the receiver, which are defined below;
\item to assess how the statistical properties of the selected signals impact the performance of the compression methods;
%\item to investigate the suitability of EMD as a tool for data compression in wireless sensor networks: in particular, our interest is on the additional benefits induced by the utilization of EMD in conjunction with the compression schemes of Section~\ref{sec:compression_methods}.  
\item to investigate whether or not data compression leads to energy savings in single- and multi-hop scenarios for: WSN) a wireless sensor network and UWN) an underwater network.
\item to obtain, through numerical fitting, close-formulas which model the considered performance metrics as a function of key parameters.
\end{itemize}

Toward the above objectives, we present simulation results obtained using synthetic signals with varying correlation length. These signals make it possible to give a fine grained description of the performance of the selected techniques, looking comprehensively at the entire range of variation of their temporal correlation statistics. Real datasets are used to validate the proposed empirical fitting formulas.

\subsection{Performance Metrics}
\label{sec:performance_metrics}

Before delving into the description of the results, in the following we give some definitions. 

\begin{mydef}{Correlation length}\\
Given a stationary discrete time series $x(n)$ with $n = 1,2,\dots,N$, we define \textbf{correlation length} of $x(n)$ the smallest value $n^\star$ such that the autocorrelation function of $x(n)$ is smaller than a predetermined threshold $\delta$. The autocorrelation is:
$$\rho_x(n) = \frac{\EV\left[(x(m)-\mu_x)(x(m+n)-\mu_x)\right]}{\sigma_x^2} \; ,$$
where $\mu_x$ and $\sigma_x^2$ are the mean ad the variance of $x(n)$, respectively. Formally, $n^\star$ is defined as: 
$$n^\star = \mathop{\argmin}_{n > 0} \left\lbrace \rho_x(n) < \delta \right\rbrace \; .$$
\end{mydef}

\begin{mydef}{Compression ratio}\\
Given a finite finite time series $x(n)$ and its compressed version $\hat{x}(n)$, we define \textbf{compression ratio} $\eta$ the 
quantity:
$$\eta = \frac{N_b(\hat{x})}{N_b(x)} \;,$$
where $N_b(\hat{x})$ and $N_b(x)$ are the number of bits used to represent the compressed time series $\hat{x}(n)$ and the original one $x(n)$, respectively.
\end{mydef}

\begin{mydef}{Energy consumption for compression}\\
For every compression method we have recorded the number of operations to process the original time series $x(n)$ accounting for the number of additions, multiplications, divisions and comparisons. Thus, depending on selected hardware architecture, we have mapped these figures into the corresponding number of clock cycles and we have subsequently mapped the latter into the corresponding energy expenditure, which is the energy drained from the battery to accomplish the compression task.
\end{mydef}

\begin{mydef}{Transmission Energy}\\
Is the energy consumed for transmission, obtained accounting for the radio chip characteristics and the protocol overhead due to physical (PHY) and medium access (MAC) layers. 
\end{mydef}

\begin{mydef}{Total Energy Consumption}\\
Is the sum of the energy consumption for compression and transmission and is expressed in [Joule]. 
\end{mydef}

In the computation of the energy consumption for compression, we only accounted for the operations performed by the CPU, without considering the possible additional costs of reading and writing in the flash memory of the sensor. For the communication cost we have only taken into consideration the transmission energy, neglecting the cost of turning on and off the radio transceiver and the energy spent at the destination to receive the data. The first are fixed costs that would also be incurred without compression, while the latter can be ignored if the receiver is not a power constrained device. Moreover, for the MAC we do not consider retransmissions due to channel errors or multi-user interference.

\subsection{Hardware Architecture}
\label{sec:architecture}

For both the WSN and the UWN scenarios we selected the TI MSP430~\cite{MSP430-TI-report} micro-controller using the corresponding $16$ bit floating point package for the calculations and for the data representation. In the active state, the MSP430 is powered by a current of $330$ $\mu$A at $2.2$ V and it has a clock rate of $1$ MHz. The resulting energy consumption per CPU cycle is $E_0 = 0.726$ nJ. In Table~\ref{tab:cpu_cycles} the number of clock cycles needed for the floating point operations are listed.

\begin{table}[h!]
\centering
%\begin{tabular}{l c}
\begin{tabular*}{0.8\columnwidth}{@{\extracolsep{\fill}} l c}
\toprule
Operation & Clock cycles \\
\midrule
Addition        	&       184 \\
Subtraction     	&       177 \\
Multiplication  	&       395 \\
Division        	&       405 \\
Comparison  	&       37 \\ 
\bottomrule
\end{tabular*}
\caption{CPU Cycles for the TI MSP430 micro-controller, see Section~5 of~\cite{MSP430-TI-report}.}
\label{tab:cpu_cycles}
\end{table}

For the WSN scenario, we selected the TI CC2420 RF transceiver~\cite{CC2420}, an IEEE~802.15.4~\cite{IEEE802.15.4}  compliant radio. The current consumption for the transmission is $17.4$ mA at $3.3$ V, for an effective data rate of $250$ kbps. Thus, the energy cost associated with the transmission of a bit is $E_{Tx}^\prime = 230$ nJ, which equals the energy spent by the micro-processor during $316$ clock cycles in the active state.

For the UWN scenario, we considered the Aquatec AquaModem~\cite{AquaModem}, an acoustic modem featuring a data rate up to $2000$ bps consuming a power of $20$ W. In this case, the energy spent for the transmission of one bit of data is $E_{Tx}^\prime = 10$~mJ. We remark that the same amount of energy is spent by the micro-processor during $13\cdot 10^{6}$ clock cycles.

\subsection{Generation of Synthetic Stationary Signals}
\label{sec:synthetic_signals}

The stationary synthetic signals have been obtained through a known method to enforce the first and second moments to a white random process, see~\cite{Davies-1987}\cite{Zordan-2011}.
Our objective is to obtain a random time series $x(n)$ with given mean $\mu_x$, variance $\sigma_x^2$ and autocorrelation function $\rho_x(n)$.
The procedure works as follow:
\begin{enumerate}
\item A random Gaussian series $G(k)$ with $k=1,2,\dots,N$ is generated in the frequency domain, where $N$ is the length of the time series $x(n)$ that we want to obtain. Every element of $G(k)$ is an independent Gaussian random variable with mean $\mu_G=0$ and variance $\sigma_G^2 = 1$.
\item The Discrete Fourier Transform (DFT) of the autocorrelation function $\rho_x(n)$ is computed, $S_x(k) = \mathcal{F}[\rho_x(n)]$, where $\mathcal{F}[\cdot]$ is the DFT operator.
\item We compute the entry-wise product $X(k) = G(k) \circ S_x(k)^{\frac{1}{2}}$.
\item The correlated time series $x(n)$ is finally obtained as $\mathcal{F}^{-1}[X(k)]$. 
\end{enumerate}
This is equivalent to filter a white random process with a linear, time invariant filter, whose transfer function is $\mathcal{F}^{-1}[S_x(k)^\frac{1}{2}]$. The stability of this procedure is ensured by a suitable choice for the correlation function, which must be square integrable.
For the simulations in this paper we have used a Gaussian correlation function~\cite{Abrahamsen-1997}, i.e., $\rho_x(n) = \exp\{-a n^2\}$, where $a$ is chosen in order to get the desired correlation length $n^\star$ as follows:
$$a = -\frac{\log(\delta)}{(n^\star)^2} \; .$$
Without loss of generality, we generate synthetic signals with $\mu_x=0$ and $\sigma_x^2=1$. In fact, applying an offset to the generated signals and a scale factor does not change the resulting correlation. For an in deep characterization of the Gaussian correlation function see~\cite{Abrahamsen-1997}. %This model is generally a good model for physical signals, etc. {\bf add a citation}. 

% additive noise
Also, in order to emulate the behavior of real WSN signals, we superimpose a noise to the synthetic signals so as to mimic random perturbations due to limited precision of the sensing hardware and random fluctuations of the observed physical phenomenon. This noise is modeled as a zero mean white Gaussian process with standard deviation $\sigma_{\rm noise}$.

\subsection{Simulation Setup}
\label{sec:simulation_setup}

For the experimental results of the following Sections~\ref{sec:performance_results} and \ref{sub:application_scenarios}, we used synthetic signals with correlation length $n^\star$ varying in $\{1,10,20,50,\dots,500\}$ time slots, where after $20$, $n^\star$ varies in steps of $30$ (we have picked $\delta=0.05$ for all the results shown in this paper). We consider time series of $N=500$ samples (time slots) at a time, progressively taken from a longer realization of the signal, so as to avoid artifacts related to the generation technique. Moreover, a Gaussian noise with standard deviation $\sigma_{\rm noise} = 0.04$ has been added to the signal, as per the signal generation method of Section~\ref{sec:synthetic_signals}. For the reconstruction accuracy, the absolute error tolerance has been set to $\varepsilon = \xi \sigma_{\rm noise}$, with $\xi \geq 0$.
In the following graphs, each point is obtained by averaging the outcomes of $10^4$ simulation runs. For a fair comparison, the same realization of the input signal $x(n)$ has been used for all the compression methods, for each simulation run and value of $n^\star$.

\subsection{Compression Ratio vs Processing Energy}
\label{sec:performance_results}

% fig.1 Adaptive modeling methods
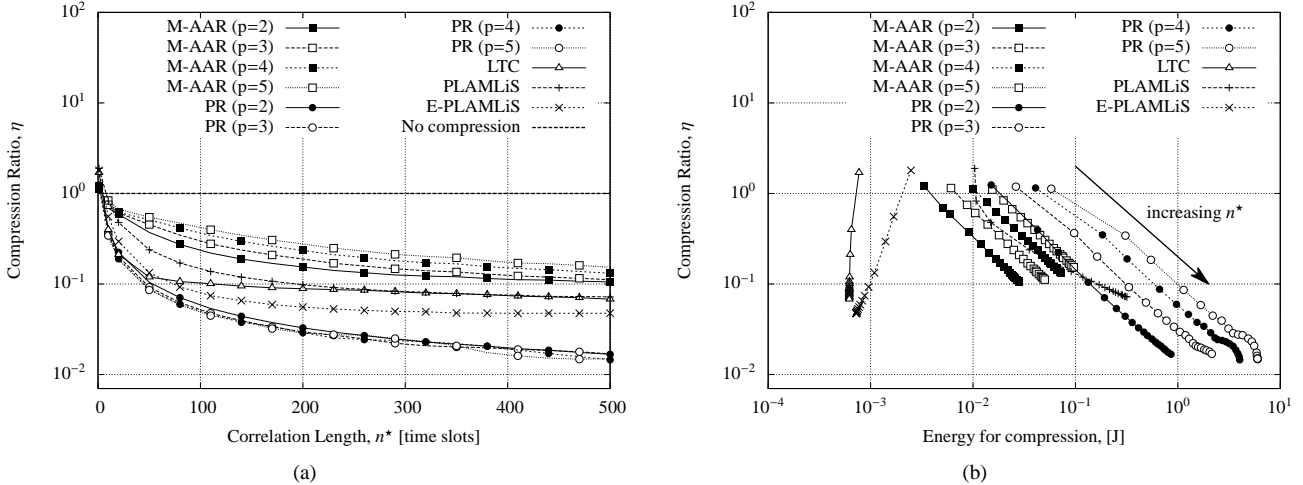
\begin{figure*}[t]
    \begin{center}
        \subfigure[]{%
            \scalebox{\scale2fig}{\input{fig_1a.tex}}%
            \label{fig:AM_perf_a}
        }
        \subfigure[]{%
            \scalebox{\scale2fig}{\input{fig_1b.tex}}%
            \label{fig:AM_perf_b}
        }
    \end{center}
    \caption{(a) $\eta$ {\it vs} Correlation Length $n^\star$ and (b) $\eta$ {\it vs} Energy consumption for compression for the Adaptive Modeling methods for fixed $\varepsilon=4 \sigma_{\rm noise}$.}
    \label{fig:AM_perf}
\end{figure*}

In the following, we analyze the performance in terms of compression effectiveness and computational complexity (energy) for the lossy compression methods of Section~\ref{sec:compression_methods}.\\

\noindent \textbf{Adaptive Modeling Methods:} in this first set of results we compare the performance of the following  compression methods: 1) Modified Adaptive Autoregressive (M-AAR); 2) Polynomial Regression (PR); 3) Piecewise Linear Approximation (PLAMLiS); 4) Enhanced Piecewise Linear Approximation (E-PLAMLiS) and 5) Lightweight Temporal Compression (LTC). For the M-AAR autoregressive filter and the polynomial regression (PR) we used four different orders, namely, $p=\{2,3,4,5\}$. 

Fig.~\ref{fig:AM_perf_a} shows the Compression Ratio achieved by the five compression methods as a function of the correlation length $n^\star$. These results reveal that for small values of $n^\star$ the compression performance is poor for all compression schemes, whereas it improves for increasing correlation length, by reaching a floor value for sufficiently large $n^\star$. This confirms that $n^\star$ is a key parameter for the performance of all schemes. Also, the compression performance differs among the different methods, with PR giving the best results. This reflects the fact that, differently from all the other methods, PR approximates $x(n)$ without requiring its fitting curves to pass from the points of the given input signal. This entails some inherent filtering, that is embedded in this scheme and makes it more robust against small and random perturbations. 

Fig.~\ref{fig:AM_perf_b} shows the energy consumption for compression. For increasing values of $n^\star$ the compression ratio becomes smaller for all schemes, but their energy expenditure substantially differs. Notably, the excellent compression capabilities of PR are counterbalanced by its demanding requirements in terms of energy. M-AAR and PLAMLiS also require a quite large  amount of processing energy, although this is almost one order of magnitude smaller than that of PR. LTC and E-PLAMLiS have the smallest energy consumption among all schemes.

We now discuss the dependence of the computational complexity (which is strictly related to the energy spent for compression) on $n^\star$. LTC encodes the input signal $x(n)$ incrementally, starting from the first sample and adding one sample at a time. Thus, the number of operations that it performs only weakly depends on the correlation length and, in turn, the energy that it spends for compression is almost constant with varying $n^\star$. E-PLAMLiS takes advantage of the increasing correlation length: as the temporal correlation increases, this method has to perform fewer ``divide and reiterate'' steps, so the number of operations required gets smaller and, consequently, also the energy spent for compression. 
For the remaining methods the complexity grows with $n^\star$. For PLAMLiS, this is due to the first step of the algorithm, where for each point the longest segment that respects the given error tolerance has to be found, see Section~\ref{sec:compression_methods}. When $x(n)$ is highly correlated, these segments become longer and PLAMLiS has to check a large number of times the tolerance constraint for each of the $N$ samples of $x(n)$.
For M-AAR and PR every time a new sample is added to a model (autoregressive for the former and polynomial for the latter), this model must be updated and the error tolerance constraint has to be checked. These tasks have a complexity that grows with the square of the length of the current model. Increasing the correlation length of the input time series also increases the length of the models, leading to smaller compression ratios and, in turn, a higher energy consumption.\\

% fig.2 Fourier-based methods
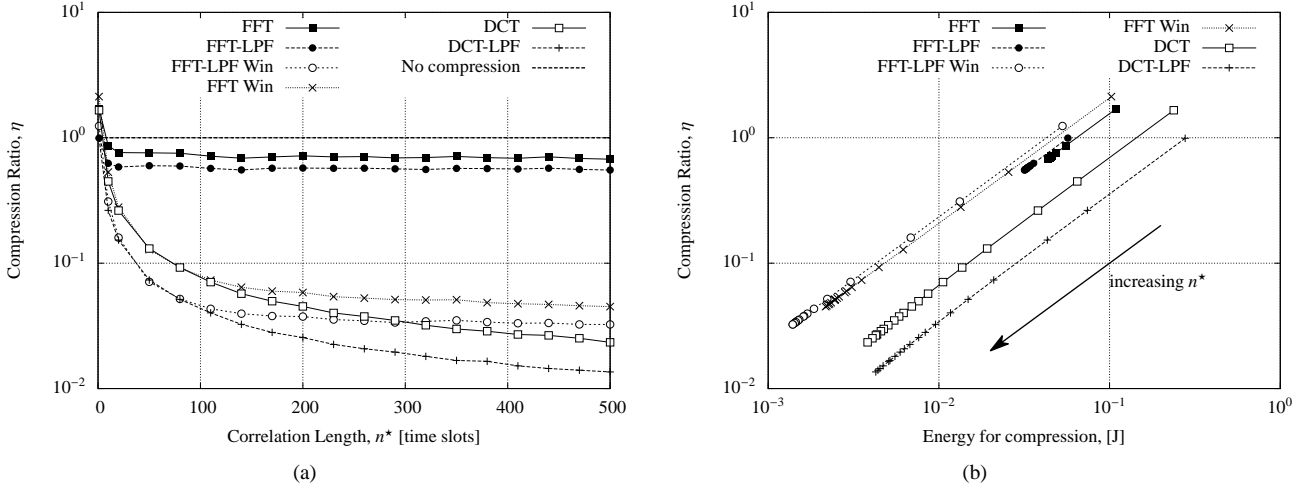
\begin{figure*}[t]
    \begin{center}
        \subfigure[]{%
            \scalebox{\scale2fig}{\input{fig_2a.tex}}%
            \label{fig:FFT_perf_a}
        }
        \subfigure[]{%
            \scalebox{\scale2fig}{\input{fig_2b.tex}}%
            \label{fig:FFT_perf_b}
        }
    \end{center}
    \caption{(a) $\eta$ {\it vs} Correlation Length $n^\star$ and (b) $\eta$ {\it vs} Energy consumption for compression for the Fourier-based methods for fixed $\varepsilon=4 \sigma_{\rm noise}$.}
    \label{fig:FFT_perf}
\end{figure*}

\begin{figure*}[t]
    \begin{center}
        \subfigure[]{%
            \scalebox{\scale2fig}{\input{fig_3a.tex}}%
            \label{fig:cr_toten_WSN}
        }
        \subfigure[]{%
            \scalebox{\scale2fig}{\input{fig_3b.tex}}%
            \label{fig:cr_toten_UWN}
        }
    \end{center}
    \caption{Compression Ratio $\eta$ {\it vs} Total Energy Consumption for the two single-hop scenarios considered: (a) WSN and (b) UWN.}
    \label{fig:cr_toten}
\end{figure*}
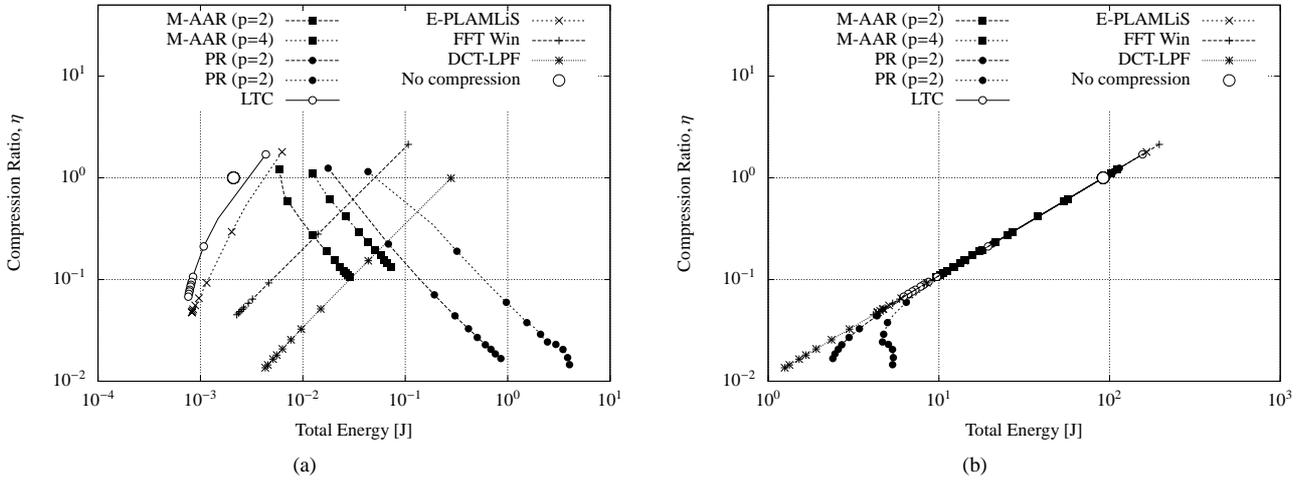

\begin{figure*}[t]
    \begin{center}
        \subfigure[]{%
            \scalebox{\scale2fig}{\input{fig_4a.tex}}%
            \label{fig:eg_cl_WSN}
        }
        \subfigure[]{%
            \scalebox{\scale2fig}{\input{fig_4b.tex}}%
            \label{fig:eg_cl_UWN}
        }
    \end{center}
    \caption{Energy Gain {\it vs} Correlation Length $\eta^\star$ for the two single-hop scenarios considered: (a) WSN and (b) UWN.}
    \label{fig:eg_cl}
\end{figure*}
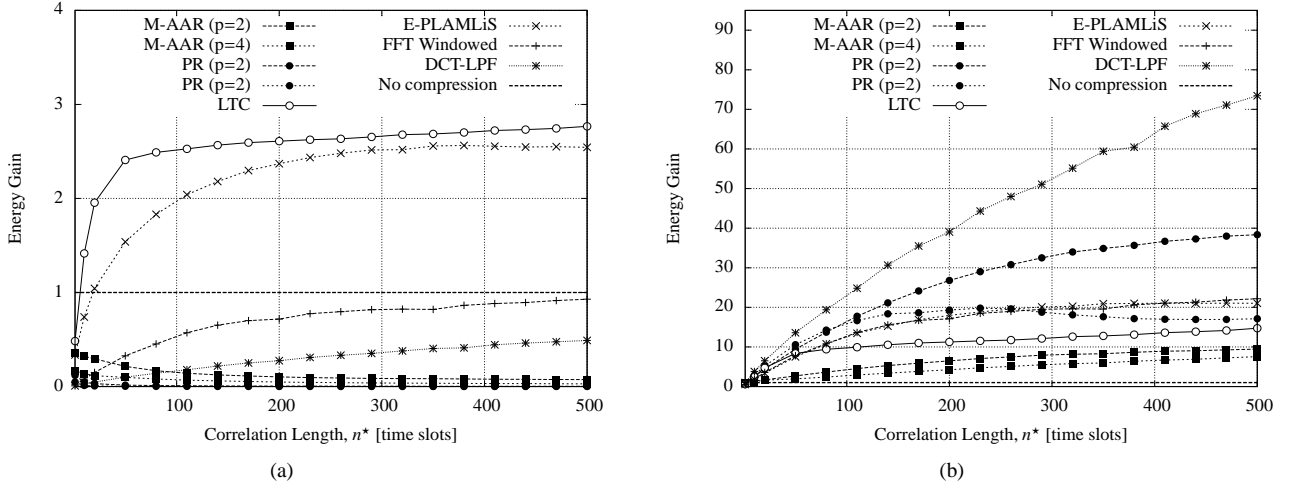

\noindent \textbf{Fourier-based Methods:} we now analyze the performance of the Fourier-based compression schemes of Section~\ref{sec:compression_methods}. We consider the same simulation setup as above. Fig.~\ref{fig:FFT_perf_a} shows that also with Fourier-based methods the compression performance improves with increasing $n^\star$. The methods that perform best are FFT Windowed, FFT-LPF Windowed and DCT-LPF, which achieve very small compression ratios, e.g., $\eta$ is around $10^{-2}$ for $n^\star \geq 300$. Conversely, FFT and FFT-LPF, due to their edge discontinuity problem (see Section~\ref{sec:compression_methods}), need to encode more coefficients to meet the prescribed error tolerance constraint and thus their compression ratio is higher, i.e., around $10^{-1}$. The energy cost for compression is reported in Fig.~\ref{fig:FFT_perf_b}, where $n^\star$ is varied as an independent parameter. The compression cost for these schemes is given by a first contribution, which represents the energy needed to evaluate the FFT/DCT of the input signal $x(n)$. Thus, there is a second contribution which depends on the number of transformation coefficients that are picked. Specifically, a decreasing $n^\star$ means that the signal is less correlated and, in this case, more coefficients are to be considered to meet a given error tolerance. Further, for each of them, an inverse transform has to be evaluated to check whether an additional coefficient is required. This leads to a decreasing  computational cost for increasing $n^\star$. 

As a last observation, we note that FFT-based methods achieve the best performance in terms of compression ratio among all schemes of Figs.~\ref{fig:AM_perf_b} and~\ref{fig:FFT_perf_b} (DCT-LPF is the best performing algorithm), whereas PLA schemes give the best performance in terms of energy consumption for compression (LTC is the best among them). 

\subsection{Application Scenarios} 
\label{sub:application_scenarios}

As discussed above, we evaluated the selected compression methods considering the energy consumed for transmission of typical radios in Wireless Sensor Networks (WSN) and an Underwater Networks (UWN). In the following, we discuss the performance for these application scenarios in single- as well as multi-hop networks.\\

\noindent \textbf{Single-hop Performance:} Fig.~\ref{fig:cr_toten} shows the performance in terms of Compression Ratio $\eta$ {\it vs} Total Energy Consumption for a set of compression methods when applied in the two selected application scenarios. PLAMLiS was not considered as its performance is always dominated by E-PLAMLiS and we only show the performance of the best Fourier-based schemes. In both graphs the large white dot represent the case where no compression is applied to the signal, which is entirely sent to the gathering node. Note that energy savings can only be obtained for those cases where the total energy lies to the left of the no compression case.

In the WSN scenario, the computational energy is comparable to the energy spent for transmission, thus, only LTC and Enhanced PLAMLiS can achieve some energy savings (see Fig.\ref{fig:cr_toten_WSN}). All the other compression methods  entail a large number of operations and, in turn, perform worse than the no compression case in terms of overall energy expenditure. For the UWN scenario, the energy spent for compression is always a negligible fraction of the energy spent for transmission. For this reason, every considered method provides some energy savings, which for PR and Fourier methods can be substantial. 

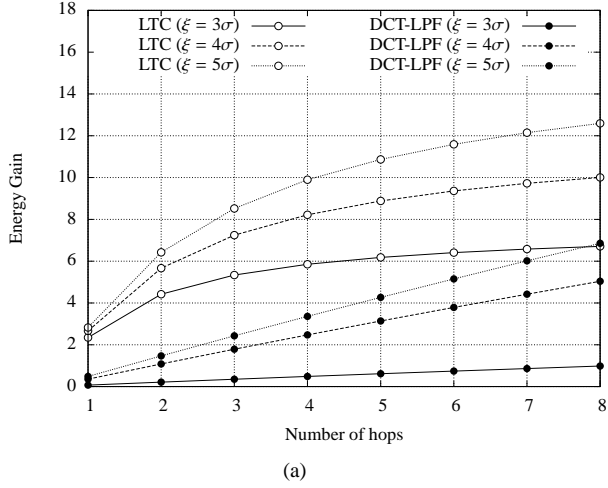
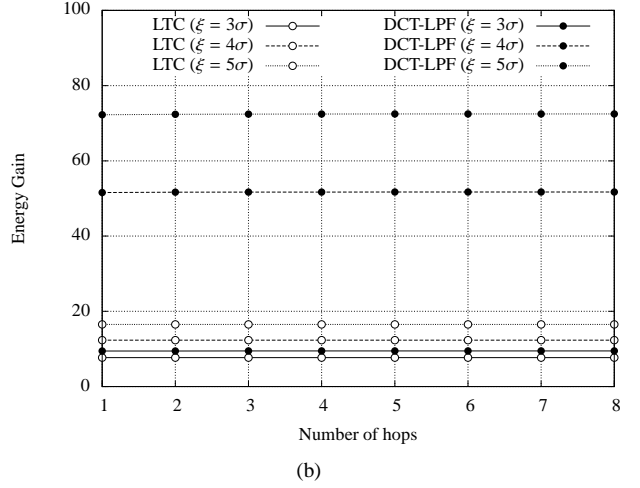
\begin{figure*}[t]
    \begin{center}
        \subfigure[]{%
            \scalebox{\scale2fig}{\input{fig_6a.tex}}%
            \label{fig:multihop_WSN}
        }
        \subfigure[]{%
            \scalebox{\scale2fig}{\input{fig_6b.tex}}%
            \label{fig:multihop_UWN}
        }
    \end{center}
    \caption{Energy Gain {\it vs} number of hops for the two multi-hop scenarios considered: (a) WSN and (b) UWN.}
    \label{fig:multihop}
\end{figure*}

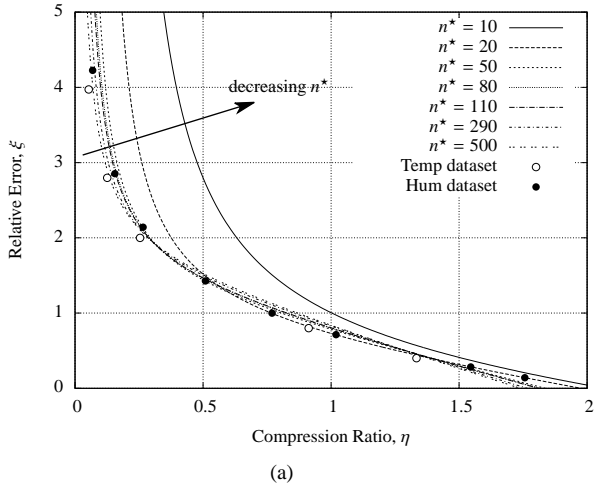
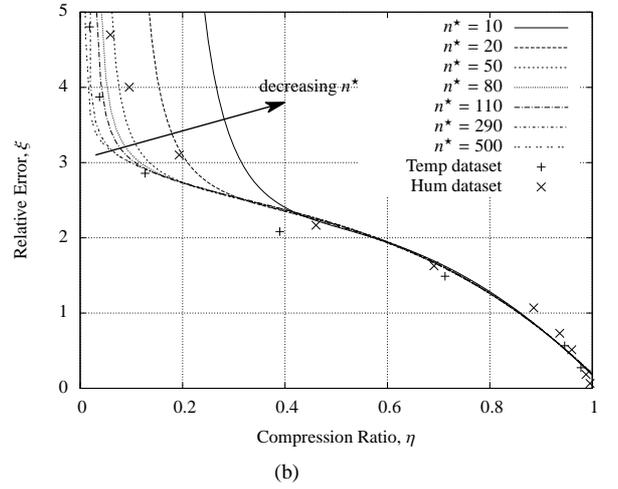
\begin{figure*}[t]
    \begin{center}
        \subfigure[]{%
            \scalebox{\scale2fig}{\input{fig_ltc_err_fit.tex}}%
            \label{fig:fitting_a}
        }
        \subfigure[]{%
            \scalebox{\scale2fig}{\input{fig_dct_err_fit.tex}}%
            \label{fig:fitting_b}
        }
    \end{center}
    \caption{Fitting functions $\xi(n^\star,\eta)$ {\it vs} experimental results: (a) LTC, (b) DTC-LPF.}
    \label{fig:fitting}
\end{figure*}

The total energy gain, defined as the ratio between the energy spent for transmission in the case with no compression and the total energy spent for compression and transmission using the selected compression techniques, is shown in Fig.~\ref{fig:eg_cl}.

In the WSN scenario, i.e., Fig~\ref{fig:eg_cl_WSN}, the method that offers the highest energy gain is LTC, although other methods such as DCT-LPF can achieve better compression performance (see Fig~\ref{fig:cr_toten_WSN}). Note that in this scenario the total energy is highly influenced by the computational cost. Thus, the most lightweight methods, such as LTC and enhanced PLAMLiS, perform best. In the UWN case, whose results are shown in Fig~\ref{fig:eg_cl_UWN}, the computational cost is instead negligible with respect to the energy spent for transmission. As a consequence, the energy gain is mainly driven by the achievable compression ratio and the highest energy gain is obtained with DCT-LPF. In this scenario, PR, which is computationally demanding, can lead to large energy savings too, whereas the energy gain that can be obtained with more lightweight schemes, such as LTC, is quite limited.\\

\noindent \textbf{Multi-hop Performance:} in Fig.~\ref{fig:multihop} we focus on multi-hop networks, and evaluate whether further gains are possible when the compressed information has to travel multiple hops to reach the data gathering point. Both WSN and UWN scenarios are considered. In this case, both transmitting and receiving energy is accounted for at each intermediate relay node. Only LTC and DCT-LPF are shown, as these are the two methods that respectively perform best in the WSN and UWN scenarios. Their performance is computed by varying the error tolerance $\varepsilon \in \{ 3\sigma_{noise},4\sigma_{noise},5\sigma_{noise} \}$, whereas the correlation length is fixed to $n^\star = 300$.

For the WSN scenario the energy gain increases with the number of hops for both compression schemes. As we have already discussed, in this case the energy spent for the compression at the source node is comparable to the energy spent for the transmission. The compression cost (compression energy) is only incurred at the source node, whereas each additional relay node only needs to send the compressed data. This leads to an energy gain that is increasing with the number of hops involved. We also note that DCT-LPF is not energy efficient in single-hop scenarios, but it can actually provide some energy gains when the number of hops is large enough (e.g., larger than $2$ for $\varepsilon \in \{ 4\sigma_{noise},5\sigma_{noise} \}$, see Fig.~\ref{fig:multihop_WSN}).

Conversely, in the UWN scenario the energy spent for compression is a negligible fraction of the energy spent for transmission. Henceforth, the overall energy gain over multiple hops is nearly constant and equal to the energy savings achieved over the first hop.

%In the following simulation we selected four compression methods, namely LTC, PR (with $p=2$), DCT-LPF and FFT-LPF Windowed, and analyzed their performance when applied with different tolerance constraint $\varepsilon$. We fixed the correlation length of the input signals to three distinct values, namely $n^\star = \{100, 300, 500\}$ while for the tolerance we used $\varepsilon = \xi \sigma_{noise}$ with $\xi = \{0.1, 0.2, 0.5, 0.7, 1, 1.5, 2, 3, 4, 5\}$. The results in Fig.~\ref{fig:cr_toten_err}, show the performance of the four selected methods in terms of compression ratio and total energy consumption for the WSN and the UWN scenario. For all the methods considered the compression ratio becomes smaller as the tolerance increases. Small values of $\varepsilon$ force the compression methods to encode also the noise in order to met the tolerance constraint. The compression performance improves when the tolerance constraint is less stringent, i.e., when $\varepsilon > \sigma_{noise}$. In this case, the compression methods can avoid to encode the noise and retain only the useful information of the input signals.\\

\subsection{Numerical Fittings} 
\label{sec:numerical_fittings}

%In this section we provide a performance analysis for the two best performing compression methods, namely, LTC and DCT-LPF by varying $\varepsilon$. The correlation length is set as $n^\star = \{100, 300, 500\}$, while for the error tolerance we used $\varepsilon = \xi \sigma_{noise}$ with $\xi = \{0.1, 0.2, \dots , 6\}$. 

In the following, we provide close-formulas to accurately relate the achievable compression ratio $\eta$ to the relative error tolerance $\xi$ and the computational complexity, $N_c$, which is expressed in terms of number of clock cycles per bit to compress the input signal $x(n)$. These fittings have been computed for the best compression methods, namely, LTC and DCT-LPF. 

Note that until now we have been thinking of $\eta$ as a performance measure which depends on the chosen error tolerance $\varepsilon=\xi \sigma_{noise}$. This amounts to considering $\xi$ as an input parameter for the compression algorithm. In the following, we approximate the mathematical relationship between $\eta$ and $\xi$, by conversely thinking of $\xi$ as a function of $\eta$, which is now our input parameter. $N_c$ can as well be expressed as a function of $\eta$. 

We found these relationships through numerical fitting, running extensive simulations with synthetic signals. The relative error tolerance $\xi$ can be related to the compression ratio $\eta$ through the following formulas:
\be
\xi (n^\star, \eta) =  
\begin{cases} 
\displaystyle \frac{p_1\eta^2 + p_2\eta + p_3}{\eta + q_1}  & \textrm{LTC} \\ 
 \displaystyle \frac{p_1\eta^4 + p_2\eta^3 + p_3\eta^2 + p_4\eta + p_5}{\eta + q_1}  & \textrm{DCT-LPF} \, ,
\end{cases} 
\label{eq:fitting_xi}
\ee
where the fitting parameters $p_1,p_2,p_3,p_4,p_5,$ and $q_1$ depend on the correlation length $n^\star$ and are given in Table~\ref{tab:fitting_xi}  for LTC and DCT-LPF. These fitting formulas have been validated against real world signals measured from the environmental monitoring WSN testbed deployed on the ground floor of the Department of Information Engineering (DEI), University of Padova, Italy~\cite{Crepaldi-07}. This dataset consists of measures of temperature and humidity, sensed with a sampling interval of $1$ minute for $6$ days. Correlation lengths are $n^\star_T = 563$ and $n^\star_H = 355$ for temperature and humidity signals, respectively. The empirical relationships of Eq.~(\ref{eq:fitting_xi}) are shown in Fig.~\ref{fig:fitting_a} and~\ref{fig:fitting_b} through solid and dashed lines, whereas the markers indicate the performance obtained applying LTC and DCT-LPF to the considered real datasets. As can be noted from these plots, although the numerical fitting was obtained for synthetic signals,  Eq.~(\ref{eq:fitting_xi}) closely represents the actual tradeoffs. Also, with decreasing $n^\star$ the curves relating $\xi$ to $\eta$ remain nearly unchanged in terms of functional shape but are shifted toward the right. Finally, we note that the dependence on $n^\star$ is particularly pronounced at small values of $n^\star$, whereas the curves tend to converge for increasing correlation length (larger than $110$ in the figure).
\begin{table*}[t]
\centering
\begin{tabular*}{0.9\textwidth}{@{\extracolsep{\fill}}  c | c | c c c c c c}
\toprule
Compression          & \multirow{2}{*}{$n^\star$} & \multicolumn{6}{c}{Fitting coefficients}  \\
Method               &       & $p_1$      & $p_2$      & $p_3$      & $p_4$ & $p_5$ & $q_1$ \\
\midrule
\multirow{6}{*}{LTC} & $10 $ & $-0.35034$ &  $0.27640$ &  $0.92834$ & --    & --    & $-0.15003$ \\ 
                     & $20 $ & $-0.51980$ &  $0.86851$ &  $0.31368$ & --    & --    & $-0.09245$ \\ 
                     & $50 $ & $-0.80775$ &  $1.38842$ &  $0.17465$ & --    & --    & $-0.03705$ \\ 
                     & $80 $ & $-0.85691$ &  $1.45560$ &  $0.18208$ & --    & --    & $-0.02366$ \\ 
                     & $110$ & $-0.86972$ &  $1.46892$ &  $0.19112$ & --    & --    & $-0.01736$ \\ 
                     & $290$ & $-0.97242$ &  $1.61970$ &  $0.17280$ & --    & --    & $-0.00747$ \\ 
                     & $500$ & $-1.03702$ &  $1.70305$ &  $0.17466$ & --    & --    & $ 0.00267$ \\
\midrule
\multirow{6}{*}{DCT-LPF} & $10 $ & $ 2.05351$ & $-12.70381$ & $14.49624$ & $-4.52198$ & $ 0.82292$ & $-0.16165$ \\
                         & $20 $ & $-0.92752$ & $-3.07506$  & $ 3.07560$ & $ 1.06902$ & $ 0.02898$ & $-0.09025$ \\
                         & $50 $ & $-1.90344$ & $-0.17491$  & $-0.13500$ & $ 2.43821$ & $-0.03826$ & $-0.03929$ \\
                         & $80 $ & $-2.59629$ & $ 1.41404$  & $-1.40970$ & $ 2.81971$ & $-0.04122$ & $-0.02667$ \\
                         & $110$ & $-2.57150$ & $ 1.43655$  & $-1.51646$ & $ 2.87138$ & $-0.02747$ & $-0.01913$ \\
                         & $290$ & $-3.43806$ & $ 3.17964$  & $-2.67444$ & $ 3.13226$ & $-0.01531$ & $-0.00848$ \\
                         & $500$ & $-3.99007$ & $ 4.17811$  & $-3.22636$ & $ 3.22590$ & $-0.01102$ & $-0.00560$ \\
\bottomrule
\end{tabular*}
\caption{Fitting coefficients for $\xi(n^\star,\eta)$.}
\label{tab:fitting_xi}
\end{table*}

\begin{figure}[t]
    \begin{center}
            \scalebox{\scale2fig}{\input{fig_fit_b.tex}}
    \end{center}
    \caption{Fitting functions $N_c(\eta)$ {\it vs} experimental results.}
    \label{fig:Nc}
\end{figure}
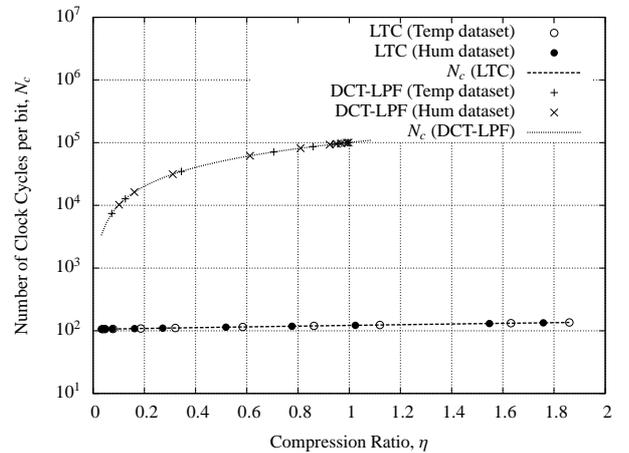

For the computational complexity, we found that $N_c$ scales linearly with $\eta$ for both LTC and DCT-LPF. Hence, $N_c$ can be expressed through a polynomial as follows:
$$
N_c (n^\star, \eta) = \alpha \eta + \gamma n^\star + \beta \; .
$$
$N_c$ exhibits a linear dependence on both $n^\star$ and $\eta$; the fitting coefficients are shown in Table~\ref{tab:fitting_Nc}. 
\begin{table}[h!]
\centering
\begin{tabular*}{\columnwidth}{@{\extracolsep{\fill}}  c c c c}
\toprule
Compression          & \multicolumn{3}{c}{Fitting coefficients}  \\
Method                   & $\alpha$ & $\beta$ & $\gamma$ \\
\midrule
LTC     		&  $16.1$ & $105.4$ &  $3.1 \cdot 10^{-16}$ \\ 
DCT-LPF     	&  $48.1 \cdot 10^3$ & $82.3$ &  $-2 \cdot 10^{-13}$ \\ 
\bottomrule
\end{tabular*}
\caption{Fitting coefficients for $N_c(n^\star,\eta)$.}
\label{tab:fitting_Nc}
\end{table}
Note that the dependence on $n^\star$ is much weaker than that on $\eta$ and for practical purposes can be neglected without loss of accuracy. In fact, for DCT-LPF there is a one-to-one mapping between any target compression ratio and the number of DCT coefficients that are to be sent to achieve this target performance (the computational complexity is directly related to this number of coefficients). Note that, differently from Fig.~\ref{fig:FFT_perf}, this reasoning entails the compression of our data without fixing beforehand the error tolerance $\varepsilon$. For LTC, the dominating term in the total number of operations performed is $\eta$, as this term is directly related to the number of segments that are to be processed. For this reason, in the remainder of this section we consider the simplified relationship:
\be
N_c (\eta) = \alpha \eta + \beta \; .
\label{eq:Nc_simplified}
\ee
The accuracy of Eq.~(\ref{eq:Nc_simplified}) is verified in Fig.~\ref{fig:Nc}, where we plot our empirical approximations against the results obtained for the real world signals described above. The overall energy consumption is obtained as $N_b(x) N_c(\eta) E_0$.\\

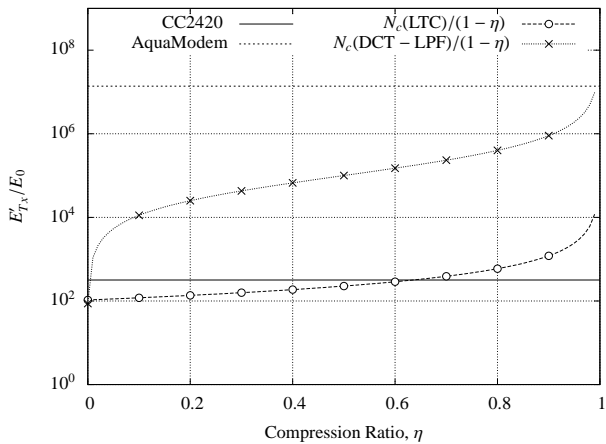
\begin{figure}[t]
    \begin{center}
            \scalebox{\scale2fig}{\input{fig_toff.tex}}
    \end{center}
    \caption{Energy savings assessment {\it vs} $\eta$ and hardware architecture.}
    \label{fig:tradeoff}
\end{figure}

\noindent \textbf{Tradeoffs:} in the following, we use the above empirical formulas to generalize our results to any processing and transmission technology, by separating out technology dependent and algorithmic dependent terms. Specifically, a compression method is energy efficient when the overall cost for compression ($E_c(x)$) and transmission of the compressed data ($E_{Tx}(\hat x)$) is strictly smaller than the cost that would be incurred in transmitting $x(n)$ uncompressed ($E_{Tx}(x)$). Mathematically, $E_c(x) + E_{Tx}(\hat x) < E_{Tx}(x)$. Dividing both sides of this inequality by $E_{Tx}(x)$ and rearranging the terms leads to:
$$
\frac{E_{Tx}(x)}{E_c(x)} = \frac{E^\prime_{Tx} N_b(x)}{E_0 N_c N_b(x)} > \frac{1}{1-\eta} \; ,
$$
where the energy for transmission $E_{Tx}(x)$ is expressed as the product of the energy expenditure for the transmission of a bit $E_{Tx}^\prime$ and the number of bits of $x(n)$, $N_b(x)$. The energy for compression is decomposed in the product of three terms: 1) the energy spent by the micro-controller in a clock cycle $E_0$, 2) the number of clock cycles performed by the compression algorithm per (uncompressed) bit of $x(n)$, $N_c$ and 3) the number of bits composing the input signal $x(n)$, $N_b(x)$. With these energy costs and the above fitting Eq.~(\ref{eq:Nc_simplified}) for $N_c$ we can rewrite the above inequality so that the quantities that depend on the selected hardware architecture appear on the left hand side, leaving those that depend on algorithmic aspects on the right hand side. The result is:
\be
\frac{E^\prime_{Tx}}{E_0} > \frac{N_c (\eta)}{1-\eta} = \frac{\alpha \eta + \beta}{1-\eta}\; ,
\label{eq:tradeoff}
\ee
where $\alpha$ and $\beta$ are the algorithmic dependent fitting parameters indicated in Table~\ref{tab:fitting_Nc}. Eq.~(\ref{eq:tradeoff}) can be used to assess whether a compression scheme is suitable for a specific device architecture. A usage example is shown in Fig.~\ref{fig:tradeoff}. In this graph, the curves with markers are obtained plotting the right hand side of Eq.~(\ref{eq:tradeoff}) for $\eta < 1$, whereas the lines refer to the expression on the left hand side of Eq.~(\ref{eq:tradeoff}) for our reference scenarios, i.e., WSN (solid line) and UWN (dashed line). In the WSN scenario, $E_{Tx}^\prime = 230$~nJ for the selected CC2420 radio, whereas for the TI MSP430 we have $E_0 = 0.726$~nJ and their ratio is $E_{Tx}^\prime/E_0 \simeq 316$. The graph indicates that, in this case, DCT-LPF is inefficient for any value of $\eta$, whereas LTC provides energy savings for $\eta \le 0.6$, that using the function $\xi(n^\star,\eta)$ for LTC can be translated into the corresponding (expected) error performance. Note that the knowledge of $n^\star$ is needed for this last evaluation. Conversely, for the UWN scenario, where $E_{Tx}^\prime = 10$~mW and $E_0 = 0.726$~nJ, both DCT-LPF and LTC curves lie below the hardware dependent ratio $E_{Tx}^\prime/E_0 \simeq 13\cdot 10^6$, indicating that energy savings are achievable by both schemes for almost all $\eta$. These results can be generalized to any other device technology, by comparing the right hand side of Eq.~(\ref{eq:tradeoff}) against the corresponding ratio $E_{Tx}^\prime/E_0$ and checking whether Eq.~(\ref{eq:tradeoff}) holds.

%% file: fig_1a.tex
% GNUPLOT: LaTeX picture with Postscript
\begingroup
  \makeatletter
  \providecommand\color[2][]{%
    \GenericError{(gnuplot) \space\space\space\@spaces}{%
      Package color not loaded in conjunction with
      terminal option `colourtext'%
    }{See the gnuplot documentation for explanation.%
    }{Either use 'blacktext' in gnuplot or load the package
      color.sty in LaTeX.}%
    \renewcommand\color[2][]{}%
  }%
  \providecommand\includegraphics[2][]{%
    \GenericError{(gnuplot) \space\space\space\@spaces}{%
      Package graphicx or graphics not loaded%
    }{See the gnuplot documentation for explanation.%
    }{The gnuplot epslatex terminal needs graphicx.sty or graphics.sty.}%
    \renewcommand\includegraphics[2][]{}%
  }%
  \providecommand\rotatebox[2]{#2}%
  \@ifundefined{ifGPcolor}{%
    \newif\ifGPcolor
    \GPcolorfalse
  }{}%
  \@ifundefined{ifGPblacktext}{%
    \newif\ifGPblacktext
    \GPblacktexttrue
  }{}%
  % define a \g@addto@macro without @ in the name:
  \let\gplgaddtomacro\g@addto@macro
  % define empty templates for all commands taking text:
  \gdef\gplbacktext{}%
  \gdef\gplfronttext{}%
  \makeatother
  \ifGPblacktext
    % no textcolor at all
    \def\colorrgb#1{}%
    \def\colorgray#1{}%
  \else
    % gray or color?
    \ifGPcolor
      \def\colorrgb#1{\color[rgb]{#1}}%
      \def\colorgray#1{\color[gray]{#1}}%
      \expandafter\def\csname LTw\endcsname{\color{white}}%
      \expandafter\def\csname LTb\endcsname{\color{black}}%
      \expandafter\def\csname LTa\endcsname{\color{black}}%
      \expandafter\def\csname LT0\endcsname{\color[rgb]{1,0,0}}%
      \expandafter\def\csname LT1\endcsname{\color[rgb]{0,1,0}}%
      \expandafter\def\csname LT2\endcsname{\color[rgb]{0,0,1}}%
      \expandafter\def\csname LT3\endcsname{\color[rgb]{1,0,1}}%
      \expandafter\def\csname LT4\endcsname{\color[rgb]{0,1,1}}%
      \expandafter\def\csname LT5\endcsname{\color[rgb]{1,1,0}}%
      \expandafter\def\csname LT6\endcsname{\color[rgb]{0,0,0}}%
      \expandafter\def\csname LT7\endcsname{\color[rgb]{1,0.3,0}}%
      \expandafter\def\csname LT8\endcsname{\color[rgb]{0.5,0.5,0.5}}%
    \else
      % gray
      \def\colorrgb#1{\color{black}}%
      \def\colorgray#1{\color[gray]{#1}}%
      \expandafter\def\csname LTw\endcsname{\color{white}}%
      \expandafter\def\csname LTb\endcsname{\color{black}}%
      \expandafter\def\csname LTa\endcsname{\color{black}}%
      \expandafter\def\csname LT0\endcsname{\color{black}}%
      \expandafter\def\csname LT1\endcsname{\color{black}}%
      \expandafter\def\csname LT2\endcsname{\color{black}}%
      \expandafter\def\csname LT3\endcsname{\color{black}}%
      \expandafter\def\csname LT4\endcsname{\color{black}}%
      \expandafter\def\csname LT5\endcsname{\color{black}}%
      \expandafter\def\csname LT6\endcsname{\color{black}}%
      \expandafter\def\csname LT7\endcsname{\color{black}}%
      \expandafter\def\csname LT8\endcsname{\color{black}}%
    \fi
  \fi
  \setlength{\unitlength}{0.0500bp}%
  \begin{picture}(7200.00,5040.00)%
    \gplgaddtomacro\gplbacktext{%
      \csname LTb\endcsname%
      \put(1188,815){\makebox(0,0)[r]{\strut{}$10^{-2}$}}%
      \csname LTb\endcsname%
      \put(1188,1816){\makebox(0,0)[r]{\strut{}$10^{-1}$}}%
      \csname LTb\endcsname%
      \put(1188,2817){\makebox(0,0)[r]{\strut{}$10^{0}$}}%
      \csname LTb\endcsname%
      \put(1188,3818){\makebox(0,0)[r]{\strut{}$10^{1}$}}%
      \csname LTb\endcsname%
      \put(1188,4819){\makebox(0,0)[r]{\strut{}$10^{2}$}}%
      \csname LTb\endcsname%
      \put(1320,440){\makebox(0,0){\strut{} 0}}%
      \csname LTb\endcsname%
      \put(2443,440){\makebox(0,0){\strut{} 100}}%
      \csname LTb\endcsname%
      \put(3566,440){\makebox(0,0){\strut{} 200}}%
      \csname LTb\endcsname%
      \put(4689,440){\makebox(0,0){\strut{} 300}}%
      \csname LTb\endcsname%
      \put(5812,440){\makebox(0,0){\strut{} 400}}%
      \csname LTb\endcsname%
      \put(6935,440){\makebox(0,0){\strut{} 500}}%
      \put(418,2739){\rotatebox{-270}{\makebox(0,0){\strut{}Compression Ratio, $\eta$}}}%
      \put(4127,110){\makebox(0,0){\strut{}Correlation Length, $n^\star$ [time slots]}}%
    }%
    \gplgaddtomacro\gplfronttext{%
      \csname LTb\endcsname%
      \put(3245,4646){\makebox(0,0)[r]{\strut{}M-AAR (p=2)}}%
      \csname LTb\endcsname%
      \put(3245,4426){\makebox(0,0)[r]{\strut{}M-AAR (p=3)}}%
      \csname LTb\endcsname%
      \put(3245,4206){\makebox(0,0)[r]{\strut{}M-AAR (p=4)}}%
      \csname LTb\endcsname%
      \put(3245,3986){\makebox(0,0)[r]{\strut{}M-AAR (p=5)}}%
      \csname LTb\endcsname%
      \put(3245,3766){\makebox(0,0)[r]{\strut{}PR (p=2)}}%
      \csname LTb\endcsname%
      \put(3245,3546){\makebox(0,0)[r]{\strut{}PR (p=3)}}%
      \csname LTb\endcsname%
      \put(5948,4646){\makebox(0,0)[r]{\strut{}PR (p=4)}}%
      \csname LTb\endcsname%
      \put(5948,4426){\makebox(0,0)[r]{\strut{}PR (p=5)}}%
      \csname LTb\endcsname%
      \put(5948,4206){\makebox(0,0)[r]{\strut{}LTC}}%
      \csname LTb\endcsname%
      \put(5948,3986){\makebox(0,0)[r]{\strut{}PLAMLiS}}%
      \csname LTb\endcsname%
      \put(5948,3766){\makebox(0,0)[r]{\strut{}E-PLAMLiS}}%
      \csname LTb\endcsname%
      \put(5948,3546){\makebox(0,0)[r]{\strut{}No compression}}%
    }%
    \gplbacktext
    \put(0,0){\includegraphics{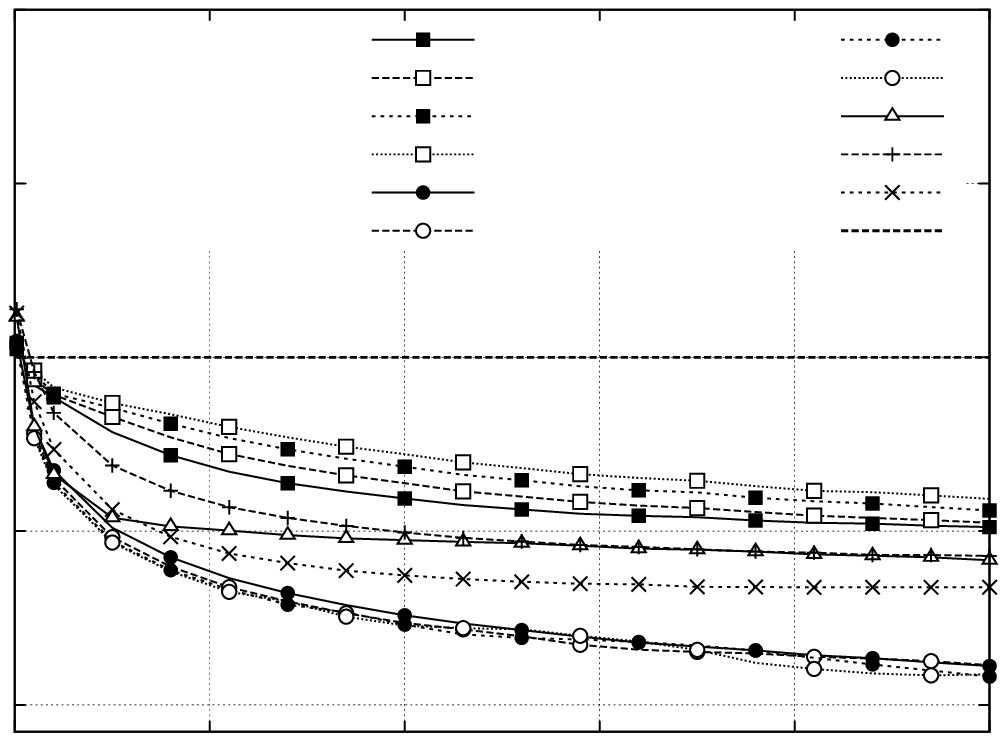}}%
    \gplfronttext
  \end{picture}%
\endgroup

%% file: fig_1b.tex
% GNUPLOT: LaTeX picture with Postscript
\begingroup
  \makeatletter
  \providecommand\color[2][]{%
    \GenericError{(gnuplot) \space\space\space\@spaces}{%
      Package color not loaded in conjunction with
      terminal option `colourtext'%
    }{See the gnuplot documentation for explanation.%
    }{Either use 'blacktext' in gnuplot or load the package
      color.sty in LaTeX.}%
    \renewcommand\color[2][]{}%
  }%
  \providecommand\includegraphics[2][]{%
    \GenericError{(gnuplot) \space\space\space\@spaces}{%
      Package graphicx or graphics not loaded%
    }{See the gnuplot documentation for explanation.%
    }{The gnuplot epslatex terminal needs graphicx.sty or graphics.sty.}%
    \renewcommand\includegraphics[2][]{}%
  }%
  \providecommand\rotatebox[2]{#2}%
  \@ifundefined{ifGPcolor}{%
    \newif\ifGPcolor
    \GPcolorfalse
  }{}%
  \@ifundefined{ifGPblacktext}{%
    \newif\ifGPblacktext
    \GPblacktexttrue
  }{}%
  % define a \g@addto@macro without @ in the name:
  \let\gplgaddtomacro\g@addto@macro
  % define empty templates for all commands taking text:
  \gdef\gplbacktext{}%
  \gdef\gplfronttext{}%
  \makeatother
  \ifGPblacktext
    % no textcolor at all
    \def\colorrgb#1{}%
    \def\colorgray#1{}%
  \else
    % gray or color?
    \ifGPcolor
      \def\colorrgb#1{\color[rgb]{#1}}%
      \def\colorgray#1{\color[gray]{#1}}%
      \expandafter\def\csname LTw\endcsname{\color{white}}%
      \expandafter\def\csname LTb\endcsname{\color{black}}%
      \expandafter\def\csname LTa\endcsname{\color{black}}%
      \expandafter\def\csname LT0\endcsname{\color[rgb]{1,0,0}}%
      \expandafter\def\csname LT1\endcsname{\color[rgb]{0,1,0}}%
      \expandafter\def\csname LT2\endcsname{\color[rgb]{0,0,1}}%
      \expandafter\def\csname LT3\endcsname{\color[rgb]{1,0,1}}%
      \expandafter\def\csname LT4\endcsname{\color[rgb]{0,1,1}}%
      \expandafter\def\csname LT5\endcsname{\color[rgb]{1,1,0}}%
      \expandafter\def\csname LT6\endcsname{\color[rgb]{0,0,0}}%
      \expandafter\def\csname LT7\endcsname{\color[rgb]{1,0.3,0}}%
      \expandafter\def\csname LT8\endcsname{\color[rgb]{0.5,0.5,0.5}}%
    \else
      % gray
      \def\colorrgb#1{\color{black}}%
      \def\colorgray#1{\color[gray]{#1}}%
      \expandafter\def\csname LTw\endcsname{\color{white}}%
      \expandafter\def\csname LTb\endcsname{\color{black}}%
      \expandafter\def\csname LTa\endcsname{\color{black}}%
      \expandafter\def\csname LT0\endcsname{\color{black}}%
      \expandafter\def\csname LT1\endcsname{\color{black}}%
      \expandafter\def\csname LT2\endcsname{\color{black}}%
      \expandafter\def\csname LT3\endcsname{\color{black}}%
      \expandafter\def\csname LT4\endcsname{\color{black}}%
      \expandafter\def\csname LT5\endcsname{\color{black}}%
      \expandafter\def\csname LT6\endcsname{\color{black}}%
      \expandafter\def\csname LT7\endcsname{\color{black}}%
      \expandafter\def\csname LT8\endcsname{\color{black}}%
    \fi
  \fi
  \setlength{\unitlength}{0.0500bp}%
  \begin{picture}(7200.00,5040.00)%
    \gplgaddtomacro\gplbacktext{%
      \csname LTb\endcsname%
      \put(1188,815){\makebox(0,0)[r]{\strut{}$10^{-2}$}}%
      \csname LTb\endcsname%
      \put(1188,1816){\makebox(0,0)[r]{\strut{}$10^{-1}$}}%
      \csname LTb\endcsname%
      \put(1188,2817){\makebox(0,0)[r]{\strut{}$10^{0}$}}%
      \csname LTb\endcsname%
      \put(1188,3818){\makebox(0,0)[r]{\strut{}$10^{1}$}}%
      \csname LTb\endcsname%
      \put(1188,4819){\makebox(0,0)[r]{\strut{}$10^{2}$}}%
      \csname LTb\endcsname%
      \put(1320,440){\makebox(0,0){\strut{}$10^{-4}$}}%
      \csname LTb\endcsname%
      \put(2443,440){\makebox(0,0){\strut{}$10^{-3}$}}%
      \csname LTb\endcsname%
      \put(3566,440){\makebox(0,0){\strut{}$10^{-2}$}}%
      \csname LTb\endcsname%
      \put(4689,440){\makebox(0,0){\strut{}$10^{-1}$}}%
      \csname LTb\endcsname%
      \put(5812,440){\makebox(0,0){\strut{}$10^{0}$}}%
      \csname LTb\endcsname%
      \put(6935,440){\makebox(0,0){\strut{}$10^{1}$}}%
      \put(418,2739){\rotatebox{-270}{\makebox(0,0){\strut{}Compression Ratio, $\eta$}}}%
      \put(4127,110){\makebox(0,0){\strut{}Energy for compression, [J]}}%
      \put(5474,2595){\makebox(0,0)[l]{\strut{}increasing $n^\star$}}%
    }%
    \gplgaddtomacro\gplfronttext{%
      \csname LTb\endcsname%
      \put(3641,4646){\makebox(0,0)[r]{\strut{}M-AAR (p=2)}}%
      \csname LTb\endcsname%
      \put(3641,4426){\makebox(0,0)[r]{\strut{}M-AAR (p=3)}}%
      \csname LTb\endcsname%
      \put(3641,4206){\makebox(0,0)[r]{\strut{}M-AAR (p=4)}}%
      \csname LTb\endcsname%
      \put(3641,3986){\makebox(0,0)[r]{\strut{}M-AAR (p=5)}}%
      \csname LTb\endcsname%
      \put(3641,3766){\makebox(0,0)[r]{\strut{}PR (p=2)}}%
      \csname LTb\endcsname%
      \put(3641,3546){\makebox(0,0)[r]{\strut{}PR (p=3)}}%
      \csname LTb\endcsname%
      \put(5948,4646){\makebox(0,0)[r]{\strut{}PR (p=4)}}%
      \csname LTb\endcsname%
      \put(5948,4426){\makebox(0,0)[r]{\strut{}PR (p=5)}}%
      \csname LTb\endcsname%
      \put(5948,4206){\makebox(0,0)[r]{\strut{}LTC}}%
      \csname LTb\endcsname%
      \put(5948,3986){\makebox(0,0)[r]{\strut{}PLAMLiS}}%
      \csname LTb\endcsname%
      \put(5948,3766){\makebox(0,0)[r]{\strut{}E-PLAMLiS}}%
    }%
    \gplbacktext
    \put(0,0){\includegraphics{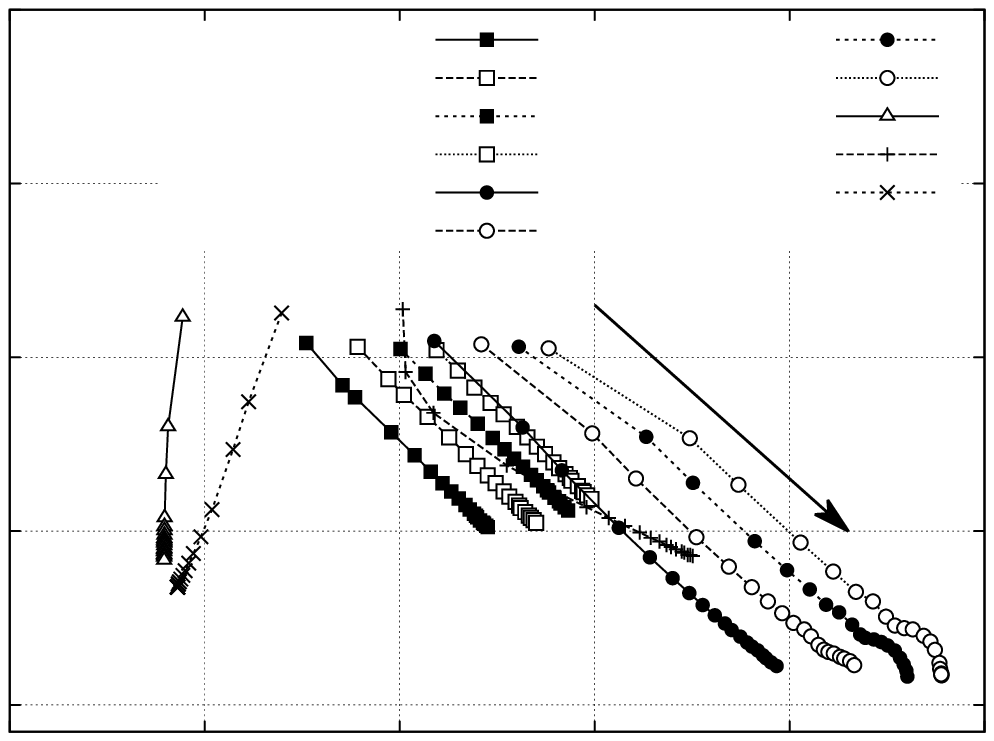}}%
    \gplfronttext
  \end{picture}%
\endgroup

%% file: fig_2a.tex
% GNUPLOT: LaTeX picture with Postscript
\begingroup
  \makeatletter
  \providecommand\color[2][]{%
    \GenericError{(gnuplot) \space\space\space\@spaces}{%
      Package color not loaded in conjunction with
      terminal option `colourtext'%
    }{See the gnuplot documentation for explanation.%
    }{Either use 'blacktext' in gnuplot or load the package
      color.sty in LaTeX.}%
    \renewcommand\color[2][]{}%
  }%
  \providecommand\includegraphics[2][]{%
    \GenericError{(gnuplot) \space\space\space\@spaces}{%
      Package graphicx or graphics not loaded%
    }{See the gnuplot documentation for explanation.%
    }{The gnuplot epslatex terminal needs graphicx.sty or graphics.sty.}%
    \renewcommand\includegraphics[2][]{}%
  }%
  \providecommand\rotatebox[2]{#2}%
  \@ifundefined{ifGPcolor}{%
    \newif\ifGPcolor
    \GPcolorfalse
  }{}%
  \@ifundefined{ifGPblacktext}{%
    \newif\ifGPblacktext
    \GPblacktexttrue
  }{}%
  % define a \g@addto@macro without @ in the name:
  \let\gplgaddtomacro\g@addto@macro
  % define empty templates for all commands taking text:
  \gdef\gplbacktext{}%
  \gdef\gplfronttext{}%
  \makeatother
  \ifGPblacktext
    % no textcolor at all
    \def\colorrgb#1{}%
    \def\colorgray#1{}%
  \else
    % gray or color?
    \ifGPcolor
      \def\colorrgb#1{\color[rgb]{#1}}%
      \def\colorgray#1{\color[gray]{#1}}%
      \expandafter\def\csname LTw\endcsname{\color{white}}%
      \expandafter\def\csname LTb\endcsname{\color{black}}%
      \expandafter\def\csname LTa\endcsname{\color{black}}%
      \expandafter\def\csname LT0\endcsname{\color[rgb]{1,0,0}}%
      \expandafter\def\csname LT1\endcsname{\color[rgb]{0,1,0}}%
      \expandafter\def\csname LT2\endcsname{\color[rgb]{0,0,1}}%
      \expandafter\def\csname LT3\endcsname{\color[rgb]{1,0,1}}%
      \expandafter\def\csname LT4\endcsname{\color[rgb]{0,1,1}}%
      \expandafter\def\csname LT5\endcsname{\color[rgb]{1,1,0}}%
      \expandafter\def\csname LT6\endcsname{\color[rgb]{0,0,0}}%
      \expandafter\def\csname LT7\endcsname{\color[rgb]{1,0.3,0}}%
      \expandafter\def\csname LT8\endcsname{\color[rgb]{0.5,0.5,0.5}}%
    \else
      % gray
      \def\colorrgb#1{\color{black}}%
      \def\colorgray#1{\color[gray]{#1}}%
      \expandafter\def\csname LTw\endcsname{\color{white}}%
      \expandafter\def\csname LTb\endcsname{\color{black}}%
      \expandafter\def\csname LTa\endcsname{\color{black}}%
      \expandafter\def\csname LT0\endcsname{\color{black}}%
      \expandafter\def\csname LT1\endcsname{\color{black}}%
      \expandafter\def\csname LT2\endcsname{\color{black}}%
      \expandafter\def\csname LT3\endcsname{\color{black}}%
      \expandafter\def\csname LT4\endcsname{\color{black}}%
      \expandafter\def\csname LT5\endcsname{\color{black}}%
      \expandafter\def\csname LT6\endcsname{\color{black}}%
      \expandafter\def\csname LT7\endcsname{\color{black}}%
      \expandafter\def\csname LT8\endcsname{\color{black}}%
    \fi
  \fi
  \setlength{\unitlength}{0.0500bp}%
  \begin{picture}(7200.00,5040.00)%
    \gplgaddtomacro\gplbacktext{%
      \csname LTb\endcsname%
      \put(1188,660){\makebox(0,0)[r]{\strut{}$10^{-2}$}}%
      \csname LTb\endcsname%
      \put(1188,2046){\makebox(0,0)[r]{\strut{}$10^{-1}$}}%
      \csname LTb\endcsname%
      \put(1188,3433){\makebox(0,0)[r]{\strut{}$10^{0}$}}%
      \csname LTb\endcsname%
      \put(1188,4819){\makebox(0,0)[r]{\strut{}$10^{1}$}}%
      \csname LTb\endcsname%
      \put(1320,440){\makebox(0,0){\strut{} 0}}%
      \csname LTb\endcsname%
      \put(2443,440){\makebox(0,0){\strut{} 100}}%
      \csname LTb\endcsname%
      \put(3566,440){\makebox(0,0){\strut{} 200}}%
      \csname LTb\endcsname%
      \put(4689,440){\makebox(0,0){\strut{} 300}}%
      \csname LTb\endcsname%
      \put(5812,440){\makebox(0,0){\strut{} 400}}%
      \csname LTb\endcsname%
      \put(6935,440){\makebox(0,0){\strut{} 500}}%
      \put(418,2739){\rotatebox{-270}{\makebox(0,0){\strut{}Compression Ratio, $\eta$}}}%
      \put(4127,110){\makebox(0,0){\strut{}Correlation Length, $n^\star$ [time slots]}}%
    }%
    \gplgaddtomacro\gplfronttext{%
      \csname LTb\endcsname%
      \put(3245,4646){\makebox(0,0)[r]{\strut{}FFT}}%
      \csname LTb\endcsname%
      \put(3245,4426){\makebox(0,0)[r]{\strut{}FFT-LPF}}%
      \csname LTb\endcsname%
      \put(3245,4206){\makebox(0,0)[r]{\strut{}FFT-LPF Win}}%
      \csname LTb\endcsname%
      \put(3245,3986){\makebox(0,0)[r]{\strut{}FFT Win}}%
      \csname LTb\endcsname%
      \put(5948,4646){\makebox(0,0)[r]{\strut{}DCT}}%
      \csname LTb\endcsname%
      \put(5948,4426){\makebox(0,0)[r]{\strut{}DCT-LPF}}%
      \csname LTb\endcsname%
      \put(5948,4206){\makebox(0,0)[r]{\strut{}No compression}}%
    }%
    \gplbacktext
    \put(0,0){\includegraphics{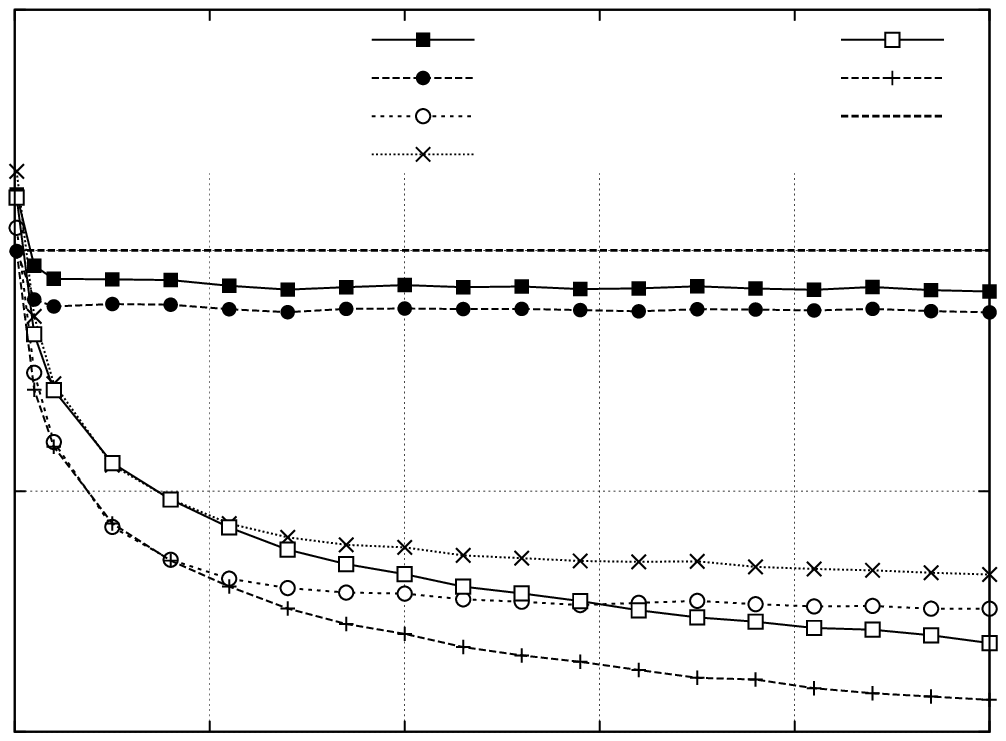}}%
    \gplfronttext
  \end{picture}%
\endgroup

%% file: fig_2b.tex
% GNUPLOT: LaTeX picture with Postscript
\begingroup
  \makeatletter
  \providecommand\color[2][]{%
    \GenericError{(gnuplot) \space\space\space\@spaces}{%
      Package color not loaded in conjunction with
      terminal option `colourtext'%
    }{See the gnuplot documentation for explanation.%
    }{Either use 'blacktext' in gnuplot or load the package
      color.sty in LaTeX.}%
    \renewcommand\color[2][]{}%
  }%
  \providecommand\includegraphics[2][]{%
    \GenericError{(gnuplot) \space\space\space\@spaces}{%
      Package graphicx or graphics not loaded%
    }{See the gnuplot documentation for explanation.%
    }{The gnuplot epslatex terminal needs graphicx.sty or graphics.sty.}%
    \renewcommand\includegraphics[2][]{}%
  }%
  \providecommand\rotatebox[2]{#2}%
  \@ifundefined{ifGPcolor}{%
    \newif\ifGPcolor
    \GPcolorfalse
  }{}%
  \@ifundefined{ifGPblacktext}{%
    \newif\ifGPblacktext
    \GPblacktexttrue
  }{}%
  % define a \g@addto@macro without @ in the name:
  \let\gplgaddtomacro\g@addto@macro
  % define empty templates for all commands taking text:
  \gdef\gplbacktext{}%
  \gdef\gplfronttext{}%
  \makeatother
  \ifGPblacktext
    % no textcolor at all
    \def\colorrgb#1{}%
    \def\colorgray#1{}%
  \else
    % gray or color?
    \ifGPcolor
      \def\colorrgb#1{\color[rgb]{#1}}%
      \def\colorgray#1{\color[gray]{#1}}%
      \expandafter\def\csname LTw\endcsname{\color{white}}%
      \expandafter\def\csname LTb\endcsname{\color{black}}%
      \expandafter\def\csname LTa\endcsname{\color{black}}%
      \expandafter\def\csname LT0\endcsname{\color[rgb]{1,0,0}}%
      \expandafter\def\csname LT1\endcsname{\color[rgb]{0,1,0}}%
      \expandafter\def\csname LT2\endcsname{\color[rgb]{0,0,1}}%
      \expandafter\def\csname LT3\endcsname{\color[rgb]{1,0,1}}%
      \expandafter\def\csname LT4\endcsname{\color[rgb]{0,1,1}}%
      \expandafter\def\csname LT5\endcsname{\color[rgb]{1,1,0}}%
      \expandafter\def\csname LT6\endcsname{\color[rgb]{0,0,0}}%
      \expandafter\def\csname LT7\endcsname{\color[rgb]{1,0.3,0}}%
      \expandafter\def\csname LT8\endcsname{\color[rgb]{0.5,0.5,0.5}}%
    \else
      % gray
      \def\colorrgb#1{\color{black}}%
      \def\colorgray#1{\color[gray]{#1}}%
      \expandafter\def\csname LTw\endcsname{\color{white}}%
      \expandafter\def\csname LTb\endcsname{\color{black}}%
      \expandafter\def\csname LTa\endcsname{\color{black}}%
      \expandafter\def\csname LT0\endcsname{\color{black}}%
      \expandafter\def\csname LT1\endcsname{\color{black}}%
      \expandafter\def\csname LT2\endcsname{\color{black}}%
      \expandafter\def\csname LT3\endcsname{\color{black}}%
      \expandafter\def\csname LT4\endcsname{\color{black}}%
      \expandafter\def\csname LT5\endcsname{\color{black}}%
      \expandafter\def\csname LT6\endcsname{\color{black}}%
      \expandafter\def\csname LT7\endcsname{\color{black}}%
      \expandafter\def\csname LT8\endcsname{\color{black}}%
    \fi
  \fi
  \setlength{\unitlength}{0.0500bp}%
  \begin{picture}(7200.00,5040.00)%
    \gplgaddtomacro\gplbacktext{%
      \csname LTb\endcsname%
      \put(1188,660){\makebox(0,0)[r]{\strut{}$10^{-2}$}}%
      \csname LTb\endcsname%
      \put(1188,2046){\makebox(0,0)[r]{\strut{}$10^{-1}$}}%
      \csname LTb\endcsname%
      \put(1188,3433){\makebox(0,0)[r]{\strut{}$10^{0}$}}%
      \csname LTb\endcsname%
      \put(1188,4819){\makebox(0,0)[r]{\strut{}$10^{1}$}}%
      \csname LTb\endcsname%
      \put(1320,440){\makebox(0,0){\strut{}$10^{-3}$}}%
      \csname LTb\endcsname%
      \put(3192,440){\makebox(0,0){\strut{}$10^{-2}$}}%
      \csname LTb\endcsname%
      \put(5063,440){\makebox(0,0){\strut{}$10^{-1}$}}%
      \csname LTb\endcsname%
      \put(6935,440){\makebox(0,0){\strut{}$10^{0}$}}%
      \put(418,2739){\rotatebox{-270}{\makebox(0,0){\strut{}Compression Ratio, $\eta$}}}%
      \put(4127,110){\makebox(0,0){\strut{}Energy for compression, [J]}}%
      \put(5063,1832){\makebox(0,0)[l]{\strut{}increasing $n^\star$}}%
    }%
    \gplgaddtomacro\gplfronttext{%
      \csname LTb\endcsname%
      \put(3641,4646){\makebox(0,0)[r]{\strut{}FFT}}%
      \csname LTb\endcsname%
      \put(3641,4426){\makebox(0,0)[r]{\strut{}FFT-LPF}}%
      \csname LTb\endcsname%
      \put(3641,4206){\makebox(0,0)[r]{\strut{}FFT-LPF Win}}%
      \csname LTb\endcsname%
      \put(5948,4646){\makebox(0,0)[r]{\strut{}FFT Win}}%
      \csname LTb\endcsname%
      \put(5948,4426){\makebox(0,0)[r]{\strut{}DCT}}%
      \csname LTb\endcsname%
      \put(5948,4206){\makebox(0,0)[r]{\strut{}DCT-LPF}}%
    }%
    \gplbacktext
    \put(0,0){\includegraphics{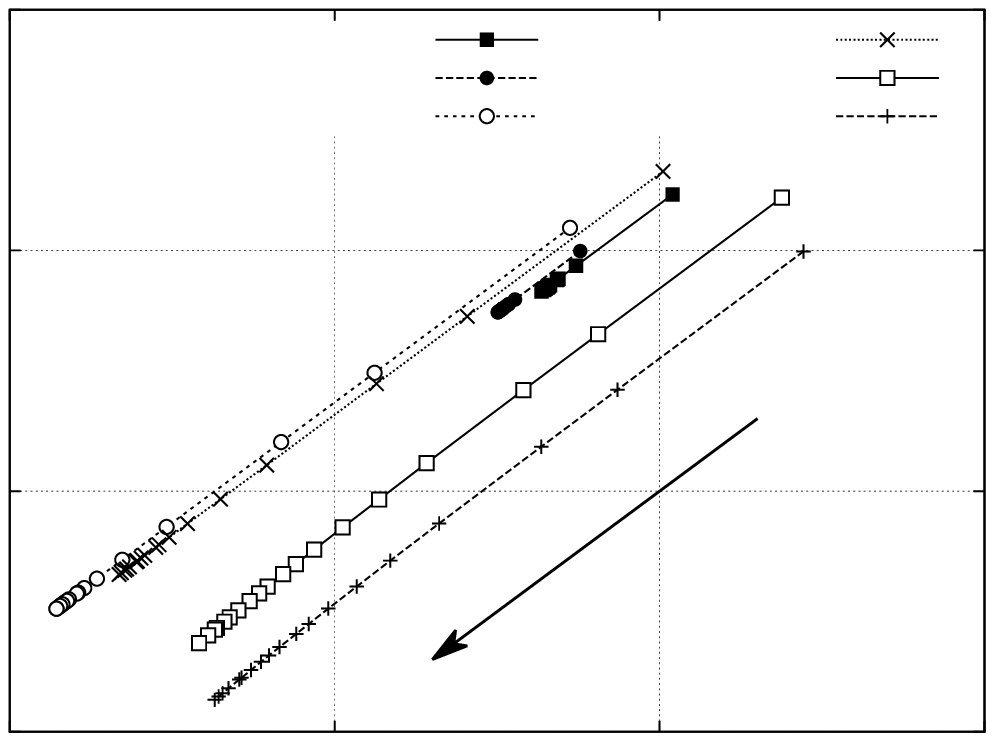}}%
    \gplfronttext
  \end{picture}%
\endgroup

%% file: fig_3a.tex
% GNUPLOT: LaTeX picture with Postscript
\begingroup
  \makeatletter
  \providecommand\color[2][]{%
    \GenericError{(gnuplot) \space\space\space\@spaces}{%
      Package color not loaded in conjunction with
      terminal option `colourtext'%
    }{See the gnuplot documentation for explanation.%
    }{Either use 'blacktext' in gnuplot or load the package
      color.sty in LaTeX.}%
    \renewcommand\color[2][]{}%
  }%
  \providecommand\includegraphics[2][]{%
    \GenericError{(gnuplot) \space\space\space\@spaces}{%
      Package graphicx or graphics not loaded%
    }{See the gnuplot documentation for explanation.%
    }{The gnuplot epslatex terminal needs graphicx.sty or graphics.sty.}%
    \renewcommand\includegraphics[2][]{}%
  }%
  \providecommand\rotatebox[2]{#2}%
  \@ifundefined{ifGPcolor}{%
    \newif\ifGPcolor
    \GPcolorfalse
  }{}%
  \@ifundefined{ifGPblacktext}{%
    \newif\ifGPblacktext
    \GPblacktexttrue
  }{}%
  % define a \g@addto@macro without @ in the name:
  \let\gplgaddtomacro\g@addto@macro
  % define empty templates for all commands taking text:
  \gdef\gplbacktext{}%
  \gdef\gplfronttext{}%
  \makeatother
  \ifGPblacktext
    % no textcolor at all
    \def\colorrgb#1{}%
    \def\colorgray#1{}%
  \else
    % gray or color?
    \ifGPcolor
      \def\colorrgb#1{\color[rgb]{#1}}%
      \def\colorgray#1{\color[gray]{#1}}%
      \expandafter\def\csname LTw\endcsname{\color{white}}%
      \expandafter\def\csname LTb\endcsname{\color{black}}%
      \expandafter\def\csname LTa\endcsname{\color{black}}%
      \expandafter\def\csname LT0\endcsname{\color[rgb]{1,0,0}}%
      \expandafter\def\csname LT1\endcsname{\color[rgb]{0,1,0}}%
      \expandafter\def\csname LT2\endcsname{\color[rgb]{0,0,1}}%
      \expandafter\def\csname LT3\endcsname{\color[rgb]{1,0,1}}%
      \expandafter\def\csname LT4\endcsname{\color[rgb]{0,1,1}}%
      \expandafter\def\csname LT5\endcsname{\color[rgb]{1,1,0}}%
      \expandafter\def\csname LT6\endcsname{\color[rgb]{0,0,0}}%
      \expandafter\def\csname LT7\endcsname{\color[rgb]{1,0.3,0}}%
      \expandafter\def\csname LT8\endcsname{\color[rgb]{0.5,0.5,0.5}}%
    \else
      % gray
      \def\colorrgb#1{\color{black}}%
      \def\colorgray#1{\color[gray]{#1}}%
      \expandafter\def\csname LTw\endcsname{\color{white}}%
      \expandafter\def\csname LTb\endcsname{\color{black}}%
      \expandafter\def\csname LTa\endcsname{\color{black}}%
      \expandafter\def\csname LT0\endcsname{\color{black}}%
      \expandafter\def\csname LT1\endcsname{\color{black}}%
      \expandafter\def\csname LT2\endcsname{\color{black}}%
      \expandafter\def\csname LT3\endcsname{\color{black}}%
      \expandafter\def\csname LT4\endcsname{\color{black}}%
      \expandafter\def\csname LT5\endcsname{\color{black}}%
      \expandafter\def\csname LT6\endcsname{\color{black}}%
      \expandafter\def\csname LT7\endcsname{\color{black}}%
      \expandafter\def\csname LT8\endcsname{\color{black}}%
    \fi
  \fi
  \setlength{\unitlength}{0.0500bp}%
  \begin{picture}(7200.00,5040.00)%
    \gplgaddtomacro\gplbacktext{%
      \csname LTb\endcsname%
      \put(1188,660){\makebox(0,0)[r]{\strut{}$10^{-2}$}}%
      \csname LTb\endcsname%
      \put(1188,1784){\makebox(0,0)[r]{\strut{}$10^{-1}$}}%
      \csname LTb\endcsname%
      \put(1188,2909){\makebox(0,0)[r]{\strut{}$10^{0}$}}%
      \csname LTb\endcsname%
      \put(1188,4033){\makebox(0,0)[r]{\strut{}$10^{1}$}}%
      \csname LTb\endcsname%
      \put(1320,440){\makebox(0,0){\strut{}$10^{-4}$}}%
      \csname LTb\endcsname%
      \put(2443,440){\makebox(0,0){\strut{}$10^{-3}$}}%
      \csname LTb\endcsname%
      \put(3566,440){\makebox(0,0){\strut{}$10^{-2}$}}%
      \csname LTb\endcsname%
      \put(4689,440){\makebox(0,0){\strut{}$10^{-1}$}}%
      \csname LTb\endcsname%
      \put(5812,440){\makebox(0,0){\strut{}$10^{0}$}}%
      \csname LTb\endcsname%
      \put(6935,440){\makebox(0,0){\strut{}$10^{1}$}}%
      \put(418,2739){\rotatebox{-270}{\makebox(0,0){\strut{}Compression Ratio, $\eta$}}}%
      \put(4127,110){\makebox(0,0){\strut{}Total Energy [J]}}%
    }%
    \gplgaddtomacro\gplfronttext{%
      \csname LTb\endcsname%
      \put(3245,4646){\makebox(0,0)[r]{\strut{}M-AAR (p=2)}}%
      \csname LTb\endcsname%
      \put(3245,4426){\makebox(0,0)[r]{\strut{}M-AAR (p=4)}}%
      \csname LTb\endcsname%
      \put(3245,4206){\makebox(0,0)[r]{\strut{}PR (p=2)}}%
      \csname LTb\endcsname%
      \put(3245,3986){\makebox(0,0)[r]{\strut{}PR (p=2)}}%
      \csname LTb\endcsname%
      \put(3245,3766){\makebox(0,0)[r]{\strut{}LTC}}%
      \csname LTb\endcsname%
      \put(5948,4646){\makebox(0,0)[r]{\strut{}E-PLAMLiS}}%
      \csname LTb\endcsname%
      \put(5948,4426){\makebox(0,0)[r]{\strut{}FFT Win}}%
      \csname LTb\endcsname%
      \put(5948,4206){\makebox(0,0)[r]{\strut{}DCT-LPF}}%
      \csname LTb\endcsname%
      \put(5948,3986){\makebox(0,0)[r]{\strut{}No compression}}%
    }%
    \gplbacktext
    \put(0,0){\includegraphics{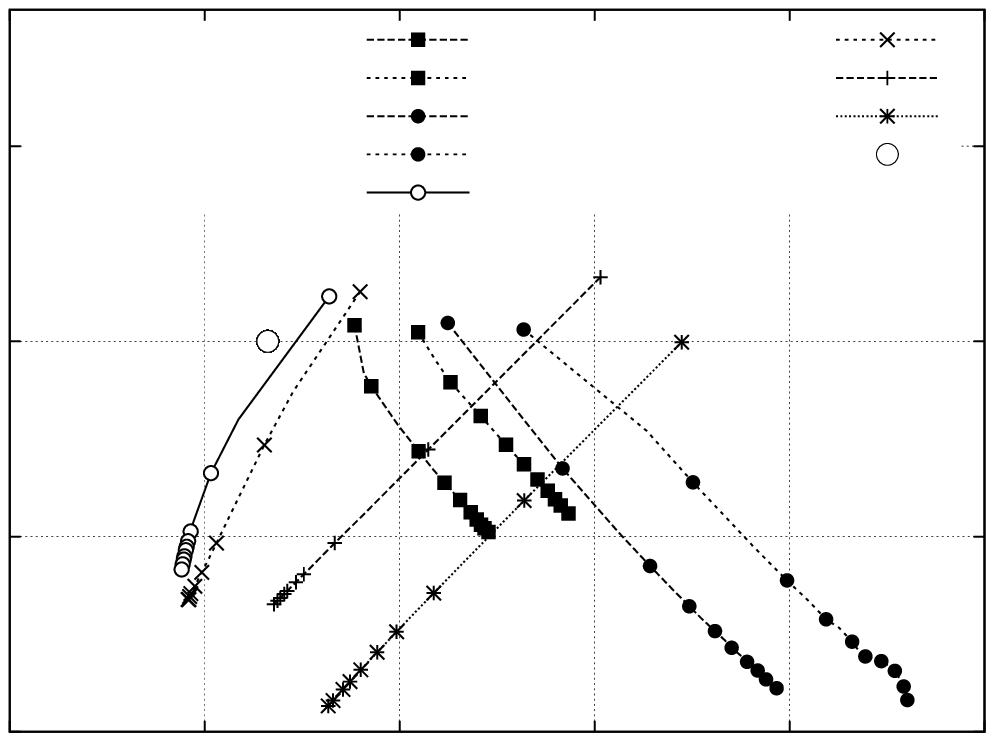}}%
    \gplfronttext
  \end{picture}%
\endgroup

%% file: fig_3b.tex
% GNUPLOT: LaTeX picture with Postscript
\begingroup
  \makeatletter
  \providecommand\color[2][]{%
    \GenericError{(gnuplot) \space\space\space\@spaces}{%
      Package color not loaded in conjunction with
      terminal option `colourtext'%
    }{See the gnuplot documentation for explanation.%
    }{Either use 'blacktext' in gnuplot or load the package
      color.sty in LaTeX.}%
    \renewcommand\color[2][]{}%
  }%
  \providecommand\includegraphics[2][]{%
    \GenericError{(gnuplot) \space\space\space\@spaces}{%
      Package graphicx or graphics not loaded%
    }{See the gnuplot documentation for explanation.%
    }{The gnuplot epslatex terminal needs graphicx.sty or graphics.sty.}%
    \renewcommand\includegraphics[2][]{}%
  }%
  \providecommand\rotatebox[2]{#2}%
  \@ifundefined{ifGPcolor}{%
    \newif\ifGPcolor
    \GPcolorfalse
  }{}%
  \@ifundefined{ifGPblacktext}{%
    \newif\ifGPblacktext
    \GPblacktexttrue
  }{}%
  % define a \g@addto@macro without @ in the name:
  \let\gplgaddtomacro\g@addto@macro
  % define empty templates for all commands taking text:
  \gdef\gplbacktext{}%
  \gdef\gplfronttext{}%
  \makeatother
  \ifGPblacktext
    % no textcolor at all
    \def\colorrgb#1{}%
    \def\colorgray#1{}%
  \else
    % gray or color?
    \ifGPcolor
      \def\colorrgb#1{\color[rgb]{#1}}%
      \def\colorgray#1{\color[gray]{#1}}%
      \expandafter\def\csname LTw\endcsname{\color{white}}%
      \expandafter\def\csname LTb\endcsname{\color{black}}%
      \expandafter\def\csname LTa\endcsname{\color{black}}%
      \expandafter\def\csname LT0\endcsname{\color[rgb]{1,0,0}}%
      \expandafter\def\csname LT1\endcsname{\color[rgb]{0,1,0}}%
      \expandafter\def\csname LT2\endcsname{\color[rgb]{0,0,1}}%
      \expandafter\def\csname LT3\endcsname{\color[rgb]{1,0,1}}%
      \expandafter\def\csname LT4\endcsname{\color[rgb]{0,1,1}}%
      \expandafter\def\csname LT5\endcsname{\color[rgb]{1,1,0}}%
      \expandafter\def\csname LT6\endcsname{\color[rgb]{0,0,0}}%
      \expandafter\def\csname LT7\endcsname{\color[rgb]{1,0.3,0}}%
      \expandafter\def\csname LT8\endcsname{\color[rgb]{0.5,0.5,0.5}}%
    \else
      % gray
      \def\colorrgb#1{\color{black}}%
      \def\colorgray#1{\color[gray]{#1}}%
      \expandafter\def\csname LTw\endcsname{\color{white}}%
      \expandafter\def\csname LTb\endcsname{\color{black}}%
      \expandafter\def\csname LTa\endcsname{\color{black}}%
      \expandafter\def\csname LT0\endcsname{\color{black}}%
      \expandafter\def\csname LT1\endcsname{\color{black}}%
      \expandafter\def\csname LT2\endcsname{\color{black}}%
      \expandafter\def\csname LT3\endcsname{\color{black}}%
      \expandafter\def\csname LT4\endcsname{\color{black}}%
      \expandafter\def\csname LT5\endcsname{\color{black}}%
      \expandafter\def\csname LT6\endcsname{\color{black}}%
      \expandafter\def\csname LT7\endcsname{\color{black}}%
      \expandafter\def\csname LT8\endcsname{\color{black}}%
    \fi
  \fi
  \setlength{\unitlength}{0.0500bp}%
  \begin{picture}(7200.00,5040.00)%
    \gplgaddtomacro\gplbacktext{%
      \csname LTb\endcsname%
      \put(1188,660){\makebox(0,0)[r]{\strut{}$10^{-2}$}}%
      \csname LTb\endcsname%
      \put(1188,1784){\makebox(0,0)[r]{\strut{}$10^{-1}$}}%
      \csname LTb\endcsname%
      \put(1188,2909){\makebox(0,0)[r]{\strut{}$10^{0}$}}%
      \csname LTb\endcsname%
      \put(1188,4033){\makebox(0,0)[r]{\strut{}$10^{1}$}}%
      \csname LTb\endcsname%
      \put(1320,440){\makebox(0,0){\strut{}$10^{0}$}}%
      \csname LTb\endcsname%
      \put(3192,440){\makebox(0,0){\strut{}$10^{1}$}}%
      \csname LTb\endcsname%
      \put(5063,440){\makebox(0,0){\strut{}$10^{2}$}}%
      \csname LTb\endcsname%
      \put(6935,440){\makebox(0,0){\strut{}$10^{3}$}}%
      \put(418,2739){\rotatebox{-270}{\makebox(0,0){\strut{}Compression Ratio, $\eta$}}}%
      \put(4127,110){\makebox(0,0){\strut{}Total Energy [J]}}%
    }%
    \gplgaddtomacro\gplfronttext{%
      \csname LTb\endcsname%
      \put(3245,4646){\makebox(0,0)[r]{\strut{}M-AAR (p=2)}}%
      \csname LTb\endcsname%
      \put(3245,4426){\makebox(0,0)[r]{\strut{}M-AAR (p=4)}}%
      \csname LTb\endcsname%
      \put(3245,4206){\makebox(0,0)[r]{\strut{}PR (p=2)}}%
      \csname LTb\endcsname%
      \put(3245,3986){\makebox(0,0)[r]{\strut{}PR (p=2)}}%
      \csname LTb\endcsname%
      \put(3245,3766){\makebox(0,0)[r]{\strut{}LTC}}%
      \csname LTb\endcsname%
      \put(5948,4646){\makebox(0,0)[r]{\strut{}E-PLAMLiS}}%
      \csname LTb\endcsname%
      \put(5948,4426){\makebox(0,0)[r]{\strut{}FFT Win}}%
      \csname LTb\endcsname%
      \put(5948,4206){\makebox(0,0)[r]{\strut{}DCT-LPF}}%
      \csname LTb\endcsname%
      \put(5948,3986){\makebox(0,0)[r]{\strut{}No compression}}%
    }%
    \gplbacktext
    \put(0,0){\includegraphics{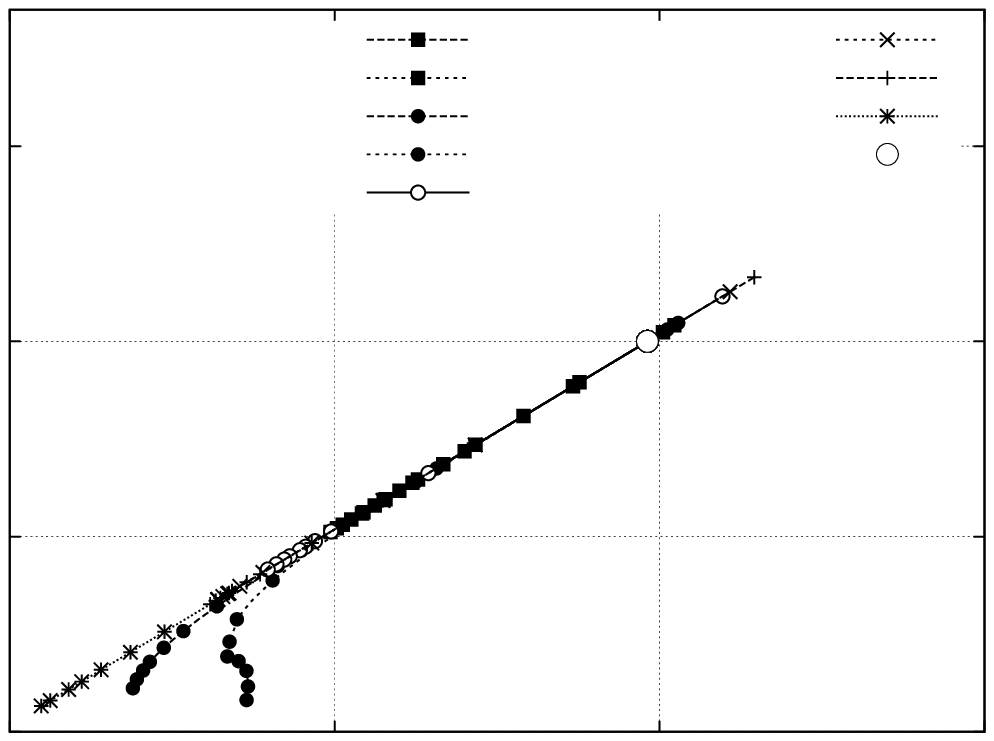}}%
    \gplfronttext
  \end{picture}%
\endgroup

%% file: fig_4a.tex
% GNUPLOT: LaTeX picture with Postscript
\begingroup
  \makeatletter
  \providecommand\color[2][]{%
    \GenericError{(gnuplot) \space\space\space\@spaces}{%
      Package color not loaded in conjunction with
      terminal option `colourtext'%
    }{See the gnuplot documentation for explanation.%
    }{Either use 'blacktext' in gnuplot or load the package
      color.sty in LaTeX.}%
    \renewcommand\color[2][]{}%
  }%
  \providecommand\includegraphics[2][]{%
    \GenericError{(gnuplot) \space\space\space\@spaces}{%
      Package graphicx or graphics not loaded%
    }{See the gnuplot documentation for explanation.%
    }{The gnuplot epslatex terminal needs graphicx.sty or graphics.sty.}%
    \renewcommand\includegraphics[2][]{}%
  }%
  \providecommand\rotatebox[2]{#2}%
  \@ifundefined{ifGPcolor}{%
    \newif\ifGPcolor
    \GPcolorfalse
  }{}%
  \@ifundefined{ifGPblacktext}{%
    \newif\ifGPblacktext
    \GPblacktexttrue
  }{}%
  % define a \g@addto@macro without @ in the name:
  \let\gplgaddtomacro\g@addto@macro
  % define empty templates for all commands taking text:
  \gdef\gplbacktext{}%
  \gdef\gplfronttext{}%
  \makeatother
  \ifGPblacktext
    % no textcolor at all
    \def\colorrgb#1{}%
    \def\colorgray#1{}%
  \else
    % gray or color?
    \ifGPcolor
      \def\colorrgb#1{\color[rgb]{#1}}%
      \def\colorgray#1{\color[gray]{#1}}%
      \expandafter\def\csname LTw\endcsname{\color{white}}%
      \expandafter\def\csname LTb\endcsname{\color{black}}%
      \expandafter\def\csname LTa\endcsname{\color{black}}%
      \expandafter\def\csname LT0\endcsname{\color[rgb]{1,0,0}}%
      \expandafter\def\csname LT1\endcsname{\color[rgb]{0,1,0}}%
      \expandafter\def\csname LT2\endcsname{\color[rgb]{0,0,1}}%
      \expandafter\def\csname LT3\endcsname{\color[rgb]{1,0,1}}%
      \expandafter\def\csname LT4\endcsname{\color[rgb]{0,1,1}}%
      \expandafter\def\csname LT5\endcsname{\color[rgb]{1,1,0}}%
      \expandafter\def\csname LT6\endcsname{\color[rgb]{0,0,0}}%
      \expandafter\def\csname LT7\endcsname{\color[rgb]{1,0.3,0}}%
      \expandafter\def\csname LT8\endcsname{\color[rgb]{0.5,0.5,0.5}}%
    \else
      % gray
      \def\colorrgb#1{\color{black}}%
      \def\colorgray#1{\color[gray]{#1}}%
      \expandafter\def\csname LTw\endcsname{\color{white}}%
      \expandafter\def\csname LTb\endcsname{\color{black}}%
      \expandafter\def\csname LTa\endcsname{\color{black}}%
      \expandafter\def\csname LT0\endcsname{\color{black}}%
      \expandafter\def\csname LT1\endcsname{\color{black}}%
      \expandafter\def\csname LT2\endcsname{\color{black}}%
      \expandafter\def\csname LT3\endcsname{\color{black}}%
      \expandafter\def\csname LT4\endcsname{\color{black}}%
      \expandafter\def\csname LT5\endcsname{\color{black}}%
      \expandafter\def\csname LT6\endcsname{\color{black}}%
      \expandafter\def\csname LT7\endcsname{\color{black}}%
      \expandafter\def\csname LT8\endcsname{\color{black}}%
    \fi
  \fi
  \setlength{\unitlength}{0.0500bp}%
  \begin{picture}(7200.00,5040.00)%
    \gplgaddtomacro\gplbacktext{%
      \csname LTb\endcsname%
      \put(1188,660){\makebox(0,0)[r]{\strut{} 0}}%
      \csname LTb\endcsname%
      \put(1188,1700){\makebox(0,0)[r]{\strut{} 1}}%
      \csname LTb\endcsname%
      \put(1188,2740){\makebox(0,0)[r]{\strut{} 2}}%
      \csname LTb\endcsname%
      \put(1188,3779){\makebox(0,0)[r]{\strut{} 3}}%
      \csname LTb\endcsname%
      \put(1188,4819){\makebox(0,0)[r]{\strut{} 4}}%
      \csname LTb\endcsname%
      \put(2434,440){\makebox(0,0){\strut{} 100}}%
      \csname LTb\endcsname%
      \put(3559,440){\makebox(0,0){\strut{} 200}}%
      \csname LTb\endcsname%
      \put(4684,440){\makebox(0,0){\strut{} 300}}%
      \csname LTb\endcsname%
      \put(5810,440){\makebox(0,0){\strut{} 400}}%
      \csname LTb\endcsname%
      \put(6935,440){\makebox(0,0){\strut{} 500}}%
      \put(682,2739){\rotatebox{-270}{\makebox(0,0){\strut{}Energy Gain}}}%
      \put(4127,110){\makebox(0,0){\strut{}Correlation Length, $n^\star$ [time slots]}}%
    }%
    \gplgaddtomacro\gplfronttext{%
      \csname LTb\endcsname%
      \put(3245,4646){\makebox(0,0)[r]{\strut{}M-AAR (p=2)}}%
      \csname LTb\endcsname%
      \put(3245,4426){\makebox(0,0)[r]{\strut{}M-AAR (p=4)}}%
      \csname LTb\endcsname%
      \put(3245,4206){\makebox(0,0)[r]{\strut{}PR (p=2)}}%
      \csname LTb\endcsname%
      \put(3245,3986){\makebox(0,0)[r]{\strut{}PR (p=2)}}%
      \csname LTb\endcsname%
      \put(3245,3766){\makebox(0,0)[r]{\strut{}LTC}}%
      \csname LTb\endcsname%
      \put(5948,4646){\makebox(0,0)[r]{\strut{}E-PLAMLiS}}%
      \csname LTb\endcsname%
      \put(5948,4426){\makebox(0,0)[r]{\strut{}FFT Windowed}}%
      \csname LTb\endcsname%
      \put(5948,4206){\makebox(0,0)[r]{\strut{}DCT-LPF}}%
      \csname LTb\endcsname%
      \put(5948,3986){\makebox(0,0)[r]{\strut{}No compression}}%
    }%
    \gplbacktext
    \put(0,0){\includegraphics{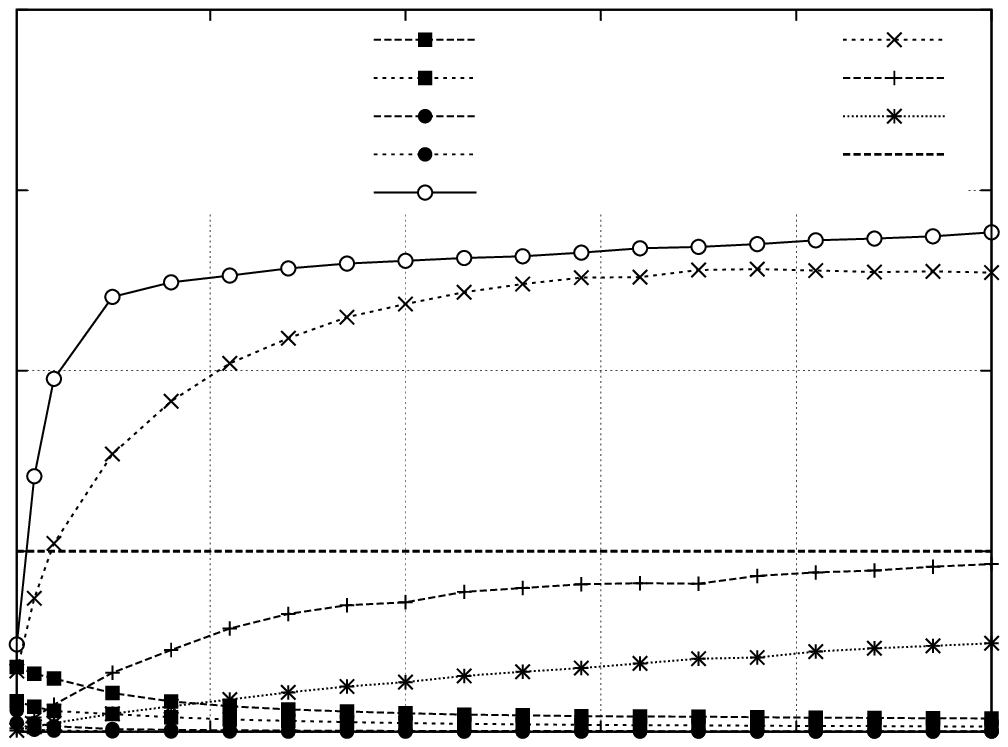}}%
    \gplfronttext
  \end{picture}%
\endgroup

%% file: fig_4b.tex
% GNUPLOT: LaTeX picture with Postscript
\begingroup
  \makeatletter
  \providecommand\color[2][]{%
    \GenericError{(gnuplot) \space\space\space\@spaces}{%
      Package color not loaded in conjunction with
      terminal option `colourtext'%
    }{See the gnuplot documentation for explanation.%
    }{Either use 'blacktext' in gnuplot or load the package
      color.sty in LaTeX.}%
    \renewcommand\color[2][]{}%
  }%
  \providecommand\includegraphics[2][]{%
    \GenericError{(gnuplot) \space\space\space\@spaces}{%
      Package graphicx or graphics not loaded%
    }{See the gnuplot documentation for explanation.%
    }{The gnuplot epslatex terminal needs graphicx.sty or graphics.sty.}%
    \renewcommand\includegraphics[2][]{}%
  }%
  \providecommand\rotatebox[2]{#2}%
  \@ifundefined{ifGPcolor}{%
    \newif\ifGPcolor
    \GPcolorfalse
  }{}%
  \@ifundefined{ifGPblacktext}{%
    \newif\ifGPblacktext
    \GPblacktexttrue
  }{}%
  % define a \g@addto@macro without @ in the name:
  \let\gplgaddtomacro\g@addto@macro
  % define empty templates for all commands taking text:
  \gdef\gplbacktext{}%
  \gdef\gplfronttext{}%
  \makeatother
  \ifGPblacktext
    % no textcolor at all
    \def\colorrgb#1{}%
    \def\colorgray#1{}%
  \else
    % gray or color?
    \ifGPcolor
      \def\colorrgb#1{\color[rgb]{#1}}%
      \def\colorgray#1{\color[gray]{#1}}%
      \expandafter\def\csname LTw\endcsname{\color{white}}%
      \expandafter\def\csname LTb\endcsname{\color{black}}%
      \expandafter\def\csname LTa\endcsname{\color{black}}%
      \expandafter\def\csname LT0\endcsname{\color[rgb]{1,0,0}}%
      \expandafter\def\csname LT1\endcsname{\color[rgb]{0,1,0}}%
      \expandafter\def\csname LT2\endcsname{\color[rgb]{0,0,1}}%
      \expandafter\def\csname LT3\endcsname{\color[rgb]{1,0,1}}%
      \expandafter\def\csname LT4\endcsname{\color[rgb]{0,1,1}}%
      \expandafter\def\csname LT5\endcsname{\color[rgb]{1,1,0}}%
      \expandafter\def\csname LT6\endcsname{\color[rgb]{0,0,0}}%
      \expandafter\def\csname LT7\endcsname{\color[rgb]{1,0.3,0}}%
      \expandafter\def\csname LT8\endcsname{\color[rgb]{0.5,0.5,0.5}}%
    \else
      % gray
      \def\colorrgb#1{\color{black}}%
      \def\colorgray#1{\color[gray]{#1}}%
      \expandafter\def\csname LTw\endcsname{\color{white}}%
      \expandafter\def\csname LTb\endcsname{\color{black}}%
      \expandafter\def\csname LTa\endcsname{\color{black}}%
      \expandafter\def\csname LT0\endcsname{\color{black}}%
      \expandafter\def\csname LT1\endcsname{\color{black}}%
      \expandafter\def\csname LT2\endcsname{\color{black}}%
      \expandafter\def\csname LT3\endcsname{\color{black}}%
      \expandafter\def\csname LT4\endcsname{\color{black}}%
      \expandafter\def\csname LT5\endcsname{\color{black}}%
      \expandafter\def\csname LT6\endcsname{\color{black}}%
      \expandafter\def\csname LT7\endcsname{\color{black}}%
      \expandafter\def\csname LT8\endcsname{\color{black}}%
    \fi
  \fi
  \setlength{\unitlength}{0.0500bp}%
  \begin{picture}(7200.00,5040.00)%
    \gplgaddtomacro\gplbacktext{%
      \csname LTb\endcsname%
      \put(1188,660){\makebox(0,0)[r]{\strut{} 0}}%
      \csname LTb\endcsname%
      \put(1188,1098){\makebox(0,0)[r]{\strut{} 10}}%
      \csname LTb\endcsname%
      \put(1188,1536){\makebox(0,0)[r]{\strut{} 20}}%
      \csname LTb\endcsname%
      \put(1188,1973){\makebox(0,0)[r]{\strut{} 30}}%
      \csname LTb\endcsname%
      \put(1188,2411){\makebox(0,0)[r]{\strut{} 40}}%
      \csname LTb\endcsname%
      \put(1188,2849){\makebox(0,0)[r]{\strut{} 50}}%
      \csname LTb\endcsname%
      \put(1188,3287){\makebox(0,0)[r]{\strut{} 60}}%
      \csname LTb\endcsname%
      \put(1188,3725){\makebox(0,0)[r]{\strut{} 70}}%
      \csname LTb\endcsname%
      \put(1188,4162){\makebox(0,0)[r]{\strut{} 80}}%
      \csname LTb\endcsname%
      \put(1188,4600){\makebox(0,0)[r]{\strut{} 90}}%
      \csname LTb\endcsname%
      \put(2434,440){\makebox(0,0){\strut{} 100}}%
      \csname LTb\endcsname%
      \put(3559,440){\makebox(0,0){\strut{} 200}}%
      \csname LTb\endcsname%
      \put(4684,440){\makebox(0,0){\strut{} 300}}%
      \csname LTb\endcsname%
      \put(5810,440){\makebox(0,0){\strut{} 400}}%
      \csname LTb\endcsname%
      \put(6935,440){\makebox(0,0){\strut{} 500}}%
      \put(550,2739){\rotatebox{-270}{\makebox(0,0){\strut{}Energy Gain}}}%
      \put(4127,110){\makebox(0,0){\strut{}Correlation Length, $n^\star$ [time slots]}}%
    }%
    \gplgaddtomacro\gplfronttext{%
      \csname LTb\endcsname%
      \put(3245,4646){\makebox(0,0)[r]{\strut{}M-AAR (p=2)}}%
      \csname LTb\endcsname%
      \put(3245,4426){\makebox(0,0)[r]{\strut{}M-AAR (p=4)}}%
      \csname LTb\endcsname%
      \put(3245,4206){\makebox(0,0)[r]{\strut{}PR (p=2)}}%
      \csname LTb\endcsname%
      \put(3245,3986){\makebox(0,0)[r]{\strut{}PR (p=2)}}%
      \csname LTb\endcsname%
      \put(3245,3766){\makebox(0,0)[r]{\strut{}LTC}}%
      \csname LTb\endcsname%
      \put(5948,4646){\makebox(0,0)[r]{\strut{}E-PLAMLiS}}%
      \csname LTb\endcsname%
      \put(5948,4426){\makebox(0,0)[r]{\strut{}FFT Windowed}}%
      \csname LTb\endcsname%
      \put(5948,4206){\makebox(0,0)[r]{\strut{}DCT-LPF}}%
      \csname LTb\endcsname%
      \put(5948,3986){\makebox(0,0)[r]{\strut{}No compression}}%
    }%
    \gplbacktext
    \put(0,0){\includegraphics{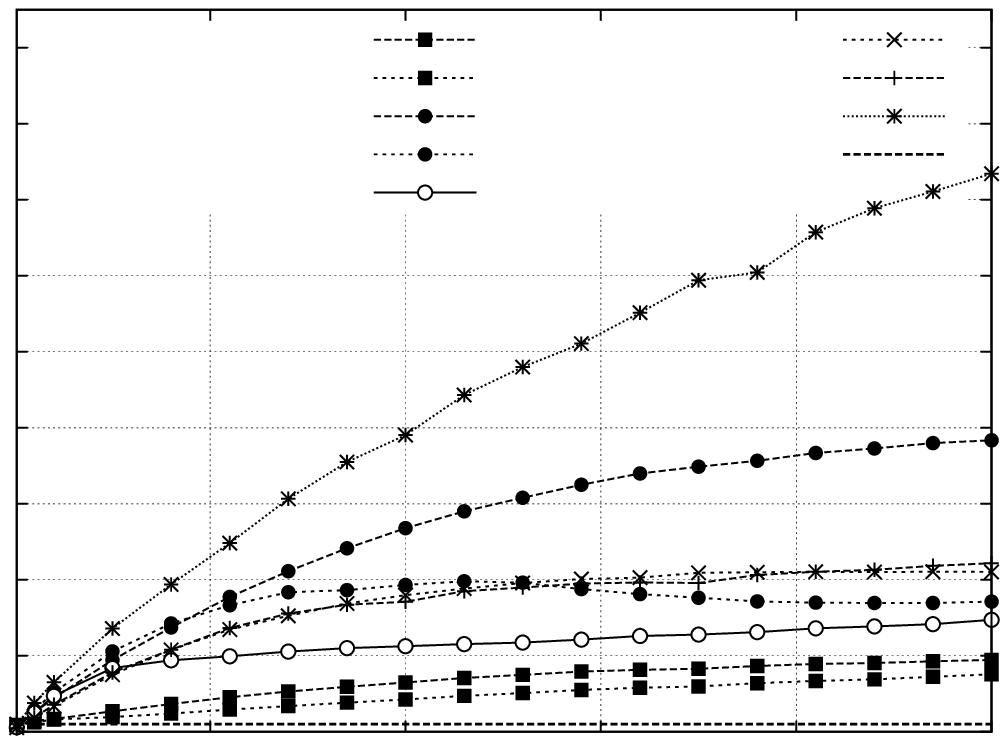}}%
    \gplfronttext
  \end{picture}%
\endgroup

%% file: fig_6a.tex
% GNUPLOT: LaTeX picture with Postscript
\begingroup
  \makeatletter
  \providecommand\color[2][]{%
    \GenericError{(gnuplot) \space\space\space\@spaces}{%
      Package color not loaded in conjunction with
      terminal option `colourtext'%
    }{See the gnuplot documentation for explanation.%
    }{Either use 'blacktext' in gnuplot or load the package
      color.sty in LaTeX.}%
    \renewcommand\color[2][]{}%
  }%
  \providecommand\includegraphics[2][]{%
    \GenericError{(gnuplot) \space\space\space\@spaces}{%
      Package graphicx or graphics not loaded%
    }{See the gnuplot documentation for explanation.%
    }{The gnuplot epslatex terminal needs graphicx.sty or graphics.sty.}%
    \renewcommand\includegraphics[2][]{}%
  }%
  \providecommand\rotatebox[2]{#2}%
  \@ifundefined{ifGPcolor}{%
    \newif\ifGPcolor
    \GPcolorfalse
  }{}%
  \@ifundefined{ifGPblacktext}{%
    \newif\ifGPblacktext
    \GPblacktexttrue
  }{}%
  % define a \g@addto@macro without @ in the name:
  \let\gplgaddtomacro\g@addto@macro
  % define empty templates for all commands taking text:
  \gdef\gplbacktext{}%
  \gdef\gplfronttext{}%
  \makeatother
  \ifGPblacktext
    % no textcolor at all
    \def\colorrgb#1{}%
    \def\colorgray#1{}%
  \else
    % gray or color?
    \ifGPcolor
      \def\colorrgb#1{\color[rgb]{#1}}%
      \def\colorgray#1{\color[gray]{#1}}%
      \expandafter\def\csname LTw\endcsname{\color{white}}%
      \expandafter\def\csname LTb\endcsname{\color{black}}%
      \expandafter\def\csname LTa\endcsname{\color{black}}%
      \expandafter\def\csname LT0\endcsname{\color[rgb]{1,0,0}}%
      \expandafter\def\csname LT1\endcsname{\color[rgb]{0,1,0}}%
      \expandafter\def\csname LT2\endcsname{\color[rgb]{0,0,1}}%
      \expandafter\def\csname LT3\endcsname{\color[rgb]{1,0,1}}%
      \expandafter\def\csname LT4\endcsname{\color[rgb]{0,1,1}}%
      \expandafter\def\csname LT5\endcsname{\color[rgb]{1,1,0}}%
      \expandafter\def\csname LT6\endcsname{\color[rgb]{0,0,0}}%
      \expandafter\def\csname LT7\endcsname{\color[rgb]{1,0.3,0}}%
      \expandafter\def\csname LT8\endcsname{\color[rgb]{0.5,0.5,0.5}}%
    \else
      % gray
      \def\colorrgb#1{\color{black}}%
      \def\colorgray#1{\color[gray]{#1}}%
      \expandafter\def\csname LTw\endcsname{\color{white}}%
      \expandafter\def\csname LTb\endcsname{\color{black}}%
      \expandafter\def\csname LTa\endcsname{\color{black}}%
      \expandafter\def\csname LT0\endcsname{\color{black}}%
      \expandafter\def\csname LT1\endcsname{\color{black}}%
      \expandafter\def\csname LT2\endcsname{\color{black}}%
      \expandafter\def\csname LT3\endcsname{\color{black}}%
      \expandafter\def\csname LT4\endcsname{\color{black}}%
      \expandafter\def\csname LT5\endcsname{\color{black}}%
      \expandafter\def\csname LT6\endcsname{\color{black}}%
      \expandafter\def\csname LT7\endcsname{\color{black}}%
      \expandafter\def\csname LT8\endcsname{\color{black}}%
    \fi
  \fi
  \setlength{\unitlength}{0.0500bp}%
  \begin{picture}(7200.00,5040.00)%
    \gplgaddtomacro\gplbacktext{%
      \csname LTb\endcsname%
      \put(1188,660){\makebox(0,0)[r]{\strut{} 0}}%
      \csname LTb\endcsname%
      \put(1188,1122){\makebox(0,0)[r]{\strut{} 2}}%
      \csname LTb\endcsname%
      \put(1188,1584){\makebox(0,0)[r]{\strut{} 4}}%
      \csname LTb\endcsname%
      \put(1188,2046){\makebox(0,0)[r]{\strut{} 6}}%
      \csname LTb\endcsname%
      \put(1188,2508){\makebox(0,0)[r]{\strut{} 8}}%
      \csname LTb\endcsname%
      \put(1188,2971){\makebox(0,0)[r]{\strut{} 10}}%
      \csname LTb\endcsname%
      \put(1188,3433){\makebox(0,0)[r]{\strut{} 12}}%
      \csname LTb\endcsname%
      \put(1188,3895){\makebox(0,0)[r]{\strut{} 14}}%
      \csname LTb\endcsname%
      \put(1188,4357){\makebox(0,0)[r]{\strut{} 16}}%
      \csname LTb\endcsname%
      \put(1188,4819){\makebox(0,0)[r]{\strut{} 18}}%
      \csname LTb\endcsname%
      \put(1320,440){\makebox(0,0){\strut{} 1}}%
      \csname LTb\endcsname%
      \put(2122,440){\makebox(0,0){\strut{} 2}}%
      \csname LTb\endcsname%
      \put(2924,440){\makebox(0,0){\strut{} 3}}%
      \csname LTb\endcsname%
      \put(3726,440){\makebox(0,0){\strut{} 4}}%
      \csname LTb\endcsname%
      \put(4529,440){\makebox(0,0){\strut{} 5}}%
      \csname LTb\endcsname%
      \put(5331,440){\makebox(0,0){\strut{} 6}}%
      \csname LTb\endcsname%
      \put(6133,440){\makebox(0,0){\strut{} 7}}%
      \csname LTb\endcsname%
      \put(6935,440){\makebox(0,0){\strut{} 8}}%
      \put(550,2739){\rotatebox{-270}{\makebox(0,0){\strut{}Energy Gain}}}%
      \put(4127,110){\makebox(0,0){\strut{}Number of hops}}%
    }%
    \gplgaddtomacro\gplfronttext{%
      \csname LTb\endcsname%
      \put(2981,4646){\makebox(0,0)[r]{\strut{}LTC ($\xi = 3\sigma$)}}%
      \csname LTb\endcsname%
      \put(2981,4426){\makebox(0,0)[r]{\strut{}LTC ($\xi = 4\sigma$)}}%
      \csname LTb\endcsname%
      \put(2981,4206){\makebox(0,0)[r]{\strut{}LTC ($\xi = 5\sigma$)}}%
      \csname LTb\endcsname%
      \put(5948,4646){\makebox(0,0)[r]{\strut{}DCT-LPF ($\xi = 3\sigma$)}}%
      \csname LTb\endcsname%
      \put(5948,4426){\makebox(0,0)[r]{\strut{}DCT-LPF ($\xi = 4\sigma$)}}%
      \csname LTb\endcsname%
      \put(5948,4206){\makebox(0,0)[r]{\strut{}DCT-LPF ($\xi = 5\sigma$)}}%
    }%
    \gplbacktext
    \put(0,0){\includegraphics{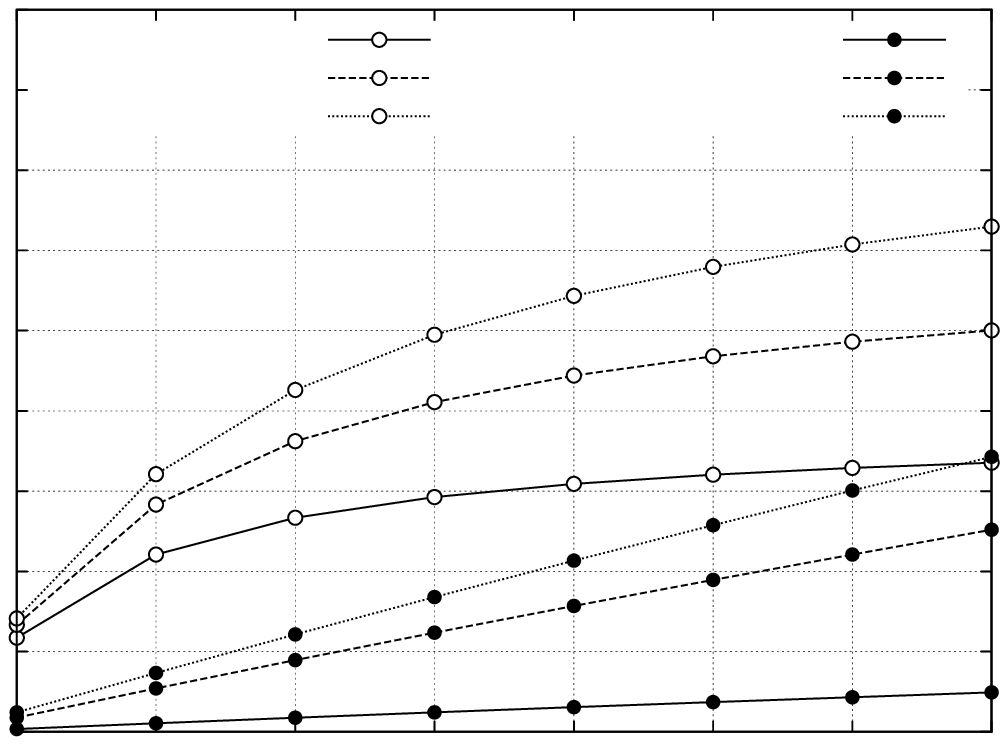}}%
    \gplfronttext
  \end{picture}%
\endgroup

%% file: fig_6b.tex
% GNUPLOT: LaTeX picture with Postscript
\begingroup
  \makeatletter
  \providecommand\color[2][]{%
    \GenericError{(gnuplot) \space\space\space\@spaces}{%
      Package color not loaded in conjunction with
      terminal option `colourtext'%
    }{See the gnuplot documentation for explanation.%
    }{Either use 'blacktext' in gnuplot or load the package
      color.sty in LaTeX.}%
    \renewcommand\color[2][]{}%
  }%
  \providecommand\includegraphics[2][]{%
    \GenericError{(gnuplot) \space\space\space\@spaces}{%
      Package graphicx or graphics not loaded%
    }{See the gnuplot documentation for explanation.%
    }{The gnuplot epslatex terminal needs graphicx.sty or graphics.sty.}%
    \renewcommand\includegraphics[2][]{}%
  }%
  \providecommand\rotatebox[2]{#2}%
  \@ifundefined{ifGPcolor}{%
    \newif\ifGPcolor
    \GPcolorfalse
  }{}%
  \@ifundefined{ifGPblacktext}{%
    \newif\ifGPblacktext
    \GPblacktexttrue
  }{}%
  % define a \g@addto@macro without @ in the name:
  \let\gplgaddtomacro\g@addto@macro
  % define empty templates for all commands taking text:
  \gdef\gplbacktext{}%
  \gdef\gplfronttext{}%
  \makeatother
  \ifGPblacktext
    % no textcolor at all
    \def\colorrgb#1{}%
    \def\colorgray#1{}%
  \else
    % gray or color?
    \ifGPcolor
      \def\colorrgb#1{\color[rgb]{#1}}%
      \def\colorgray#1{\color[gray]{#1}}%
      \expandafter\def\csname LTw\endcsname{\color{white}}%
      \expandafter\def\csname LTb\endcsname{\color{black}}%
      \expandafter\def\csname LTa\endcsname{\color{black}}%
      \expandafter\def\csname LT0\endcsname{\color[rgb]{1,0,0}}%
      \expandafter\def\csname LT1\endcsname{\color[rgb]{0,1,0}}%
      \expandafter\def\csname LT2\endcsname{\color[rgb]{0,0,1}}%
      \expandafter\def\csname LT3\endcsname{\color[rgb]{1,0,1}}%
      \expandafter\def\csname LT4\endcsname{\color[rgb]{0,1,1}}%
      \expandafter\def\csname LT5\endcsname{\color[rgb]{1,1,0}}%
      \expandafter\def\csname LT6\endcsname{\color[rgb]{0,0,0}}%
      \expandafter\def\csname LT7\endcsname{\color[rgb]{1,0.3,0}}%
      \expandafter\def\csname LT8\endcsname{\color[rgb]{0.5,0.5,0.5}}%
    \else
      % gray
      \def\colorrgb#1{\color{black}}%
      \def\colorgray#1{\color[gray]{#1}}%
      \expandafter\def\csname LTw\endcsname{\color{white}}%
      \expandafter\def\csname LTb\endcsname{\color{black}}%
      \expandafter\def\csname LTa\endcsname{\color{black}}%
      \expandafter\def\csname LT0\endcsname{\color{black}}%
      \expandafter\def\csname LT1\endcsname{\color{black}}%
      \expandafter\def\csname LT2\endcsname{\color{black}}%
      \expandafter\def\csname LT3\endcsname{\color{black}}%
      \expandafter\def\csname LT4\endcsname{\color{black}}%
      \expandafter\def\csname LT5\endcsname{\color{black}}%
      \expandafter\def\csname LT6\endcsname{\color{black}}%
      \expandafter\def\csname LT7\endcsname{\color{black}}%
      \expandafter\def\csname LT8\endcsname{\color{black}}%
    \fi
  \fi
  \setlength{\unitlength}{0.0500bp}%
  \begin{picture}(7200.00,5040.00)%
    \gplgaddtomacro\gplbacktext{%
      \csname LTb\endcsname%
      \put(1188,660){\makebox(0,0)[r]{\strut{} 0}}%
      \csname LTb\endcsname%
      \put(1188,1492){\makebox(0,0)[r]{\strut{} 20}}%
      \csname LTb\endcsname%
      \put(1188,2324){\makebox(0,0)[r]{\strut{} 40}}%
      \csname LTb\endcsname%
      \put(1188,3155){\makebox(0,0)[r]{\strut{} 60}}%
      \csname LTb\endcsname%
      \put(1188,3987){\makebox(0,0)[r]{\strut{} 80}}%
      \csname LTb\endcsname%
      \put(1188,4819){\makebox(0,0)[r]{\strut{} 100}}%
      \csname LTb\endcsname%
      \put(1320,440){\makebox(0,0){\strut{} 1}}%
      \csname LTb\endcsname%
      \put(2122,440){\makebox(0,0){\strut{} 2}}%
      \csname LTb\endcsname%
      \put(2924,440){\makebox(0,0){\strut{} 3}}%
      \csname LTb\endcsname%
      \put(3726,440){\makebox(0,0){\strut{} 4}}%
      \csname LTb\endcsname%
      \put(4529,440){\makebox(0,0){\strut{} 5}}%
      \csname LTb\endcsname%
      \put(5331,440){\makebox(0,0){\strut{} 6}}%
      \csname LTb\endcsname%
      \put(6133,440){\makebox(0,0){\strut{} 7}}%
      \csname LTb\endcsname%
      \put(6935,440){\makebox(0,0){\strut{} 8}}%
      \put(418,2739){\rotatebox{-270}{\makebox(0,0){\strut{}Energy Gain}}}%
      \put(4127,110){\makebox(0,0){\strut{}Number of hops}}%
    }%
    \gplgaddtomacro\gplfronttext{%
      \csname LTb\endcsname%
      \put(2981,4646){\makebox(0,0)[r]{\strut{}LTC ($\xi = 3\sigma$)}}%
      \csname LTb\endcsname%
      \put(2981,4426){\makebox(0,0)[r]{\strut{}LTC ($\xi = 4\sigma$)}}%
      \csname LTb\endcsname%
      \put(2981,4206){\makebox(0,0)[r]{\strut{}LTC ($\xi = 5\sigma$)}}%
      \csname LTb\endcsname%
      \put(5948,4646){\makebox(0,0)[r]{\strut{}DCT-LPF ($\xi = 3\sigma$)}}%
      \csname LTb\endcsname%
      \put(5948,4426){\makebox(0,0)[r]{\strut{}DCT-LPF ($\xi = 4\sigma$)}}%
      \csname LTb\endcsname%
      \put(5948,4206){\makebox(0,0)[r]{\strut{}DCT-LPF ($\xi = 5\sigma$)}}%
    }%
    \gplbacktext
    \put(0,0){\includegraphics{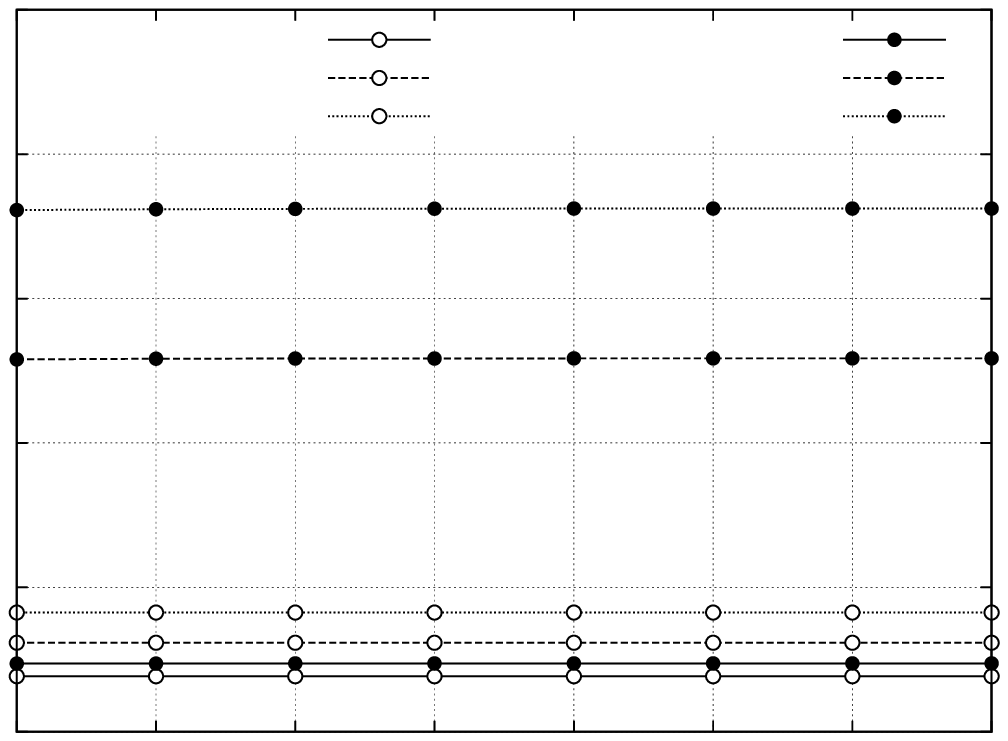}}%
    \gplfronttext
  \end{picture}%
\endgroup

%% file: fig_ltc_err_fit.tex
% GNUPLOT: LaTeX picture with Postscript
\begingroup
  \makeatletter
  \providecommand\color[2][]{%
    \GenericError{(gnuplot) \space\space\space\@spaces}{%
      Package color not loaded in conjunction with
      terminal option `colourtext'%
    }{See the gnuplot documentation for explanation.%
    }{Either use 'blacktext' in gnuplot or load the package
      color.sty in LaTeX.}%
    \renewcommand\color[2][]{}%
  }%
  \providecommand\includegraphics[2][]{%
    \GenericError{(gnuplot) \space\space\space\@spaces}{%
      Package graphicx or graphics not loaded%
    }{See the gnuplot documentation for explanation.%
    }{The gnuplot epslatex terminal needs graphicx.sty or graphics.sty.}%
    \renewcommand\includegraphics[2][]{}%
  }%
  \providecommand\rotatebox[2]{#2}%
  \@ifundefined{ifGPcolor}{%
    \newif\ifGPcolor
    \GPcolorfalse
  }{}%
  \@ifundefined{ifGPblacktext}{%
    \newif\ifGPblacktext
    \GPblacktexttrue
  }{}%
  % define a \g@addto@macro without @ in the name:
  \let\gplgaddtomacro\g@addto@macro
  % define empty templates for all commands taking text:
  \gdef\gplbacktext{}%
  \gdef\gplfronttext{}%
  \makeatother
  \ifGPblacktext
    % no textcolor at all
    \def\colorrgb#1{}%
    \def\colorgray#1{}%
  \else
    % gray or color?
    \ifGPcolor
      \def\colorrgb#1{\color[rgb]{#1}}%
      \def\colorgray#1{\color[gray]{#1}}%
      \expandafter\def\csname LTw\endcsname{\color{white}}%
      \expandafter\def\csname LTb\endcsname{\color{black}}%
      \expandafter\def\csname LTa\endcsname{\color{black}}%
      \expandafter\def\csname LT0\endcsname{\color[rgb]{1,0,0}}%
      \expandafter\def\csname LT1\endcsname{\color[rgb]{0,1,0}}%
      \expandafter\def\csname LT2\endcsname{\color[rgb]{0,0,1}}%
      \expandafter\def\csname LT3\endcsname{\color[rgb]{1,0,1}}%
      \expandafter\def\csname LT4\endcsname{\color[rgb]{0,1,1}}%
      \expandafter\def\csname LT5\endcsname{\color[rgb]{1,1,0}}%
      \expandafter\def\csname LT6\endcsname{\color[rgb]{0,0,0}}%
      \expandafter\def\csname LT7\endcsname{\color[rgb]{1,0.3,0}}%
      \expandafter\def\csname LT8\endcsname{\color[rgb]{0.5,0.5,0.5}}%
    \else
      % gray
      \def\colorrgb#1{\color{black}}%
      \def\colorgray#1{\color[gray]{#1}}%
      \expandafter\def\csname LTw\endcsname{\color{white}}%
      \expandafter\def\csname LTb\endcsname{\color{black}}%
      \expandafter\def\csname LTa\endcsname{\color{black}}%
      \expandafter\def\csname LT0\endcsname{\color{black}}%
      \expandafter\def\csname LT1\endcsname{\color{black}}%
      \expandafter\def\csname LT2\endcsname{\color{black}}%
      \expandafter\def\csname LT3\endcsname{\color{black}}%
      \expandafter\def\csname LT4\endcsname{\color{black}}%
      \expandafter\def\csname LT5\endcsname{\color{black}}%
      \expandafter\def\csname LT6\endcsname{\color{black}}%
      \expandafter\def\csname LT7\endcsname{\color{black}}%
      \expandafter\def\csname LT8\endcsname{\color{black}}%
    \fi
  \fi
  \setlength{\unitlength}{0.0500bp}%
  \begin{picture}(7200.00,5040.00)%
    \gplgaddtomacro\gplbacktext{%
      \csname LTb\endcsname%
      \put(1188,660){\makebox(0,0)[r]{\strut{} 0}}%
      \csname LTb\endcsname%
      \put(1188,1492){\makebox(0,0)[r]{\strut{} 1}}%
      \csname LTb\endcsname%
      \put(1188,2324){\makebox(0,0)[r]{\strut{} 2}}%
      \csname LTb\endcsname%
      \put(1188,3155){\makebox(0,0)[r]{\strut{} 3}}%
      \csname LTb\endcsname%
      \put(1188,3987){\makebox(0,0)[r]{\strut{} 4}}%
      \csname LTb\endcsname%
      \put(1188,4819){\makebox(0,0)[r]{\strut{} 5}}%
      \csname LTb\endcsname%
      \put(1320,440){\makebox(0,0){\strut{} 0}}%
      \csname LTb\endcsname%
      \put(2724,440){\makebox(0,0){\strut{} 0.5}}%
      \csname LTb\endcsname%
      \put(4128,440){\makebox(0,0){\strut{} 1}}%
      \csname LTb\endcsname%
      \put(5531,440){\makebox(0,0){\strut{} 1.5}}%
      \csname LTb\endcsname%
      \put(6935,440){\makebox(0,0){\strut{} 2}}%
      \put(682,2739){\rotatebox{-270}{\makebox(0,0){\strut{}Relative Error, $\xi$}}}%
      \put(4127,110){\makebox(0,0){\strut{}Compression Ratio, $\eta$}}%
      \put(3005,3987){\makebox(0,0)[l]{\strut{}decreasing $n^\star$}}%
    }%
    \gplgaddtomacro\gplfronttext{%
      \csname LTb\endcsname%
      \put(5948,4646){\makebox(0,0)[r]{\strut{}$n^\star = 10$}}%
      \csname LTb\endcsname%
      \put(5948,4426){\makebox(0,0)[r]{\strut{}$n^\star = 20$}}%
      \csname LTb\endcsname%
      \put(5948,4206){\makebox(0,0)[r]{\strut{}$n^\star = 50$}}%
      \csname LTb\endcsname%
      \put(5948,3986){\makebox(0,0)[r]{\strut{}$n^\star = 80$}}%
      \csname LTb\endcsname%
      \put(5948,3766){\makebox(0,0)[r]{\strut{}$n^\star = 110$}}%
      \csname LTb\endcsname%
      \put(5948,3546){\makebox(0,0)[r]{\strut{}$n^\star = 290$}}%
      \csname LTb\endcsname%
      \put(5948,3326){\makebox(0,0)[r]{\strut{}$n^\star = 500$}}%
      \csname LTb\endcsname%
      \put(5948,3106){\makebox(0,0)[r]{\strut{}Temp dataset}}%
      \csname LTb\endcsname%
      \put(5948,2886){\makebox(0,0)[r]{\strut{}Hum dataset}}%
    }%
    \gplbacktext
    \put(0,0){\includegraphics{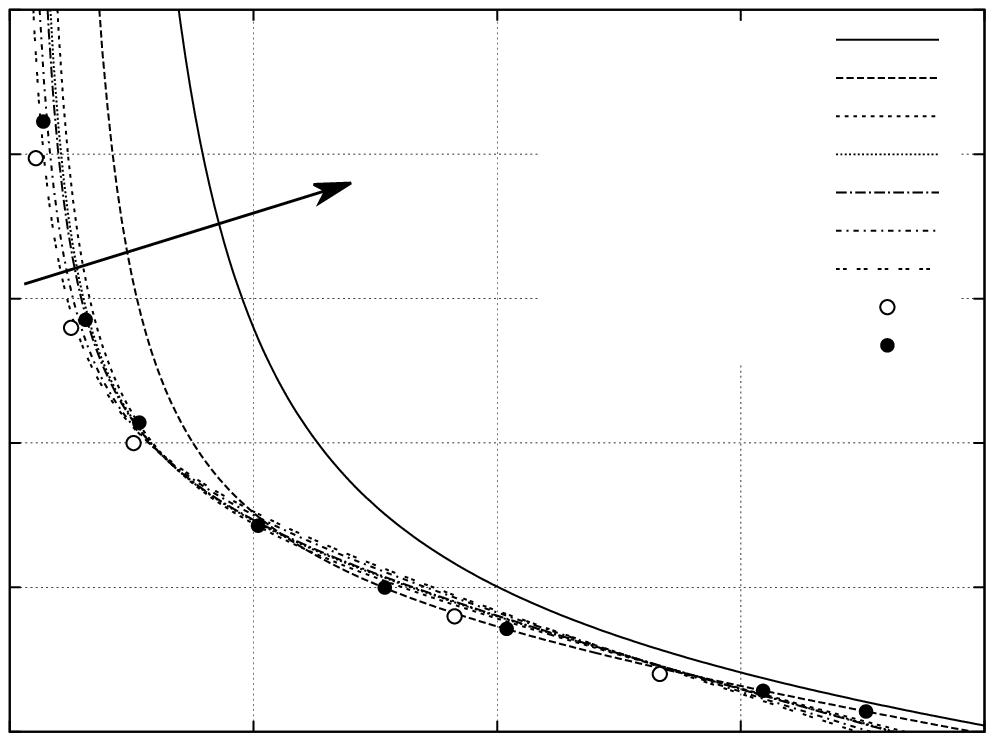}}%
    \gplfronttext
  \end{picture}%
\endgroup

%% file: fig_dct_err_fit.tex
% GNUPLOT: LaTeX picture with Postscript
\begingroup
  \makeatletter
  \providecommand\color[2][]{%
    \GenericError{(gnuplot) \space\space\space\@spaces}{%
      Package color not loaded in conjunction with
      terminal option `colourtext'%
    }{See the gnuplot documentation for explanation.%
    }{Either use 'blacktext' in gnuplot or load the package
      color.sty in LaTeX.}%
    \renewcommand\color[2][]{}%
  }%
  \providecommand\includegraphics[2][]{%
    \GenericError{(gnuplot) \space\space\space\@spaces}{%
      Package graphicx or graphics not loaded%
    }{See the gnuplot documentation for explanation.%
    }{The gnuplot epslatex terminal needs graphicx.sty or graphics.sty.}%
    \renewcommand\includegraphics[2][]{}%
  }%
  \providecommand\rotatebox[2]{#2}%
  \@ifundefined{ifGPcolor}{%
    \newif\ifGPcolor
    \GPcolorfalse
  }{}%
  \@ifundefined{ifGPblacktext}{%
    \newif\ifGPblacktext
    \GPblacktexttrue
  }{}%
  % define a \g@addto@macro without @ in the name:
  \let\gplgaddtomacro\g@addto@macro
  % define empty templates for all commands taking text:
  \gdef\gplbacktext{}%
  \gdef\gplfronttext{}%
  \makeatother
  \ifGPblacktext
    % no textcolor at all
    \def\colorrgb#1{}%
    \def\colorgray#1{}%
  \else
    % gray or color?
    \ifGPcolor
      \def\colorrgb#1{\color[rgb]{#1}}%
      \def\colorgray#1{\color[gray]{#1}}%
      \expandafter\def\csname LTw\endcsname{\color{white}}%
      \expandafter\def\csname LTb\endcsname{\color{black}}%
      \expandafter\def\csname LTa\endcsname{\color{black}}%
      \expandafter\def\csname LT0\endcsname{\color[rgb]{1,0,0}}%
      \expandafter\def\csname LT1\endcsname{\color[rgb]{0,1,0}}%
      \expandafter\def\csname LT2\endcsname{\color[rgb]{0,0,1}}%
      \expandafter\def\csname LT3\endcsname{\color[rgb]{1,0,1}}%
      \expandafter\def\csname LT4\endcsname{\color[rgb]{0,1,1}}%
      \expandafter\def\csname LT5\endcsname{\color[rgb]{1,1,0}}%
      \expandafter\def\csname LT6\endcsname{\color[rgb]{0,0,0}}%
      \expandafter\def\csname LT7\endcsname{\color[rgb]{1,0.3,0}}%
      \expandafter\def\csname LT8\endcsname{\color[rgb]{0.5,0.5,0.5}}%
    \else
      % gray
      \def\colorrgb#1{\color{black}}%
      \def\colorgray#1{\color[gray]{#1}}%
      \expandafter\def\csname LTw\endcsname{\color{white}}%
      \expandafter\def\csname LTb\endcsname{\color{black}}%
      \expandafter\def\csname LTa\endcsname{\color{black}}%
      \expandafter\def\csname LT0\endcsname{\color{black}}%
      \expandafter\def\csname LT1\endcsname{\color{black}}%
      \expandafter\def\csname LT2\endcsname{\color{black}}%
      \expandafter\def\csname LT3\endcsname{\color{black}}%
      \expandafter\def\csname LT4\endcsname{\color{black}}%
      \expandafter\def\csname LT5\endcsname{\color{black}}%
      \expandafter\def\csname LT6\endcsname{\color{black}}%
      \expandafter\def\csname LT7\endcsname{\color{black}}%
      \expandafter\def\csname LT8\endcsname{\color{black}}%
    \fi
  \fi
  \setlength{\unitlength}{0.0500bp}%
  \begin{picture}(7200.00,5040.00)%
    \gplgaddtomacro\gplbacktext{%
      \csname LTb\endcsname%
      \put(1188,660){\makebox(0,0)[r]{\strut{} 0}}%
      \csname LTb\endcsname%
      \put(1188,1492){\makebox(0,0)[r]{\strut{} 1}}%
      \csname LTb\endcsname%
      \put(1188,2324){\makebox(0,0)[r]{\strut{} 2}}%
      \csname LTb\endcsname%
      \put(1188,3155){\makebox(0,0)[r]{\strut{} 3}}%
      \csname LTb\endcsname%
      \put(1188,3987){\makebox(0,0)[r]{\strut{} 4}}%
      \csname LTb\endcsname%
      \put(1188,4819){\makebox(0,0)[r]{\strut{} 5}}%
      \csname LTb\endcsname%
      \put(1320,440){\makebox(0,0){\strut{} 0}}%
      \csname LTb\endcsname%
      \put(2443,440){\makebox(0,0){\strut{} 0.2}}%
      \csname LTb\endcsname%
      \put(3566,440){\makebox(0,0){\strut{} 0.4}}%
      \csname LTb\endcsname%
      \put(4689,440){\makebox(0,0){\strut{} 0.6}}%
      \csname LTb\endcsname%
      \put(5812,440){\makebox(0,0){\strut{} 0.8}}%
      \csname LTb\endcsname%
      \put(6935,440){\makebox(0,0){\strut{} 1}}%
      \put(682,2739){\rotatebox{-270}{\makebox(0,0){\strut{}Relative Error, $\xi$}}}%
      \put(4127,110){\makebox(0,0){\strut{}Compression Ratio, $\eta$}}%
      \put(3285,3987){\makebox(0,0)[l]{\strut{}decreasing $n^\star$}}%
    }%
    \gplgaddtomacro\gplfronttext{%
      \csname LTb\endcsname%
      \put(5948,4646){\makebox(0,0)[r]{\strut{}$n^\star = 10$}}%
      \csname LTb\endcsname%
      \put(5948,4426){\makebox(0,0)[r]{\strut{}$n^\star = 20$}}%
      \csname LTb\endcsname%
      \put(5948,4206){\makebox(0,0)[r]{\strut{}$n^\star = 50$}}%
      \csname LTb\endcsname%
      \put(5948,3986){\makebox(0,0)[r]{\strut{}$n^\star = 80$}}%
      \csname LTb\endcsname%
      \put(5948,3766){\makebox(0,0)[r]{\strut{}$n^\star = 110$}}%
      \csname LTb\endcsname%
      \put(5948,3546){\makebox(0,0)[r]{\strut{}$n^\star = 290$}}%
      \csname LTb\endcsname%
      \put(5948,3326){\makebox(0,0)[r]{\strut{}$n^\star = 500$}}%
      \csname LTb\endcsname%
      \put(5948,3106){\makebox(0,0)[r]{\strut{}Temp dataset}}%
      \csname LTb\endcsname%
      \put(5948,2886){\makebox(0,0)[r]{\strut{}Hum dataset}}%
    }%
    \gplbacktext
    \put(0,0){\includegraphics{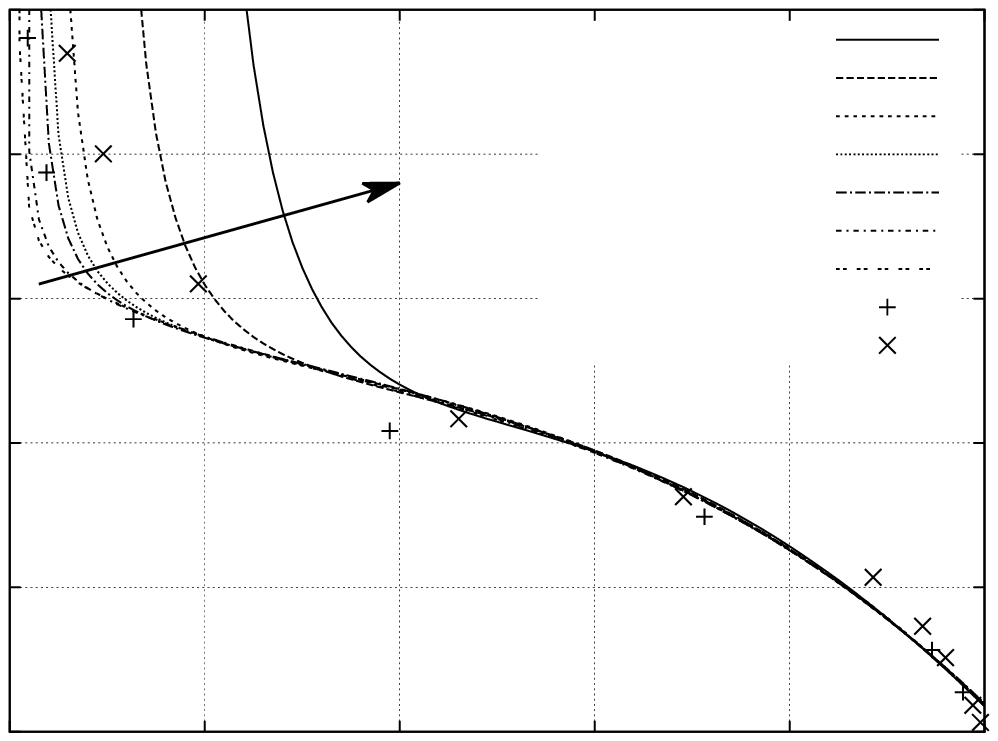}}%
    \gplfronttext
  \end{picture}%
\endgroup

%% file: fig_fit_b.tex
% GNUPLOT: LaTeX picture with Postscript
\begingroup
  \makeatletter
  \providecommand\color[2][]{%
    \GenericError{(gnuplot) \space\space\space\@spaces}{%
      Package color not loaded in conjunction with
      terminal option `colourtext'%
    }{See the gnuplot documentation for explanation.%
    }{Either use 'blacktext' in gnuplot or load the package
      color.sty in LaTeX.}%
    \renewcommand\color[2][]{}%
  }%
  \providecommand\includegraphics[2][]{%
    \GenericError{(gnuplot) \space\space\space\@spaces}{%
      Package graphicx or graphics not loaded%
    }{See the gnuplot documentation for explanation.%
    }{The gnuplot epslatex terminal needs graphicx.sty or graphics.sty.}%
    \renewcommand\includegraphics[2][]{}%
  }%
  \providecommand\rotatebox[2]{#2}%
  \@ifundefined{ifGPcolor}{%
    \newif\ifGPcolor
    \GPcolorfalse
  }{}%
  \@ifundefined{ifGPblacktext}{%
    \newif\ifGPblacktext
    \GPblacktexttrue
  }{}%
  % define a \g@addto@macro without @ in the name:
  \let\gplgaddtomacro\g@addto@macro
  % define empty templates for all commands taking text:
  \gdef\gplbacktext{}%
  \gdef\gplfronttext{}%
  \makeatother
  \ifGPblacktext
    % no textcolor at all
    \def\colorrgb#1{}%
    \def\colorgray#1{}%
  \else
    % gray or color?
    \ifGPcolor
      \def\colorrgb#1{\color[rgb]{#1}}%
      \def\colorgray#1{\color[gray]{#1}}%
      \expandafter\def\csname LTw\endcsname{\color{white}}%
      \expandafter\def\csname LTb\endcsname{\color{black}}%
      \expandafter\def\csname LTa\endcsname{\color{black}}%
      \expandafter\def\csname LT0\endcsname{\color[rgb]{1,0,0}}%
      \expandafter\def\csname LT1\endcsname{\color[rgb]{0,1,0}}%
      \expandafter\def\csname LT2\endcsname{\color[rgb]{0,0,1}}%
      \expandafter\def\csname LT3\endcsname{\color[rgb]{1,0,1}}%
      \expandafter\def\csname LT4\endcsname{\color[rgb]{0,1,1}}%
      \expandafter\def\csname LT5\endcsname{\color[rgb]{1,1,0}}%
      \expandafter\def\csname LT6\endcsname{\color[rgb]{0,0,0}}%
      \expandafter\def\csname LT7\endcsname{\color[rgb]{1,0.3,0}}%
      \expandafter\def\csname LT8\endcsname{\color[rgb]{0.5,0.5,0.5}}%
    \else
      % gray
      \def\colorrgb#1{\color{black}}%
      \def\colorgray#1{\color[gray]{#1}}%
      \expandafter\def\csname LTw\endcsname{\color{white}}%
      \expandafter\def\csname LTb\endcsname{\color{black}}%
      \expandafter\def\csname LTa\endcsname{\color{black}}%
      \expandafter\def\csname LT0\endcsname{\color{black}}%
      \expandafter\def\csname LT1\endcsname{\color{black}}%
      \expandafter\def\csname LT2\endcsname{\color{black}}%
      \expandafter\def\csname LT3\endcsname{\color{black}}%
      \expandafter\def\csname LT4\endcsname{\color{black}}%
      \expandafter\def\csname LT5\endcsname{\color{black}}%
      \expandafter\def\csname LT6\endcsname{\color{black}}%
      \expandafter\def\csname LT7\endcsname{\color{black}}%
      \expandafter\def\csname LT8\endcsname{\color{black}}%
    \fi
  \fi
  \setlength{\unitlength}{0.0500bp}%
  \begin{picture}(7200.00,5040.00)%
    \gplgaddtomacro\gplbacktext{%
      \csname LTb\endcsname%
      \put(1188,660){\makebox(0,0)[r]{\strut{}$10^{1}$}}%
      \csname LTb\endcsname%
      \put(1188,1353){\makebox(0,0)[r]{\strut{}$10^{2}$}}%
      \csname LTb\endcsname%
      \put(1188,2046){\makebox(0,0)[r]{\strut{}$10^{3}$}}%
      \csname LTb\endcsname%
      \put(1188,2740){\makebox(0,0)[r]{\strut{}$10^{4}$}}%
      \csname LTb\endcsname%
      \put(1188,3433){\makebox(0,0)[r]{\strut{}$10^{5}$}}%
      \csname LTb\endcsname%
      \put(1188,4126){\makebox(0,0)[r]{\strut{}$10^{6}$}}%
      \csname LTb\endcsname%
      \put(1188,4819){\makebox(0,0)[r]{\strut{}$10^{7}$}}%
      \csname LTb\endcsname%
      \put(1320,440){\makebox(0,0){\strut{} 0}}%
      \csname LTb\endcsname%
      \put(1882,440){\makebox(0,0){\strut{} 0.2}}%
      \csname LTb\endcsname%
      \put(2443,440){\makebox(0,0){\strut{} 0.4}}%
      \csname LTb\endcsname%
      \put(3005,440){\makebox(0,0){\strut{} 0.6}}%
      \csname LTb\endcsname%
      \put(3566,440){\makebox(0,0){\strut{} 0.8}}%
      \csname LTb\endcsname%
      \put(4128,440){\makebox(0,0){\strut{} 1}}%
      \csname LTb\endcsname%
      \put(4689,440){\makebox(0,0){\strut{} 1.2}}%
      \csname LTb\endcsname%
      \put(5251,440){\makebox(0,0){\strut{} 1.4}}%
      \csname LTb\endcsname%
      \put(5812,440){\makebox(0,0){\strut{} 1.6}}%
      \csname LTb\endcsname%
      \put(6373,440){\makebox(0,0){\strut{} 1.8}}%
      \csname LTb\endcsname%
      \put(6935,440){\makebox(0,0){\strut{} 2}}%
      \put(550,2739){\rotatebox{-270}{\makebox(0,0){\strut{}Number of Clock Cycles per bit, $N_c$}}}%
      \put(4127,110){\makebox(0,0){\strut{}Compression Ratio, $\eta$}}%
    }%
    \gplgaddtomacro\gplfronttext{%
      \csname LTb\endcsname%
      \put(5948,4646){\makebox(0,0)[r]{\strut{}LTC (Temp dataset)}}%
      \csname LTb\endcsname%
      \put(5948,4426){\makebox(0,0)[r]{\strut{}LTC (Hum dataset)}}%
      \csname LTb\endcsname%
      \put(5948,4206){\makebox(0,0)[r]{\strut{}$N_{c}$ (LTC)}}%
      \csname LTb\endcsname%
      \put(5948,3986){\makebox(0,0)[r]{\strut{}DCT-LPF (Temp dataset)}}%
      \csname LTb\endcsname%
      \put(5948,3766){\makebox(0,0)[r]{\strut{}DCT-LPF (Hum dataset)}}%
      \csname LTb\endcsname%
      \put(5948,3546){\makebox(0,0)[r]{\strut{}$N_{c}$ (DCT-LPF)}}%
    }%
    \gplbacktext
    \put(0,0){\includegraphics{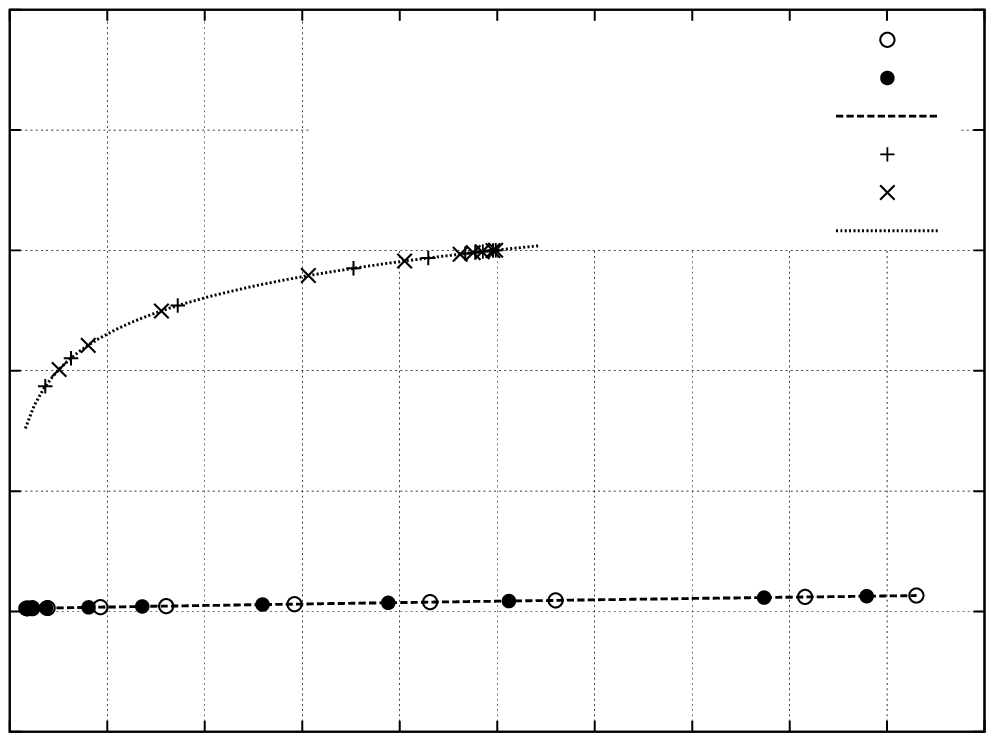}}%
    \gplfronttext
  \end{picture}%
\endgroup

%% file: fig_toff.tex
% GNUPLOT: LaTeX picture with Postscript
\begingroup
  \makeatletter
  \providecommand\color[2][]{%
    \GenericError{(gnuplot) \space\space\space\@spaces}{%
      Package color not loaded in conjunction with
      terminal option `colourtext'%
    }{See the gnuplot documentation for explanation.%
    }{Either use 'blacktext' in gnuplot or load the package
      color.sty in LaTeX.}%
    \renewcommand\color[2][]{}%
  }%
  \providecommand\includegraphics[2][]{%
    \GenericError{(gnuplot) \space\space\space\@spaces}{%
      Package graphicx or graphics not loaded%
    }{See the gnuplot documentation for explanation.%
    }{The gnuplot epslatex terminal needs graphicx.sty or graphics.sty.}%
    \renewcommand\includegraphics[2][]{}%
  }%
  \providecommand\rotatebox[2]{#2}%
  \@ifundefined{ifGPcolor}{%
    \newif\ifGPcolor
    \GPcolorfalse
  }{}%
  \@ifundefined{ifGPblacktext}{%
    \newif\ifGPblacktext
    \GPblacktexttrue
  }{}%
  % define a \g@addto@macro without @ in the name:
  \let\gplgaddtomacro\g@addto@macro
  % define empty templates for all commands taking text:
  \gdef\gplbacktext{}%
  \gdef\gplfronttext{}%
  \makeatother
  \ifGPblacktext
    % no textcolor at all
    \def\colorrgb#1{}%
    \def\colorgray#1{}%
  \else
    % gray or color?
    \ifGPcolor
      \def\colorrgb#1{\color[rgb]{#1}}%
      \def\colorgray#1{\color[gray]{#1}}%
      \expandafter\def\csname LTw\endcsname{\color{white}}%
      \expandafter\def\csname LTb\endcsname{\color{black}}%
      \expandafter\def\csname LTa\endcsname{\color{black}}%
      \expandafter\def\csname LT0\endcsname{\color[rgb]{1,0,0}}%
      \expandafter\def\csname LT1\endcsname{\color[rgb]{0,1,0}}%
      \expandafter\def\csname LT2\endcsname{\color[rgb]{0,0,1}}%
      \expandafter\def\csname LT3\endcsname{\color[rgb]{1,0,1}}%
      \expandafter\def\csname LT4\endcsname{\color[rgb]{0,1,1}}%
      \expandafter\def\csname LT5\endcsname{\color[rgb]{1,1,0}}%
      \expandafter\def\csname LT6\endcsname{\color[rgb]{0,0,0}}%
      \expandafter\def\csname LT7\endcsname{\color[rgb]{1,0.3,0}}%
      \expandafter\def\csname LT8\endcsname{\color[rgb]{0.5,0.5,0.5}}%
    \else
      % gray
      \def\colorrgb#1{\color{black}}%
      \def\colorgray#1{\color[gray]{#1}}%
      \expandafter\def\csname LTw\endcsname{\color{white}}%
      \expandafter\def\csname LTb\endcsname{\color{black}}%
      \expandafter\def\csname LTa\endcsname{\color{black}}%
      \expandafter\def\csname LT0\endcsname{\color{black}}%
      \expandafter\def\csname LT1\endcsname{\color{black}}%
      \expandafter\def\csname LT2\endcsname{\color{black}}%
      \expandafter\def\csname LT3\endcsname{\color{black}}%
      \expandafter\def\csname LT4\endcsname{\color{black}}%
      \expandafter\def\csname LT5\endcsname{\color{black}}%
      \expandafter\def\csname LT6\endcsname{\color{black}}%
      \expandafter\def\csname LT7\endcsname{\color{black}}%
      \expandafter\def\csname LT8\endcsname{\color{black}}%
    \fi
  \fi
  \setlength{\unitlength}{0.0500bp}%
  \begin{picture}(7200.00,5040.00)%
    \gplgaddtomacro\gplbacktext{%
      \csname LTb\endcsname%
      \put(1188,660){\makebox(0,0)[r]{\strut{}$10^{0}$}}%
      \csname LTb\endcsname%
      \put(1188,1584){\makebox(0,0)[r]{\strut{}$10^{2}$}}%
      \csname LTb\endcsname%
      \put(1188,2508){\makebox(0,0)[r]{\strut{}$10^{4}$}}%
      \csname LTb\endcsname%
      \put(1188,3433){\makebox(0,0)[r]{\strut{}$10^{6}$}}%
      \csname LTb\endcsname%
      \put(1188,4357){\makebox(0,0)[r]{\strut{}$10^{8}$}}%
      \csname LTb\endcsname%
      \put(1320,440){\makebox(0,0){\strut{} 0}}%
      \csname LTb\endcsname%
      \put(2443,440){\makebox(0,0){\strut{} 0.2}}%
      \csname LTb\endcsname%
      \put(3566,440){\makebox(0,0){\strut{} 0.4}}%
      \csname LTb\endcsname%
      \put(4689,440){\makebox(0,0){\strut{} 0.6}}%
      \csname LTb\endcsname%
      \put(5812,440){\makebox(0,0){\strut{} 0.8}}%
      \csname LTb\endcsname%
      \put(6935,440){\makebox(0,0){\strut{} 1}}%
      \put(550,2739){\rotatebox{-270}{\makebox(0,0){\strut{}$E_{Tx}^\prime/E_0$}}}%
      \put(4127,110){\makebox(0,0){\strut{}Compression Ratio, $\eta$}}%
    }%
    \gplgaddtomacro\gplfronttext{%
      \csname LTb\endcsname%
      \put(2849,4646){\makebox(0,0)[r]{\strut{}CC2420 }}%
      \csname LTb\endcsname%
      \put(2849,4426){\makebox(0,0)[r]{\strut{}AquaModem}}%
      \csname LTb\endcsname%
      \put(5948,4646){\makebox(0,0)[r]{\strut{}$N_c{\rm (LTC)}/(1-\eta)$ }}%
      \csname LTb\endcsname%
      \put(5948,4426){\makebox(0,0)[r]{\strut{}$N_c{\rm (DCT-LPF)}/(1-\eta)$}}%
    }%
    \gplbacktext
    \put(0,0){\includegraphics{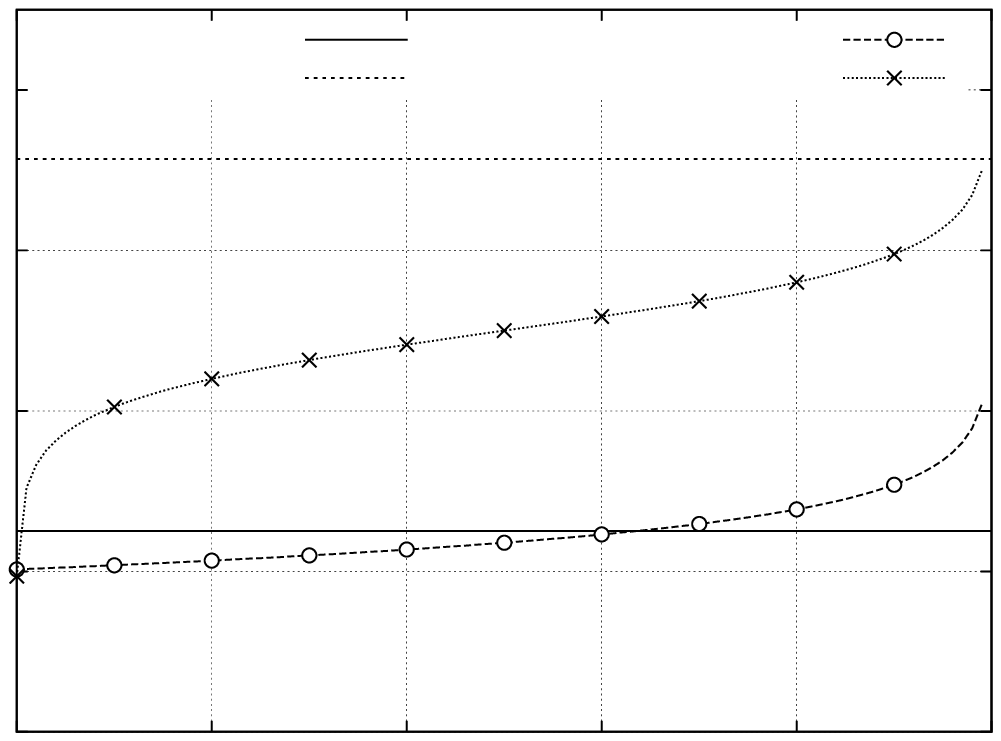}}%
    \gplfronttext
  \end{picture}%
\endgroup

%% file: conclusion.tex
\section{Conclusions}
\label{sec:conclusions}

In this paper we have systematically compared lossy compression algorithms for constrained sensor networking, by investigating whether energy savings are possible depending on signal statistics, compression performance and hardware characteristics. Our results revealed that, for wireless transmission scenarios, the energy required by compression algorithms is of the same order of magnitude of that spent in the transmission at the physical layer. In this case, the only class of algorithms that provides some energy savings is that based on piecewise linear approximations, as these algorithms have the smaller computational cost. In addition, we have also considered underwater channels which, due to the nature of the transmission medium, require more energy demanding acoustic modems. In this scenario, techniques based on Fourier transforms are the algorithms of choice, as these provide the highest compression performance. Finally, we have obtained fitting formulas for the best compression methods to relate their computational complexity, approximation accuracy and compression ratio performance. These have been validated against real datasets and can be used to assess the effectiveness of the selected compression schemes for further hardware architectures.

%% file: biographies.tex
\vspace{0.3cm}
\parpic{\includegraphics[width=1in,clip,keepaspectratio]{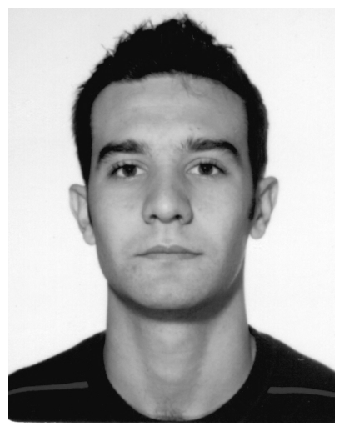}}
\noindent {\bf Davide Zordan} received the B.Sc. and the M.Sc. in Telecommunications Engineering from the University of Padova (Italy) in 2007 and 2010, respectively. He is currently a Ph.D. student at the University of Padova, under the supervision of Michele Rossi. His research interests include in network processing techniques for WSNs (including Compressive Sensing), stochastic modeling and optimization, protocol design and performance evaluation for wireless networks, energy efficient protocols for WSNs and energy harvesting techniques.

\vspace{0.3cm}
\parpic{\includegraphics[width=1in,clip,keepaspectratio]{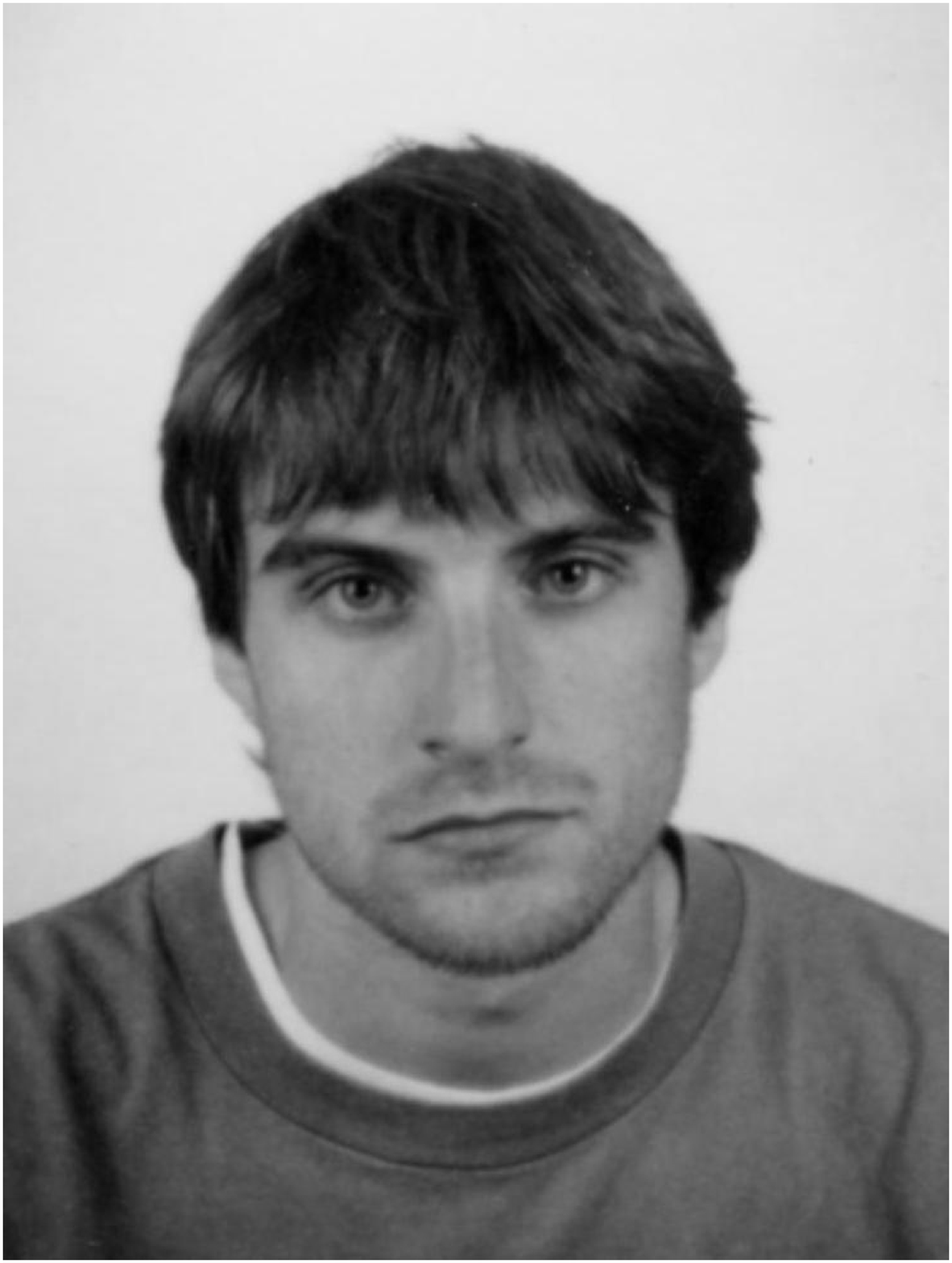}}
\noindent {\bf Borja Martinez} received the B.Sc. degree in Physics in 2000 and the B.Sc. degree in Electronics Engineering in 2003, both from the Universitat Aut\`onoma de Barcelona. In 2007, he obtained the M.Sc. degree in Microelectronics and Electronics Systems, whereas from 2004 to 2010 he has been a member of the Technology Transfer Group of the Microelectronics Department of the same university. During this period, Borja participated in several projects related to Wireless Smart Devices. Close cooperation with industrial partners, allowed him to acquire a strong background on robust algorithm design and firmware implementation. Recently, he started his Ph.D. on data compression techniques, focusing on techniques for constrained Wireless Sensor Networks. In 2010, he joined WordSensing with the purpose of implementing algorithms in real sensor network deployments. Since 2007 Borja has been Assistant Professor at the High School of Engineering of the Universitat Aut\`onoma de Barcelona.

\vspace{0.3cm}
\parpic{\includegraphics[width=1in,clip,keepaspectratio]{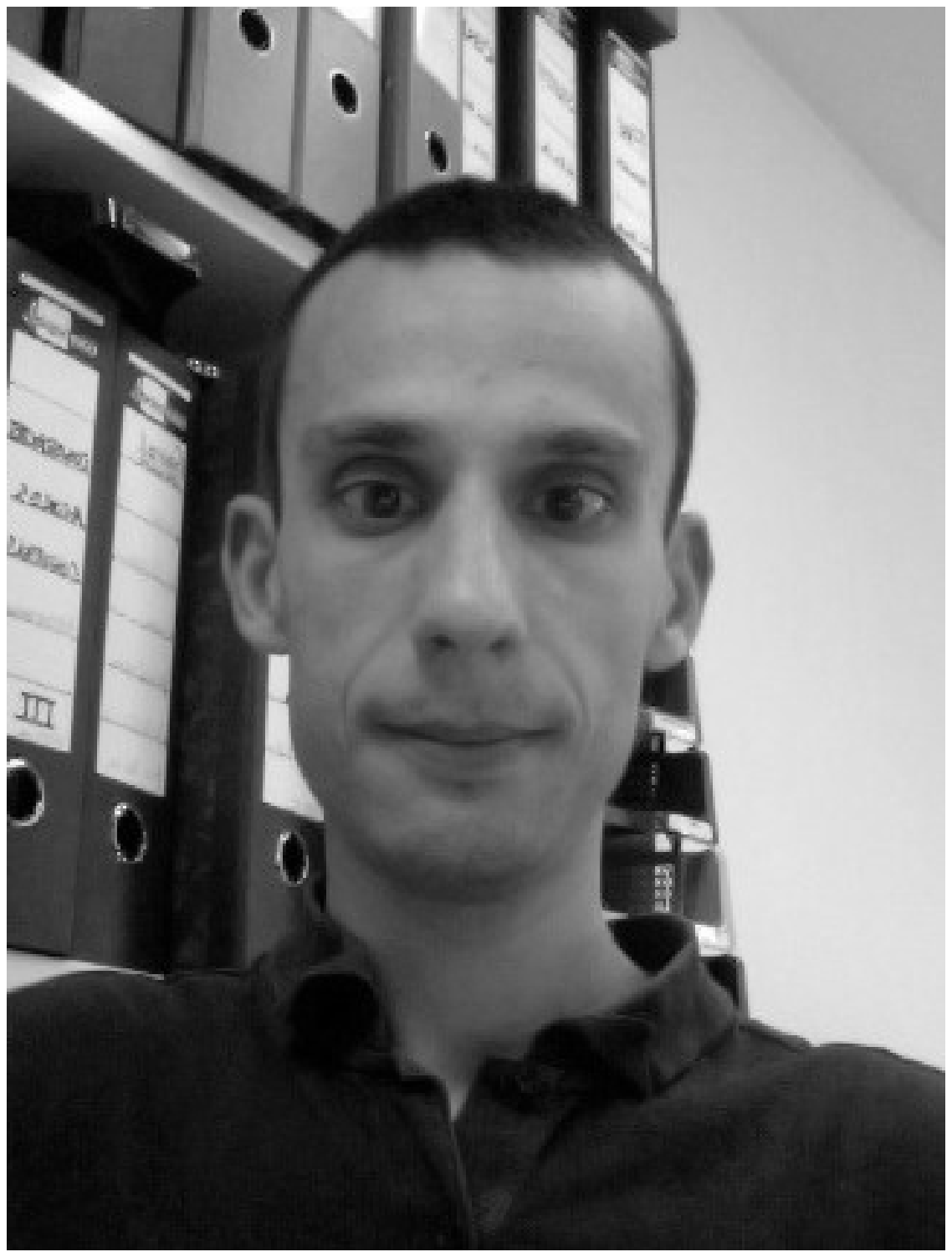}}
\noindent {\bf Ignasi Vilajosana}, CEO of Worldsensing, obtained the Ph.D. in Physics in 2008. He is author of more than 20 high impact publications in the geophysical field. He also has a strong background in signal processing techniques and experience in project management. He has been a visiting researcher at the ``Norwegian Geothecnical Institute'' (Oslo, 2007) and at the ``Swiss Federal Institute For Snow and Avalanche Research'' (Davos, Switzerland, 2008). From 2008 to 2009, he has been professor at the Politechnical University of Calalonia (UPC).

\vspace{0.3cm}
\parpic{\includegraphics[width=1in,clip,keepaspectratio]{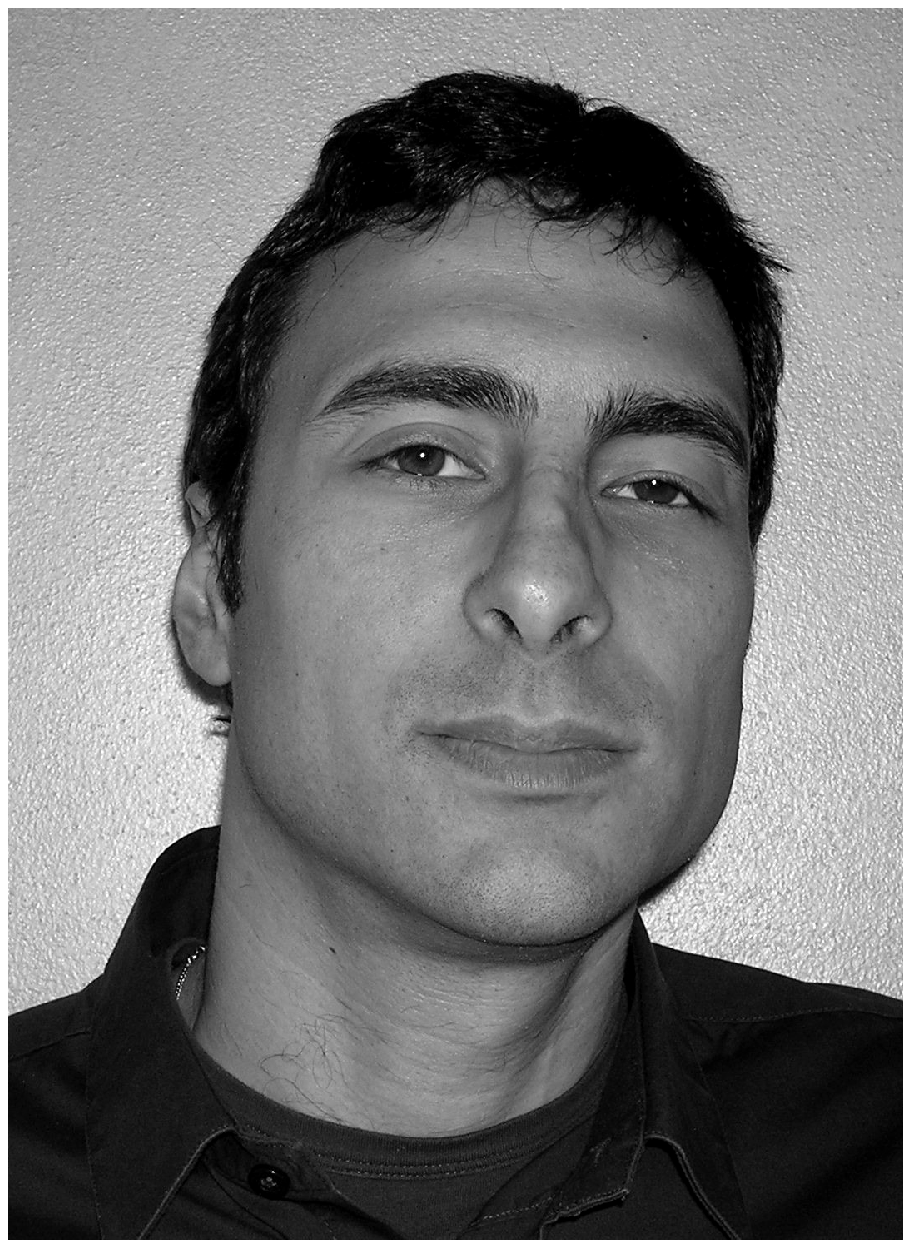}}
\noindent {\bf Michele Rossi} received the Laurea degree in Electrical Engineering and the Ph.D.  degree in Information Engineering from the University of Ferrara in 2000 and 2004, respectively. From March 2000 to October 2005 he has been a Research Fellow at the Department of Engineering of the University of Ferrara. During 2003 he was on leave at the Center for Wireless Communications (CWC) at the University of California San Diego (UCSD), where he performed research on wireless sensor networks. In November 2005 he joined the Department of Information Engineering of the University of Padova, Italy, where he is an Assistant Professor. Dr. Rossi is currently involved in the EU-funded IoT-A and SWAP projects, both on wireless sensor networking. His research interests are centered around the dissemination of data in distributed ad hoc and wireless sensor networks, including protocol design through stochastic optimization, integrated MAC/routing schemes, compression techniques for wireless sensor networks and cooperative routing policies for ad hoc networks. Dr. Rossi currently serves on the Editorial Board of the IEEE TRANSACTIONS ON WIRELESS COMMUNICATIONS.